\documentclass[12pt, a4paper]{report}

\usepackage{amsmath}
\usepackage{graphicx}
\usepackage{a4wide}
\usepackage[colorlinks, breaklinks=true, bookmarks=true]{hyperref}

\newcommand{\sech}{\mathrm{sech}}
\newcommand{\Real}{\mathrm{Re}}
\newcommand{\Imag}{\mathrm{Im}}

\newcommand{\deriv}[2]{\frac{d#1}{d#2}}
\newcommand{\nderiv}[3]{\frac{d^#3#1}{{d#2}^#3}}
\newcommand{\pderiv}[2]{\frac{\partial#1}{\partial#2}}

\newcommand{\pderivdd}[3]{\frac{\partial^2#1}{\partial#2\partial#3}}
\newcommand{\fnlderiv}[2]{\frac{\delta#1}{\delta#2}}
\newcommand{\fnlderivdd}[3]{\frac{\delta^2#1}{\delta#2\delta#3}}

\newenvironment{bsmatrix}{\left[\begin{smallmatrix}}{\end{smallmatrix}\right]}

\newcommand{\BigFig}[1]{\parbox{12pt}{\Huge #1}}
\newcommand{\BigZero}{\BigFig{0}}
\newcommand{\ket}[1]{\left|#1\right>}
\newcommand{\bra}[1]{\left<#1\right|}
\newcommand{\eigs}[1]{\left|\varphi_{#1}\right>}
\newcommand{\bracketsO}[3]{\left<#1 \left|#2 \right|#3 \right>}
\newcommand{\bracketsObigm}[3]{\left<#1\! \bigm|\!#2 \!\bigm|\!#3 \right>}
\newcommand{\bracketsObiggm}[3]{\left<#1\! \biggm|\!#2 \!\biggm|\!#3 \right>}
\newcommand{\brackets}[2]{\left<#1 | #2\right>}

\newcommand{\bracketsbiggm}[2]{\left<#1 \biggm| #2\right>}
\newcommand{\exval}[1]{\left<#1 \right>}
\newcommand{\operator}[1]{\mathbf{\hat#1}}
\newcommand{\commut}[2]{\left[#1, #2\right]}

\newcommand{\etal}{\textit{et al.}\ }
\newcommand{\ie}{i.\,e.\ }
\newcommand{\eg}{e.\,g.\ }

\newcommand{\epsw}{\bar{\epsilon}(\omega)}
\newcommand{\etaw}{\bar{\eta}(\omega)}
\newcommand{\muw}{\overline{\exval{\operator{\mu}}}(\omega)}
\newcommand{\Ow}{\overline{\exval{\operator{O}}}(\omega)}
\newcommand{\Oaw}{\overline{\exval{\operator{O}_a}}(\omega)}

\newcommand{\feps}{f_{\epsilon}(\omega)}
\newcommand{\tfeps}{\tilde{f}_{\epsilon}(\omega)}
\newcommand{\fmu}{f_{\mu}(\omega)}
\newcommand{\tfmu}{\tilde{f}_{\mu}(\omega)}
\newcommand{\fO}{f_{O}(\omega)}
\newcommand{\tfO}{\tilde{f}_{O}(\omega)}

\newcommand{\evmu}{\exval{\operator{\mu}}\!(t)}

\newcommand{\opq}[1]{\operator{q}^{(#1)}}
\newcommand{\opdo}[1]{\operator{d}_o^{(#1)}}
\newcommand{\opdO}[1]{\operator{d}_O^{(#1)}}
\newcommand{\opO}[1]{\operator{O}^{(#1)}}
\newcommand{\opmu}[1]{\operator{\mu}^{(#1)}}
\newcommand{\muwn}[1]{\overline{\exval{\operator{\mu}^{(#1)}}}(\omega)}

\begin{document}

	\title{Quantum Optimal Control Theory\\
	of Harmonic Generation\\
	\vspace{0.5cm}
	\Large{Master's Thesis}\\
	\vspace{0.7cm}
	\large{Fritz Haber Center\\
The Institute of Chemistry\\
The Hebrew University of Jerusalem}}
	\author
	{\emph{Supervisor:} \\
	Prof.~Ronnie Kosloff \\
	ronnie@fh.huji.ac.il
	\and
	\emph{Author:} \\
	Ido Schaefer\\
	ido.schaefer@mail.huji.ac.il}
	\vspace{0.7cm}
	\date
	{{Submitted:}\\
	March 29, 2012\\
	{Revised:}\\
	\today}
	\maketitle

	\begin{abstract}
A new method for controlling harmonic generation, in the framework of quantum optimal control theory (QOCT), is developed. The problem is formulated in the frequency domain using a new maximization functional. The relaxation method is used as the optimization procedure. The new formulation is generalized to other control problems with requirements in the frequency domain. The method is applied to several simple problems. The results are analysed and discussed. General conclusions on harmonic generation mechanisms are obtained.	
	\end{abstract}

	\tableofcontents
	\chapter*{Preface}\label{pref}
%
%
\emph{General remark}: the eigenstates of the Hamiltonian, in the various problems, will be denoted by $\ket{\varphi_n}$; the index $n$ represents the eigenstate with the eigenenergy $E_n$, where:
\[
	E_i<E_{i+1} \qquad i=0,1,\ldots
\]

\emph{General remarks for all numerical results:}
\begin{itemize}
	\item Atomic units are used throughout.
	\item The important details of the problem and the computational process are presented in a table. Most of the notations are defined in the text. The meaning of all notations is described in the following table: 
\end{itemize}

\begin{center}
\renewcommand{\arraystretch}{1.5}
\begin{tabular}{|c||c|}	\hline
	\textbf{Notation} & \textbf{Description} \\ \hline \hline
	$\operator{H}_0$ & the unperturbed Hamiltonian \\ \hline
	$\operator{\mu}$ & the dipole moment operator \\ \hline
	$\ket{\psi_0}$ & the initial state vector \\ \hline
	$\psi_0(x)$ &  the initial wave function in the $x$ domain \\ \hline			
	$\ket{\phi}$ & the target state vector \\ \hline
	$\phi(x)$ & the target wave function in the $x$ domain\\ \hline
	$T$ & the final time \\ \hline
	$\alpha$ & the penalty factor \\ \hline
	$\tfeps$ & the filter function of the forcing field\\ \hline
	$\tfmu$ & the filter function of the dipole moment expectation value \\ \hline
	$\epsilon^{0}(t)$ & the initial guess of the field, in the time domain \\ \hline
	$\bar{\epsilon}^{0}(\omega)$ & the initial guess of the field, in the frequency domain \\ \hline
	$L$ & $n=L$ is the index of the maximal allowed eigenstate \\ \hline
	$\gamma_n$ & the penalty factor of the forbidden state $\ket{\varphi_n}$ \\ \hline
	$\kappa$ & the penalty factor of $\left(\deriv{\exval{\operator{\mu}}\!(T)}{t}\right)^2$ \\ \hline
	$K_i$ & the initial guess of $K$, for the relaxation method \\ \hline
	$x \text{ domain}$ & the domain of the $x$ grid \\ \hline
	$N_{grid}$ & the number of points in the $x$ grid \\ \hline
	tolerance & the tolerance of the convergence of the field (see App.~\ref{ap:num}) \\ \hline
\end{tabular}
\end{center}
$u(x)$ denotes the Heaviside step function:
\[
	u(x) =
	\begin{cases}
		0 & \qquad x<0 \\
		1 & \qquad 0 \leq x
	\end{cases}
\]	

	\chapter{Introduction}\label{ch:int}
%
%
%
\emph{Harmonic generation} is a process in which the frequency of an electromagnetic radiation is multiplied by a quantum system. This phenomenon occurs when the system emits radiation at a frequency higher than that of the incident radiation.

In the harmonic generation process, the incident forcing electromagnetic field produces oscillations in the dipole moment expectation value $\evmu$ of the system. According to Maxwell equations, an accelerated charge emits electromagnetic radiation, proportional to the acceleration of the charge. Hence, an oscillating quantum system emits radiation, proportional to the acceleration of $\evmu$. The spectrum of the field emitted by the system consists of the same frequencies as the spectrum of the $\evmu$ oscillations. Harmonic generation occurs when the $\evmu$ spectrum contains frequencies higher than that of the forcing field.

Typically, the most important component of the $\evmu$ spectrum is the frequency of the forcing field. This component represents the \emph{linear response} of the system to the radiation. The harmonic generation phenomenon originates from \emph{non-linear} effects. Typically, these are small in magnitude compared to the linear response.

The harmonic generation phenomenon has been utilized since the invention of the laser to convert radiation from a laser source to a frequency higher than that available from this source. Most commonly, the laser radiation frequency is multiplied by a factor of 2 (second harmonic generation) or 3 (third harmonic generation).

In the late 1980's, a new harmonic generation phenomenon was discovered. In this phenomenon the incident laser radiation is multiplied by factors of tens, even hundreds. This phenomenon is known as \emph{``high-harmonic generation''}. The high-harmonic generation phenomenon is the key element in the production of attosecond laser pulses; these are of extreme importance for investigation and control of electronic processes \cite{HHGcol}.

As mentioned, the harmonic generation phenomenon is small in magnitude. In addition, the harmonic generation spectrum often consists of many frequencies, while interest is in only a single frequency, or a region in the spectrum. It is of great interest to \emph{control} the harmonic generation phenomenon in order to increase the intensity of the emitted field at the frequencies of interest.

The problem may be addressed by means of the \emph{quantum control} discipline. It deals with the task of controlling  quantum systems by an electromagnetic radiation using designed field sequence shapes. The control of the harmonic generation phenomenon may be achieved by designing an appropriate time-shape of the incident laser field. This field sequence will extend over a \emph{spectrum} of relatively low frequencies instead of being limited to a monochromatic radiation.

In order to control a quantum system it is desirable to find the \emph{optimal field} for the problem of interest. There are two approaches for seeking the optimal field for quantum control problems:
\begin{enumerate}
	\item Experimentally, by a sophisticated trial and error process using a genetic algorithm (see~\cite{David});
	\item By theoretical calculation, using our knowledge about the quantum system.
\end{enumerate}
In the experimental approach all the aspects of the problem are taken into account, unlike in the theoretical approach. This makes the experimental approach much more accurate and efficient. The advantage of the theoretical approach is the possibility of investigating the control mechanism. This is the main importance for providing a theoretical method of calculation. The theoretical calculation may also provide a good starting point for the experimental genetic algorithm search. Frequently, a good starting point is necessary for the success of this method.

Quantum optimal control theory (OCT) is the most successful theoretical method available for finding an optimal field in quantum control problems. It is sometimes shortened as QOCT\@. In the framework of QOCT, the problem is formulated as a maximization problem using the calculus of variation formalism.

Great progress in the task of controlling high harmonic generation has been achieved in the last decade by using the experimental approach \cite{HHGcol}. However, a theoretical method of calculation for harmonic generation is still missing. The main purpose of the present work is to fill this gap.

The goals of our research are:
\begin{enumerate}
	\item Developing a procedure for finding an optimal field for harmonic generation, in the framework of QOCT;
	\item Finding a way for dealing with control problems with frequency requirements, in the framework of QOCT;
	\item Exploring new mechanisms of harmonic generation.
\end{enumerate}

In Ch.~\ref{ch:theo}, we present the relevant theoretical background for the present work. In Ch.~\ref{ch:new}, the new method is developed. In Ch.~\ref{ch:RD}, the new method is applied to several harmonic generation problems. The results are discussed, and general conclusions on mechanisms of harmonic generation are obtained.

	\chapter{Background}\label{ch:theo}
%
%
In this chapter we present the basic background in QOCT, needed to understand the present work. We also present existing works that are related to harmonic generation task.

We can divide our task into two distinct parts:
\begin{enumerate}
	\item Imposing a restriction on the spectrum of the forcing field
	
	\item Controlling the spectrum of the oscillating dipole, which is the spectrum of the emitted field
\end{enumerate}
In this chapter, and also in the following one, we shall treat these two parts separately.

In sections~\ref{sec:OCTf}-\ref{sec:theonum}, we present the basic relevant background in QOCT\@. Section~\ref{sec:theolim} deals with the existing works related to the first part of our task, mentioned above. Section~\ref{sec:GrossHG} deals with a work related to the second part, and the failure of an attempt to combine the two parts for harmonic generation control.

\section{OCT of a target operator in the final time}\label{sec:OCTf}
The simplest and most common task in QOCT, is seeking a time dependent forcing field that maximizes the expectation value of an operator in a final time (see~\cite{David, tutorial, Rabitz, Ronnie89}, in more detail). Hence, the formulation of this problem is the natural starting point for any discussion in QOCT.

Consider a quantum system in an initial state: $\ket{\psi(0)}=\ket{\psi_0}$, with the unperturbed Hamiltonian: $\operator{H}_0$, under a forcing field: $\epsilon(t)$; the OCT formulation translates the above mentioned maximization requirement into the maximization of the following functional:
\begin{equation}\label{eq:Jmaxs2s}
	J_{max}[\epsilon(t)] \equiv \bracketsO{\psi(T)}{\operator{O}}{\psi(T)}
\end{equation}
where $\operator{O}$ is an arbitrary operator, and $T$ is the final time. $ \ket{\psi(T)}$ depends on $\epsilon(t)$ in a rather complicated way, through the Schr\"odinger equation, under the given initial condition:
\begin{align}\label{eq:Schrodinger}
	&\pderiv{\ket{\psi(t)}}{t} = -i\operator{H}(t)\ket{\psi(t)}, & 	\ket{\psi(0)} & = \ket{\psi_0} \\
	&\operator{H}(t) = \operator{H}_0 - \operator{\mu}\epsilon(t) \nonumber
\end{align}
where $\operator{\mu}$ is the dipole moment operator. (Atomic units are used throughout, so we set: $\hbar=1$.) We have used the dipole approximation for $\operator{H}(t)$.

The most common target operator is the projection operator on a given target state, which will be denoted as $\ket{\phi}$:
\begin{equation}\label{eq:POf}
	\operator P_{\phi} = \ket{\phi}\bra{\phi}
\end{equation}
In this case, $J_{max}$ will attain its maximal value when: $\ket{\psi(T)} = e^{i\theta}\ket{\phi}$ ($\theta$ is an arbitrary phase). This kind of target operator is used when we want to find an optimal field for a state-to-state transition.


If we try to seek a maximum of $J_{max}$, we obtain non-physical fields of very large, short timed pulses; the reason is that the solution of this maximization problem is singular, with an infinite field. To get a physical solution, we must restrict the intensity of the field. It is also desirable, practically, to use lasers of as small an intensity as possible. This requirement can be achieved, by adding a ``penalty'' term to the functional object of maximization:
\begin{equation}\label{eq:Jpenals2s}
	J_{penal}[\epsilon(t)] \equiv -\alpha\int_0^T\epsilon^2(t)\,dt \qquad\qquad \alpha>0
\end{equation}
This term will be maximized when $\epsilon(t)$ is minimal. It ``penalizes" the object of maximization for using high intensity fields. $\alpha$ is a ``penalty factor'' - a positive constant, whose value determines the ``cost" of large intensities. The suitable value for the problem usually has to be determined by a trial and error process.

Let us define the functional:
\begin{equation}\label{eq:Jmps2s}
	J_{mp}[\epsilon(t)] \equiv J_{max}[\epsilon(t)] + J_{penal}[\epsilon(t)]
\end{equation}
Our problem is to find a function $\epsilon(t)$ that maximizes $J_{mp}$. This kind of a problem belongs to the category of variational problems. It can be handled most conveniently by the functional derivative formalism. The condition for an extremal is:
\begin{equation}\label{eq:dJmp}
	\fnlderiv{J_{mp}}{\epsilon(t)} = 0
\end{equation}
The complicated explicit dependence of $J_{max}$ on $\epsilon(t)$, makes dealing with Eq.~\eqref{eq:dJmp} rather inconvenient (see~\cite{Degani} for such an approach). The most convenient way to handle the problem is by using the Lagrange-multiplier method; it enables us to treat implicitly dependent variables as independent, in the first stage. Eq.~\eqref{eq:Schrodinger} will be treated as a constraint, which will be enforced later.

Before we proceed, we have to choose the variables that will be treated as independent. We should notice, that $\ket{\psi(t)}$ is complex, and is composed of two distinct variables: the real and imaginary parts. Moreover, the constraint equation,~\eqref{eq:Schrodinger}, consists of two conjugate constraint equations --- the equivalence relations between the real and imaginary parts of the expressions on both sides of the equation. There are also two initial conditions. We could have treated $\Real{\ket{\psi(t)}}$ and $\Imag{\ket{\psi(t)}}$ as our independent variables; nevertheless, it appears to be much more convenient to treat $\ket{\psi(t)}$ and $\bra{\psi(t)}$ as independent. Now, we can use Eq.~\eqref{eq:Schrodinger} as is, as our first constraint equation and initial condition; the additional constraint equation and initial condition is given by the complex conjugate of Eq.~\eqref{eq:Schrodinger}:
\begin{align}\label{eq:conjSchrodinger}
	\pderiv{\bra{\psi(t)}}{t} & = i\bra{\psi(t)}\operator{H}(t), & \bra{\psi(0)} & =\bra{\psi_0}
\end{align}
\eqref{eq:conjSchrodinger} ensures that $\bra{\psi(t)}$ will always be the complex conjugate of $\ket{\psi(t)}$, as required.
 
Now, we modify $J_{mp}$ by adding a constraint functional term. First, we write equations~\eqref{eq:Schrodinger},~\eqref{eq:conjSchrodinger} in the following way:
\begin{align}
	\pderiv{\ket{\psi(t)}}{t} + i\operator{H}(t)\ket{\psi(t)}& = 0 \label{eq:Schr0}\\
	\pderiv{\bra{\psi(t)}}{t} - i\bra{\psi(t)}\operator{H}(t)& = 0 \label{eq:conjSchr0}
\end{align}
Note that equations~\eqref{eq:Schr0},~\eqref{eq:conjSchr0} refer to all $t$ in the interval: $0\leq t \leq T$; hence, they impose distinct constraints on the state at all time points of the interval. According to the Lagrange-multiplier method, we have to add to the functional a distinct term for every constraint. Let us start with the constraints imposed by \eqref{eq:Schr0}: We add to $J_{mp}$ a continuous summation over all the LHS expressions of the constraints in all $t$, each multiplied by its own Lagrange-multiplier, $\bra{\chi(t)}$; the sequence of Lagrange-multipliers forms a continuous function of $t$ in the interval (see \cite[Ch.~2]{AOC}). The additional term for the constraint in \eqref{eq:Schr0} takes the form:
\begin{equation}\label{eq:Jcket}
	-\int_0^T\bracketsO{\chi(t)}{\pderiv{}{t}+i\operator H(t)}{\psi(t)}\,dt
\end{equation}
We take $\bra{\chi(t)}$ to be a complex function, like a wave function. $\ket{\chi(t)}$ is sometimes called: ``the conjugate function" of $\ket{\psi(t)}$.

We treat the additional term for the constraint in \eqref{eq:conjSchr0} in the same way. We have to use another Lagrange-multiplier function of $t$ for this term; since we took $\bra{\chi(t)}$ to be complex, we may treat $\ket{\chi(t)}$ as an independent function. The additional term is:
\begin{equation}\label{eq:Jcbra}
	-\int_0^T\bracketsbiggm{\left(\pderiv{}{t}-i\operator H(t)\right)\psi(t)}{\chi(t)}\,dt
\end{equation}
Note that \eqref{eq:Jcbra} is the complex conjugate of \eqref{eq:Jcket}.

The overall additional constraint term for the functional object is:
\begin{equation}\label{eq:Jcons2s}
	J_{con} = -2\Real{\int_0^T\bracketsO{\chi(t)}{\pderiv{}{t}+i\operator H(t)}{\psi(t)}\,dt}
\end{equation}

The overall new object of maximization is the following functional:
\begin{align}
	J & \equiv J_{max} + J_{penal} + J_{con} \nonumber \\
	  & =  \bracketsO{\psi(T)}{\operator{O}}{\psi(T)} -\alpha\int_0^T\epsilon^2(t)\,dt
	      -2\Real{\int_0^T\bracketsO{\chi(t)}{\pderiv{}{t}+i\operator H(t)}{\psi(t)}\,dt} \label{eq:Js2s}
\end{align}
The conditions for an extremal are given by the following equations:
\begin{align}
	&\fnlderiv{J}{\epsilon(t)} = 0 \label{eq:dJdeps} \\ 
	&\fnlderiv{J}{\ket{\psi(t)}} = 0 \label{eq:dJdpsitk} \\ 
	&\fnlderiv{J}{\bra{\psi(t)}} = 0 \label{eq:dJdpsitb} \\	
	&\fnlderiv{J}{\ket{\psi(T)}} = 0 \label{eq:dJdpsiTk} \\ 
	&\fnlderiv{J}{\bra{\psi(T)}} = 0 \label{eq:dJdpsiTb}
\end{align}
together with the constraints, \eqref{eq:Schrodinger} and \eqref{eq:conjSchrodinger}. All functional derivatives are taken while holding all other variables constant. 

This set of equations forms the basis for the derivation of the so called: ``Euler-Lagrange equations'' of the problem . The resulting equations are (see \cite{David, tutorial} for example, for more details):
\begin{align}
	&\pderiv{\ket{\psi(t)}}{t} = -i\operator{H}(t)\ket{\psi(t)}, & 	\ket{\psi(0)} & = \ket{\psi_0} \label{eq:ELs2spsi} \\ 
	&\pderiv{\ket{\chi(t)}}{t} = -i\operator{H}(t)\ket{\chi(t)}, & 	\ket{\chi(T)} & = \operator{O}\ket{\psi(T)} \label{eq:ELs2schi} \\ 
	&\operator{H}(t) = \operator{H}_0 - \operator{\mu}\epsilon(t) \nonumber \\
	&\epsilon(t) = -\frac{\Imag{\bracketsO{\chi(t)}{\operator{\mu}}{\psi(t)}}}{\alpha} \label{eq:ELs2seps}
\end{align}
A solution $\epsilon(t)$ which satisfies all these equations is an extremal field, which gives a maximum for the functional $J$, and hence, also for $J_{mp}$ (a minimum is almost never encountered in this kind of problems).

There is no general method to solve this set of equations analytically. Hence, we have to employ numerical methods to find a solution. Several methods will be presented in section~\ref{sec:theonum}.

\section{OCT of a time dependent target operator}\label{sec:OCTt}
The task that was described in the last section refers to the control of the system state in a single time point: $t=T$. A more difficult problem is to control the state of the system over an interval of time. This is the kind of problem that we deal with in the present work. In this section, we describe the existing method for the control of such kind of problems (see \cite{Serban, tutorial} for more details). 

The object of maximization is the expectation value of a time dependent operator, during the time interval: $0\leq t \leq T$. The $J_{max}$ from Eq.~\eqref{eq:Jmaxs2s} is replaced by (see \cite{tutorial}):
\begin{equation}\label{eq:JmaxTD}
	J_{max} \equiv \int_0^T \bracketsO{\psi(t)}{\operator{O}(t)}{\psi(t)} w(t)\,dt
\end{equation}
where $w(t)$ is a weight function, which satisfies:
\[
	\int_0^T w(t)\,dt = 1
\]
$w(t)$ determines the relative importance of the maximization target during different parts of the time interval. The functional will be maximized when the integrand is maximized at all time points in the interval.

The method is intended for a positive-semidefinite $\operator{O}(t)$, in which a greater expectation value indicates greater success. This is not always the case --- it may occur that we want the expectation value to vary in a predefined path, and not to be maximized at all $t$. In such a case, this method is not helpful.

A common example for a time dependent target operator is a time dependent projection operator:
\begin{equation} \label{eq:POt}
	\operator P_{\phi(t)} = \ket{\phi(t)}\bra{\phi(t)}
\end{equation}
This target operator should be used when we want to force the system to follow a predefined path, defined by
the sequence of wave-functions: $\ket{\phi(t)}$.

The other parts of the object of maximization $J$, \ie $J_{penal}$ and $J_{con}$, are the same as in Eqs.~\eqref{eq:Jpenals2s},~\eqref{eq:Jcons2s}. The overall object of maximization is the following functional:
\begin{align}
	J  \equiv & J_{max} + J_{penal} + J_{con} \nonumber \\
	   =  & \int_0^T \bracketsO{\psi(t)}{\operator{O}(t)}{\psi(t)} w(t)\,dt -\alpha\int_0^T\epsilon^2(t)\,dt - 2\Real{\int_0^T\bracketsO{\chi(t)}{\pderiv{}{t}+i\operator H(t)}{\psi(t)}\,dt} 	  
\label{eq:JTD1}
\end{align}

The conditions for an extremal are the same as in Eqs.~\eqref{eq:dJdeps}-\eqref{eq:dJdpsitb}, together with the constraints: \eqref{eq:Schrodinger} and \eqref{eq:conjSchrodinger}. The resulting Euler-Lagrange equations are (see \cite{tutorial} for details):
\begin{align}
	&\pderiv{\ket{\psi(t)}}{t} = -i\operator{H}(t)\ket{\psi(t)}, & 	\ket{\psi(0)} & = \ket{\psi_0} \label{eq:ELTDpsi} \\ 
	&\pderiv{\ket{\chi(t)}}{t} = -i\operator{H}(t)\ket{\chi(t)} - w(t)\operator{O}(t)\ket{\psi(t)}, & 	\ket{\chi(T)} & = 0 \label{eq:ELTDchi} \\ 
	&\operator{H}(t) = \operator{H}_0 - \operator{\mu}\epsilon(t) \nonumber \\
	&\epsilon(t) = -\frac{\Imag{\bracketsO{\chi(t)}{\operator{\mu}}{\psi(t)}}}{\alpha} \label{eq:ELTDeps}
\end{align}
Note that \eqref{eq:ELTDpsi} and \eqref{eq:ELTDeps} are the same as \eqref{eq:ELs2spsi} and \eqref{eq:ELs2seps}. However, \eqref{eq:ELTDchi} is different from \eqref{eq:ELs2schi}; we have here an additional, inhomogeneous, $\ket{\psi(t)}$ dependent term on the RHS of the equation. This form of equation is known as ``the inhomogeneous Schr\"odinger equation". The boundary condition is also different. It does not come from the conditions for maximization in Eq.~\eqref{eq:dJdeps}-\eqref{eq:dJdpsitb}; this boundary condition represents the ``natural boundary conditions" of the problem (see, for example, \cite[Ch. 2, Sec. 3]{AOC}). The fact that the Lagrange-multiplier $\ket{\chi(T)}$ vanishes originates from the independence of $J_{max}$ on $\ket{\psi(T)}$ (more precisely, the dependence is infinitesimal, and does not contribute to the boundary condition in $T$). This should be compared with the boundary condition in \eqref{eq:ELs2schi}, where $J_{max}$ does depend on $\ket{\psi(T)}$.

This boundary condition is problematic because it imposes on the field the condition of vanishing at the final time:
\[
	\epsilon(T) = 0
\]
Our experience shows that this condition is the source of difficulties in achieving control in the neighbourhood of $t=T$.

Another version of the formulation of the problem is available in the literature (\cite[Ch. 2]{AOC},~\cite{Jose}), in which this difficulty is eliminated; $J$ is modified  by adding the following boundary term:
\begin{equation}\label{eq:JbTD}
	J_{bound} \equiv \kappa\bracketsO{\psi(T)}{\operator{O}(T)}{\psi(T)} \qquad\qquad \kappa>0
\end{equation}
$\kappa$ is a positive parameter, that determines the relative importance of the boundary term. The modified object of optimization is:
\begin{equation}\label{eq:JTD2}
	J  \equiv J_{max} + J_{bound} + J_{penal} + J_{con}
\end{equation}
Now we have additional conditions for extremum: \eqref{eq:dJdpsiTk} and \eqref{eq:dJdpsiTb}.

The resulting Euler-Lagrange equations are the same as \eqref{eq:ELTDpsi}-\eqref{eq:ELTDeps}, except the boundary condition in \eqref{eq:ELTDchi}, which is replaced by:
\begin{equation}\label{eq:TDchiT}
	\ket{\chi(T)} = \kappa\operator{O}(T)\ket{\psi(T)}
\end{equation}

The disadvantage of this approach is that it gives exaggerated importance to the target at $T$, compared to the other time points.

\section{Numerical methods for the maximization of the functional}\label{sec:theonum}
In this section, we present a number of numerical methods available for the maximization of the $J$ functionals  mentioned in the two previous sections.
	\subsection{The naive approach}\label{ssec:naive}
As a first thought, we may propose the iterative scheme, described in the following steps \cite{David, Ronnie89}:
\begin{enumerate}
	\item Guess a field sequence: $\epsilon(t)$.
	\item Repeat the following steps, until convergence:\label{pr:ndo}
	\begin{enumerate}	
		\item Propagate $\ket{\psi(t)}$ forward from $t=0$ to $t=T$, using \eqref{eq:ELs2spsi}, with $\epsilon(t)$.\label{pr:npsi}
		\item Set $\ket{\chi(T)}$ according to the boundary condition in \eqref{eq:ELs2schi}, \eqref{eq:ELTDchi} or \eqref{eq:TDchiT}.
		\item Propagate $\ket{\chi(t)}$ backward from $t=T$ to $t=0$, using \eqref{eq:ELs2schi} or \eqref{eq:ELTDchi}, with $\epsilon(t)$ (and $\ket{\psi(t)}$ from step~\ref{pr:npsi}, for \eqref{eq:ELTDchi}).\label{pr:nchi}
		\item Update $\epsilon(t)$ according to \eqref{eq:ELs2seps}, using the sequences of $\ket{\psi(t)}$ and $\ket{\chi(t)}$ from steps~\ref{pr:npsi},~\ref{pr:nchi}.
	\end{enumerate}
\end{enumerate}

Unfortunately, this simple scheme seldom converges. More elaborate methods are needed for control problems.

	\subsection{The gradient methods}\label{ssec:grad}
The gradient methods are a family of general methods of optimization; they use information about the derivatives of the object of minimization/maximization with respect to the variable to be optimized (see~\cite{AOC, mathworks}). In our case, we talk about functional derivatives of $J$ with respect to the control field $\epsilon(t)$. This information is used to ``climb up" in the hypersurface of $J$ vs. $\epsilon(t)$ (at all time points).

The simplest gradient method is the so called: ``first-order gradient method", or ``steepest descent/accent method". As its name indicates it consists of the information from the first order derivative, or the gradient, of the object of maximization. In our case, the gradient is:
\begin{equation}\label{eq:gradt}
	\nabla_{\epsilon(t)}J = \fnlderiv{J}{\epsilon(t)} = -2\left[\alpha\epsilon(t) +\Imag{\bracketsO{\chi(t)}{\operator{\mu}}{\psi(t)}}\right]
\end{equation}
The gradient is a vector (in our case, a continuous vector, \ie a function) that points in the direction of the maximal increase of the object of maximization. The method is based on following this direction. In our case, we update $\epsilon(t)$ according to the following rule \cite{Rabitz}:
\begin{equation}\label{eq:updateg}
	\epsilon^{new}(t) = \epsilon^{old}(t) + K\,\nabla_{\epsilon(t)}J\biggm|_{\epsilon(t)=\epsilon^{old}(t)} \qquad K>0
\end{equation}
where $K$ is a positive parameter, which determines the rate of propagation in the direction specified by $\nabla_{\epsilon(t)}J$. We repeat the process until convergence.

$K$ must be small enough for the first order approximation to be satisfactory; if $K$ is too large, the sign of the gradient might change in the way to the new $\epsilon(t)$, and the value of $J$ will not necessarily increase. On the other hand, if $K$ is too small, the rate of convergence will be slow. An optimal value of $K$ has to be found by a trial and error process. In addition, it should be varied during the process of optimization, from greater values in the early iterations, to smaller values close to the maximum. A commonly used method for searching for an optimal value of $K$ is called ``line search". The line search has to be reimplemented in every iteration.

Let $k$ denote the iteration index; the whole procedure is summarized in the following scheme: 
\begin{enumerate}
	\item Guess a field sequence: $\epsilon^{(0)}(t)$.
	\item Propagate $\ket{\psi^{(0)}(t)}$ forward from $t=0$ to $t=T$, with $\epsilon^{(0)}(t)$.
	\item Calculate $J^{(0)}$ with $\ket{\psi^{(0)}(t)}$ and $\epsilon^{(0)}(t)$ (note that $J_{con}=0$, since the constraint is satisfied).
	\item (k = 0)
	\item Repeat the following steps, until convergence:\label{pr:gdo}
	\begin{enumerate}
		\item Set $\ket{\chi^{(k)}(T)}$ according to the suitable boundary condition, using  $\ket{\psi^{(k)}(T)}$.
		\item Propagate $\ket{\chi^{(k)}(t)}$ backward from $t=T$ to $t=0$, with $\epsilon^{(k)}(t)$.
		\item Perform a line search, to find an optimal value for $K$; the line search involves $J$ evaluations for various values of $K$, to be compared with $J^{(k)}$; these require the following steps:
		\begin{enumerate}
			\item Set a new field, using Eq.~\eqref{eq:gradt}:
			\begin{equation}\label{eq:tryg}
				\epsilon^{trial}(t) = \epsilon^{(k)}(t) - 2K\,\left[\alpha\epsilon^{(k)}(t) +\Imag{\bracketsObiggm{\chi^{(k)}(t)}{\operator{\mu}}{\psi^{(k)}(t)}}\right]
			\end{equation}
			\item Propagate $\ket{\psi^{trial}(t)}$ forward from $t=0$ to $t=T$, with $\epsilon^{trial}(t)$.
			\item Calculate $J^{trial}$ with $\ket{\psi^{trial}(t)}$ and $\epsilon^{trial}(t)$, and compare with $J^{(k)}$.
		\end{enumerate}
		\item When an optimal $K$ was found, update all the variables according to this $K$:
		\begin{equation}
			\epsilon^{(k+1)}(t) = \epsilon^{trial}(t) \qquad \ket{\psi^{(k+1)}(t)} = \ket{\psi^{trial}(t)} \qquad J^{(k+1)} = J^{trial} \label{eq:prgup}
		\end{equation}		
					
		\item (k = k + 1)
	\end{enumerate}
\end{enumerate}

The first order gradient method usually shows considerable improvement in the first iterations. However, it is known to have a very slow convergence rate when getting close to the maximum. Hence, we should avoid using it if possible.

The ``second order gradient method", or ``Newton method", uses second order derivative information about the object of maximization, in addition to the gradient. It has good convergence characteristics near the maximum. This method is not suitable for QOCT problems, for a reason that will be explained later; however, approximate versions of this method --- ``quasi-Newton methods" --- can be employed successfully \cite{Ruvi}. 

We start with the description of the regular Newton method for a nonlinear function of a single variable --- $f(x)$. Suppose we have an initial guess $x_0$, which is known to be close enough to the maximum; in ``close" we mean that the negative sign of $f''(x)$ does not change from the maximum to $x_0$. Now, we adjust a parabola to this point, which is the second order approximation for $f(x)$:
\[
	f(x) \approx p(x) = f(x_0) + f'(x_0)(x-x_0) + \frac{1}{2}f''(x_0)(x-x_0)^2
\]
Then, we move to the $x$ value of the parabola's maximum; it is easily found to be:
\begin{equation}\label{eq:Newton1}
	x^* = x_0 - [f''(x_0)]^{-1}f'(x_0)
\end{equation}
We update the value of $x$ to $x^*$. We repeat the process, until convergence.

If we deal with a function of a vector, the procedure is very similar; let $F(\mathbf{v})$ be a function of the vector:
\[
	\mathbf{v} = 
		\begin{bmatrix}
			v_1 \\
			\vdots \\
			v_N
		\end{bmatrix}
\]
Suppose we have an initial guess, $\mathbf{v}_0$, which is close enough to the maximum; in this case, the word ``close" refers to the negative-definiteness of the Hessian matrix $S$, which is defined by:
\begin{equation}\label{eq:Hess}
	[S]_{ij} = \pderivdd{F}{v_i}{v_j}
\end{equation}
We adjust a multidimensional paraboloid to the point $\mathbf{v}_0$, and we find its maximum. The expression for  $\mathbf{v}$ at the maximum is very similar to~\eqref{eq:Newton1}:
\begin{equation}\label{eq:Newtonv}
	\mathbf{v}^* = \mathbf{v}_0 - S^{-1}\nabla F\biggm|_{\mathbf{v}=\mathbf{v}_0}
\end{equation}
$\mathbf{v}$ is updated to $\mathbf{v}^*$, and the process is repeated until convergence.

In our case, the Hessian is a ``continuous matrix", defined by a two variable function:
\begin{equation}\label{eq:HessJ}
	s(t, t') = \fnlderivdd{J}{\epsilon(t)}{\epsilon(t')}
\end{equation}
The operation of the Hessian matrix on a vector in the discrete case, is replaced by the operation of an operator on a function in the continuous case; the Hessian operator is defined by:
\begin{equation}\label{eq:HessopJ}
	\operator{S}g(t) = \int_0^Ts(t, t')g(t')\,dt'
\end{equation}
where $g(t)$ is an arbitrary function of $t$. The update rule is:
\begin{equation}\label{eq:Newtoneps}
	\epsilon^{new}(t) = \epsilon^{old}(t) - \operator{S}^{-1}\nabla_{\epsilon(t)}J\biggm|_{\epsilon(t)=\epsilon^{old}(t)}
\end{equation}

The main problem with the Newton-method is that often the Hessian is too complex to be computed easily. The expression for the first derivative in Eq.~\eqref{eq:gradt} is dependent on $\epsilon(t)$ explicitly, and implicitly, through $\ket{\chi(t)}$ and $\ket{\psi(t)}$. The implicit dependence is very complicated; it follows, that the expression of the second derivatives, $s(t, t')$, is also very complicated. Hence, practically, the Hessian cannot be computed directly for all $t$, $t'$, in every iteration.\footnote{You may ask: why do we treat $\ket{\chi(t)}$ and $\ket{\psi(t)}$ as $\epsilon(t)$ dependent when dealing with the second derivatives, while they were treated as $\epsilon(t)$ independent when we dealt with the first derivative (in Eq.~\eqref{eq:gradt})? The answer is, that in fact, $\ket{\chi(t)}$ and $\ket{\psi(t)}$ do depend on $\epsilon(t)$; we ignore this dependence when handling with the first derivative, because this is exactly the role of the Lagrange-multipliers --- they are adjusted in a way that makes Eqs.~\eqref{eq:dJdpsitk}-\eqref{eq:dJdpsiTb} true. Then, the full derivative of $J$ with respect to $\epsilon(t)$, coincides with the partial derivative in Eq.~\eqref{eq:gradt}. However, when treating the second derivatives of $J$, this choice of the Lagrange-multipliers has no special significance, and the implicit dependence on $\epsilon(t)$ cannot be ignored (see~\cite{Lanczos}).}

The quasi-Newton methods make use of an approximated Hessian. The most commonly used method is the BFGS method. It approximates the Hessian using an information from the gradient (see~\cite{mathworks, Ruvi} for more details). We will denote the approximated Hessian as: $S_{ap}$.

When using an approximate Hessian, Eq.~\eqref{eq:Newtonv} (in the discrete case) should not be used as is. Instead, we treat the vector $-S_{ap}^{-1}\nabla F$ as the direction of search for the new $\mathbf{v}$. This vector has exactly the same role as the gradient vector in the first order gradient method. The update rule in the discrete case is:
\begin{equation}\label{eq:qNtnv}
	\mathbf{v}^{new} = \mathbf{v}^{old} - K\,S_{ap}^{-1}\nabla F\biggm|_{\mathbf{v}=\mathbf{v}_0} \qquad K>0
\end{equation}
and in the continuous case:
\begin{equation}\label{eq:qNtneps}
	\epsilon^{new}(t) = \epsilon^{old}(t) - K\,\operator{S}_{ap}^{-1}\nabla_{\epsilon(t)}J\biggm|_{\epsilon(t)=\epsilon^{old}(t)} \qquad K>0
\end{equation}
A line search is made to find an optimal value for $K$.

The full process of the quasi-Newton methods, in the context of QOCT, is very similar to the scheme that was presented for the first order gradient method. The only difference is that we use another direction for search;  Eq.~\eqref{eq:tryg} is replaced by the following:
\begin{equation}\label{eq:tryN}
	\epsilon^{trial}(t) = \epsilon^{(k)}(t) - K\,\operator{S}_{ap}^{(k)-1}[\nabla_{\epsilon(t)}J]^{(k)}
\end{equation}
where the Hessian approximation $\operator{S}_{ap}^{(k)}$, and the gradient $[\nabla_{\epsilon(t)}J]^{(k)}$, are computed using: $\epsilon^{(k)}$, $\ket{\psi^{(k)}(t)}$, $\ket{\chi^{(k)}(t)}$.

	\subsection{The Krotov method}\label{ssec:Krotov}
The most popular method for QOCT problems is called: ``the Krotov method". It was first introduced, in the context of QOCT, in~\cite{Orlov}. There are several variants of the method; here we present the variant that is used in the examples of this thesis \cite{tutorial}.

The idea is not very different from the one of the naive approach; the difference between the methods is in the stage where the update of the field takes place. In the naive approach, the field is updated after the propagations of $\ket{\psi(t)}$ and $\ket{\chi(t)}$ were completed; in the Krotov method, the field is updated \emph{during the propagation}, at every new time point, and the propagation proceeds using the new field.

The procedure is presented in the following scheme:
\begin{enumerate}
	\item Guess a field sequence: $\epsilon^{(0)}(t)$.
	\item Propagate $\ket{\psi^{(0)}(t)}$ forward from $t=0$ to $t=T$, with $\epsilon^{(0)}(t)$.
	\item (k = 0)
	\item Repeat the following steps, until convergence:
	\begin{enumerate}
		\item Set $\ket{\chi^{(k)}(T)}$ according to the suitable boundary condition, using $\ket{\psi^{(k)}(T)}$.
		\item Propagate $\ket{\chi^{(k)}(t)}$ backward from $t=T$ to $t=0$, with the field defined by:
		\begin{equation}\label{eq:Krotovteps}
			\tilde{\epsilon}^{(k)}(t) = -\frac{\Imag{\bracketsObigm{\chi^{(k)}(t)}{\operator{\mu}}{\psi^{(k)}(t)}}}{\alpha}		
		\end{equation}
		\item Propagate $\ket{\psi^{(k+1)}(t)}$ forward from $t=0$ to $t=T$, with the new field, defined by:
		\begin{equation}\label{eq:Krotoveps}
			\epsilon^{(k+1)}(t) = -\frac{\Imag{\bracketsObigm{\chi^{(k)}(t)}{\operator{\mu}}{\psi^{(k+1)}(t)}}}{\alpha}		
		\end{equation}
		\item (k = k + 1)
	\end{enumerate}
\end{enumerate}
As can be seen, the procedure is considerably simpler than that of the gradient methods.

The most important property of the Krotov algorithm is that it is \emph{monotonically convergent} (with no need of a trial and error process, as in the gradient methods). The convergence analysis is presented in~\cite{genKrotov}.

Contrary to the gradient methods, which are general methods, the Krotov method utilizes the special structure of the problem. The update rule of $\epsilon(t)$ depends only on the functions in time $t$; this allows the update of $\epsilon(t)$ during the propagations, in such a way that the resulting sequence of $\epsilon(t)$ is consistent with the resulting sequence of the wave functions, and with itself.

The rate of convergence in the Krotov method is known to be faster than that of the gradient methods, in the early iterations. However, when getting very close to the maximum, the convergence rate becomes slow. Hence, the Krotov method is problematic when a high fidelity solution is required. In~\cite{Ruvi} it was shown that the Krotov method degenerates into a first order gradient method close to the the maximum.

\section{QOCT with a restriction on the field spectrum}\label{sec:theolim}
The task of restricting the spectrum of the forcing field has considerable importance in QOCT; the reason is that most of the computed fields turn out to be too oscillatory to be produced experimentally. Several methods have been proposed to overcome this problem~\cite{Degani}. Most of the methods are based on limiting the spectrum of the field. In this section we present the important methods.

The methods were formulated in the context of the problem of Sec.~\ref{sec:OCTf}. They can be used for other problems without additional mathematical complications. However, the rate of success of the method might not be the same.

   
	\subsection{The method of Werschnik and Gross: Brute force}\label{ssec:brute}
Werschnik and Gross \cite{tutorial} use the Krotov method as a basis for the optimization process. However, the process is interrupted in every iteration by a spectral filtration of the resulting field. The filtered field is used in the next propagation. This amounts to replacing the update rule in~\eqref{eq:Krotoveps} by the following:
\begin{equation}\label{eq:Grossfil}
	\epsilon^{(k+1)}(t) = \mathcal{F}^{-1}\left\lbrace f(\omega)\mathcal{F}\left[\tilde{\epsilon}^{(k)}(t)\right]\right\rbrace
\end{equation}
where $\mathcal{F}$ stands for the Fourier transform, $\mathcal{F}^{-1}$ for its inverse, $\omega$ is the frequency variable, and $f(\omega)$ is a filter function.

This kind of interruption in the process is called ``brute force". The brute force method inserts into the process of optimization something that is not self-consistent with its reasoning. This destroys the monotonic convergence property of the Krotov algorithm. However, useful results can be obtained by storing in memory during the process the best field obtained so far. The field that is stored after a predefined number of iterations is the result of the optimization process.

	\subsection{The method of Degani \etal: A non-diagonal pe\-nal\-ty term}\label{ssec:Degani}
Degani \etal \cite{Degani} developed a method of restricting the spectrum by a more complicated penalty term, $J_{penal}$. This enables the inclusion of more information on the desired field.\footnote{The formulation of the control problem in \cite{Degani} is discrete, due to the introduction of control by a piecewise constant field, in the same paper. To avoid complications and inhomogeneity in the present text, we reformulate the problem in a continuous context. However, note that there is indeed a need for a discrete formulation, when dealing with the numerical method of maximization.\label{fn:Deg}}

As an introduction, we first mention a simpler use of the penalty factor, in order to control the properties of the field (see, for example,~\cite{tutorial}). A time dependent penalty factor, $\alpha(t)$ can be introduced, in order to control the time-shape of the field. The modified $J_{penal}$ is given by:
\begin{equation}\label{eq:penalt}
	J_{penal}[\epsilon(t)] \equiv - \int_0^T\alpha(t)\epsilon^2(t)\,dt
\end{equation}
For instance, we can force a Gaussian envelope on the field profile, by choosing an appropriate $\alpha(t)$.

The $\alpha(t)$ function is one dimensional, and contains only one dimensional information. Degani \etal use a two dimensional penalty term, which contains greater wealth of information about the properties of the desired field. We introduce the two dimensional penalty function $\beta(t, t')$; the new penalty term takes the form:
\begin{equation}\label{eq:penalttp}
	J_{penal}[\epsilon(t)] \equiv - \int_0^T\negthickspace\int_0^T\epsilon(t)\beta(t, t')\epsilon(t')\,dt\,dt'
\end{equation}
An alternative formulation is in the operator language, where $\operator{B}$ stands for an arbitrary linear operator:
\begin{align}
	& J_{penal}[\epsilon(t)] \equiv - \int_0^T\epsilon(t)\operator{B}\epsilon(t)\,dt\label{eq:penalttpO}\\
	& \operator{B}\epsilon(t) = \int_0^T\beta(t, t')\epsilon(t')\,dt'\label{eq:Bdef}
\end{align}
In this formulation, the role of $\beta(t, t')$ is more apparent.

The $\operator{B}$ operator (or matrix, in the original context; see footnote~\ref{fn:Deg}) used by the authors is the following:
\begin{align}
	& \operator{B} \equiv \alpha_{good}\operator{P}_{good} + \alpha_{bad}\operator{P}_{bad}\label{eq:DegB} \\
	& \alpha_{good}<0 \qquad \alpha_{bad}>0 \nonumber
\end{align}
$\operator{P}_{good}$ is the projection operator on the subspace of desired control functions; $\operator{P}_{bad}$ is the projection operator on the subspace of undesired control function, where:
\[
	\operator{P}_{bad} = \operator{I} - \operator{P}_{good}
\]
($\operator{I}$ is the identity operator). The $\alpha$'s have a role similar to a penalty factor. The idea is to ``penalize" the optimization process for undesired control functions and to encourage the appearance of desired control functions. In this case, the desired functions are fields in the desired frequency domain. In this way, we can control the spectral properties of the field.\footnote{The authors include an additional term in $\operator{B}$: $\alpha\operator{I}$, maybe for convenience. We didn't include it, because: $\operator{I} = \operator{P}_{good} + \operator{P}_{bad}$, so this term is unnecessary.} (Another suggestion for $\operator{B}$ is introduced by the authors; we will not discuss it here.)

The resulting Euler-Lagrange equations are the same as in Sec.~\ref{sec:OCTf}, apart from the replacement of Eq.~\eqref{eq:ELs2seps} by the following:
\begin{equation}\label{eq:ELDeg}
	\operator{B}\epsilon(t) = \int_0^T\beta(t, t')\epsilon(t')\,dt' = -\Imag{\bracketsO{\chi(t)}{\operator{\mu}}{\psi(t)}} 
\end{equation}

The new equation does not have the special property mentioned in subsection~\ref{ssec:Krotov} -- $\epsilon(t)$ does not depend only on other objects at the same time $t$, but also in the field at all other time points. This prevents the possibility of using the Krotov method in the way it appears in subsection~\ref{ssec:Krotov}. Instead, the authors present a Krotov-like method, which requires the use of a discretized version of Eq.~\eqref{eq:ELDeg} (of course, discretization of the problem is required anyway for a numerical solution). Let us define a time grid (the authors use an equidistant grid; here we define a grid which is not necessarily equidistant):
\[
	t_1,t_2,\ldots, t_N
\]
Now, we approximate Eq.~\eqref{eq:ELDeg} by a discretized form of the equation:
\begin{equation}\label{eq:ELDegd}
	T\sum_{j=1}^{N}\beta(t_i, t_j)\epsilon(t_j)w_j = -\Imag{\bracketsO{\chi(t_i)}{\operator{\mu}}{\psi(t_i)}}
\end{equation}
where $w_j$ is the integration weight for time point $t_j$. Let us define:
\[
	h_j=Tw_j
\]
Rearrangement of Eq.~\eqref{eq:ELDegd} gives:
\begin{equation}\label{eq:ELDegepst}
	\epsilon(t_i) = \frac{-\Imag{\bracketsO{\chi(t_i)}{\operator{\mu}}{\psi(t_i)}} - \sum_{j=1}^{i-1}\beta(t_i, t_j)\epsilon(t_j)h_j - \sum_{j=i+1}^{N}\beta(t_i, t_j)\epsilon(t_j)h_j}{\beta(t_i, t_i)h_i}
\end{equation}
This equation is the basis for the update rule of $\epsilon(t_i)$ in the Krotov-like algorithm. The idea is to compute in each time point the field with the newest values available of the variables. Eq.~\eqref{eq:Krotovteps} is replaced by:
\begin{align}
	&\tilde{\epsilon}^{(k)}(t_i) = \frac{-\Imag{\bracketsObigm{\chi^{(k)}(t_i)}{\operator{\mu}}{\psi^{(k)}(t_i)}} - \sum_{j=1}^{i-1}\beta(t_i, t_j)\epsilon^{(k)}(t_j)h_j - \sum_{j=i+1}^{N}\beta(t_i, t_j)\tilde{\epsilon}^{(k)}h_j}{\beta(t_i, t_i)h_i}
	\label{eq:DegKrotovteps}
\end{align}
Eq.~\eqref{eq:Krotoveps} is replaced by:
\begin{align}
	&\epsilon^{(k+1)}(t_i) = \frac{-\Imag{\bracketsObigm{\chi^{(k)}(t_i)}{\operator{\mu}}{\psi^{(k+1)}(t_i)}} - \sum_{j=1}^{i-1}\beta(t_i, t_j)\epsilon^{(k+1)}(t_j)h_j - \sum_{j=i+1}^{N}\beta(t_i, t_j)\tilde{\epsilon}^{(k)}h_j}{\beta(t_i, t_i)h_i}
	\label{eq:DegKrotoveps}
\end{align}

Unlike the standard Krotov algorithm, this algorithm lacks the self consistency of the sequence of $\epsilon(t)$, because of the use of data from another sequence to compute $\epsilon(t_i)$. Nevertheless, the authors report monotonic convergence of the algorithm. However, a convergence analysis is not supplied.

	\subsection{The method of Skinner and Gershenzon: controlling a list of frequency terms}
Skinner and Gershenzon \cite{Skinner} present an approach of controlling the properties of the field by defining the field as an analytic function of a desired form with several adjustable parameters. The parameters are the optimized variables, instead of $\epsilon(t)$. In our case, the field is defined by a cosine series:
\begin{equation}\label{eq:epscosseries}
	\epsilon(t) = \sum_{n=0}^{N}a_n\cos(n\,\Delta\omega\,t)
\end{equation}
where $\Delta\omega$ is defined by the resolution of $\omega$, at a time interval of $T$: \text{$\Delta\omega = \pi/T$}. The maximal frequency that may be present in the resulting field is: $N\Delta\omega$. The $a_n$ parameters are optimized to maximize $J$.

The optimization of different variables requires an alteration in the maximization condition; Eq.~\eqref{eq:dJdeps} is replaced by the following set of equations:
\begin{equation}
	\pderiv{J}{a_n}=0 \qquad n=0,1,\ldots ,N
\end{equation}

The maximization problem that is dealt with by the authors is somewhat different from ours; hence, we avoid presenting the resulting Euler-Lagrange equations.

The optimized variables are not defined in the time domain, like $\epsilon(t)$. This rules out the possibility of using the Krotov method. The authors use a first order gradient method.

\section{OCT of a time dependent dipole moment}\label{sec:GrossHG}
In order to generate an emitted radiation of a desired frequency, we have to be able to control the frequency of the oscillations of the time dependent dipole moment expectation value:
\[
	\exval{\operator{\mu}}\!(t) = \bracketsO{\psi(t)}{\operator{\mu}}{\psi(t)}
\]
Hence, the development of methods of controlling $\exval{\operator{\mu}}\!(t)$, are of utmost importance in the task of harmonic generation control.

The main problem in controlling the path of $\exval{\operator{\mu}}\!(t)$ by OCT is that this problem is not a maximization problem, when formulated in the terms of $\exval{\operator{\mu}}\!(t)$. Hence, we obviously cannot use the method of Sec.~\ref{sec:OCTt} by simply setting: \text{$\operator{O}(t) = \operator{\mu}$}.

Serban, Werschnik and Gross \cite{Serban} propose to solve this problem by maximizing an alternative time dependent operator, instead of dealing with $\operator{\mu}$ directly.  Then, the method of Sec.~\ref{sec:OCTt} is applicable. Consider a case of a quantum system with one spatial variable, $x$, and a charge: $q=1$ for convenience. In this case: \text{$\operator{\mu}=\operator{X}$}. The time-dependent operator is:
\begin{equation}\label{eq:deltaO}
	\operator{O}(t) = \delta\left[\operator{X}-\xi(t)\operator{I}\right]
\end{equation}
where $\delta(x)$ is the Dirac delta function. Practically, it is approximated by a sharp Gaussian:
\[
	\delta(x) \approx \sqrt{\frac{b}{\pi}}\exp(-bx^2)
\]
where $b$ is large. $\xi(t)$ represents the desired path of $\exval{\operator{\mu}}\!(t)$. When $\exval{\delta[\operator{X}-\xi(t)\operator{I}]}$ is maximized, the density of the wave function in the neighbourhood of $x=\xi(t)$ is also maximized. Hence:
\[
	 \exval{\operator{\mu}}\!(t) = \exval{\operator{X}}\!(t) \approx \xi(t)
\]
By specifying the desired path $\xi(t)$, we will hopefully be able to control the path of $\exval{\operator{\mu}}\!(t)$.
 
In this approach the problem is treated \emph{semi-classically} by considering the wave-function as a localized object. This restricts considerably the possible mechanisms for varying $\exval{\operator{\mu}}\!(t)$ along a predefined path. Moreover, nothing ensures that the semi-classical mechanism is at all possible for a given path, while other mechanisms may be possible.

The method was employed rather successfully for a path that was known in advance to be possible.

If the method is to be used for harmonic generation, $\xi(t)$ has to be defined as an oscillating function with the desired frequency. This has to be combined with the restriction of the forcing field spectrum.

The authors reported (in 2005) on a research in high harmonic generation control. We assume, that the brute force method (introduced by two of the authors, Werschnik and Gross) was employed for the restriction of the forcing field spectrum. The results of this research were not published. We know from private communications that the reason is that this attempt has failed.

A possible explanation for the failure of this attempt is the one mentioned above: the method enforces a semi-classical mechanism, which is not necessarily possible. If this is the reason for the failure of this method, another formulation for controlling $\evmu$ is necessary for the harmonic generation problem. Another possible explanation is that the brute force method is applicable only for simpler targets, like in Sec.~\ref{sec:OCTf}. If this is the reason for the failure of this attempt, other methods for restricting the field spectrum may be attempted, combined with the formulation described in this section.

%

	\chapter{The new method}\label{ch:new}
%
%
%
%
In this chapter, we present the new method for harmonic generation by QOCT\@. We present a new formulation for the maximization problem, and discuss the numerical methods that can be applied in this formulation.

As an introduction, we point out the main difficulty in formulating the harmonic generation problem in the context of QOCT\@. The quantum dynamics is formulated in the \emph{time domain}. In the common QOCT formulation the quantum dynamics is represented by the time-dependent Schr\"odinger equation. In the regular control problems (that were presented in the previous chapter, Sec.~\ref{sec:OCTf}, \ref{sec:OCTt}), the maximization requirements are related to one, or many time points, each with its own requirement. Hence, the whole problem is naturally formulated in the time domain. As for the harmonic generation problem, the requirements on the forcing and emitted fields are related to the \emph{frequency domain}. Such requirements are naturally formulated in the frequency domain, but not in the time domain. A requirement on a single frequency is related to the whole time sequence simultaneously and not to a single time point. When we deal with requirements on a spectrum sequence, with many frequencies, the situation is even more complex.

The difficulty is to create a unified formulation that takes into account the quantum dynamics and satisfies also the frequency requirements. We have already seen several ways to deal with this difficulty, in the discussion on the methods for restricting the forcing field spectrum (Sec.~\ref{sec:theolim}).

The approach that was adopted here to deal with this problem is very simple: Each part of the problem is handled in its natural domain --- the quantum dynamics is formulated in the time domain, by the time-dependent Schr\"odinger equation, while the frequency requirements are formulated in the frequency domain, by using appropriate functionals. The switch between the time and frequency domains is made by the cosine transform (we adopted the cosine transform, instead of the more commonly used Fourier transform, for reasons that will be mentioned). This makes the resulting equations to look somewhat cumbersome; nevertheless, it is possible to recognize the meaning of the resulting forms. 

The treatment of the problem of harmonic generation is divided in this chapter into the two parts of our task, mentioned at the beginning of the previous chapter: in Sec.~\ref{sec:methlim}, we propose a new formulation for handling the general problem of imposing a restriction on the forcing field spectrum. Sec.~\ref{sec:methnum} deals with the numerical methods that are appropriate for this formulation. In Sec.~\ref{sec:methHG}, we introduce a new maximization functional for the requested spectrum of the dipole, and present the full maximization problem for harmonic generation.

\section{New formulation for imposing restrictions on the forcing field spectrum in QOCT}\label{sec:methlim}
In this section, we present a new formulation for dealing with the general problem of restricting the spectrum of the forcing field. In Sec.~\ref{sec:methHG}, this formulation will be used for the harmonic generation problem.

As will be seen, there is a close relationship between all the methods of Sec.~\ref{sec:theolim} and this formulation.

	\subsection{Frequency dependent penalty factor}\label{ssec:epsw}
Before we start, we introduce the \emph{cosine transform}, that will be used frequently throughout this chapter. We denote it by the symbol $\mathcal{C}$. The cosine transform of an arbitrary function $g(t)$ is defined as:
\begin{equation}\label{eq:costrans}
	\mathcal{C}[g(t)] \equiv \sqrt{\frac{2}{\pi}}\int_0^\infty g(t)\cos(\omega t)\,dt
\end{equation}
The function $\bar{g}(\omega)$ is defined as the result of the operation of the cosine transform on $g(t)$:
\begin{equation}\label{eq:bardef}
	\bar{g}(\omega) \equiv \mathcal{C}[g(t)]
\end{equation}
We call it: ``the cosine transform of $g(t)$''.

The inverse of the cosine transform is the \emph{inverse cosine transform}, that will be denoted as $\mathcal{C}^{-1}$:
\begin{equation}\label{eq:invcostrans}
	\mathcal{C}^{-1}[\bar{g}(\omega)] \equiv \sqrt{\frac{2}{\pi}}\int_0^\infty \bar{g}(\omega)\cos(\omega t)\,d\omega = g(t)
\end{equation}
Note that the inverse cosine transform is the same as the cosine transform, except the variable names.

We prefer the cosine transform as a spectral tool over the more commonly used Fourier transform for two reasons:
\begin{enumerate}
	\item The cosine transform of a real function is also a real function. This is not true for the Fourier transform of a real function. By using the cosine transform for a real function, we avoid complications due to the appearance of a complex transformed function (we have already encountered such complications in the previous chapter, Sec.~\ref{sec:OCTf}, with $\ket{\psi(t)}$).
	\item When handling with the resulting equations numerically, the cosine transform is approximated by a discrete cosine transform (DCT)\@. In our case, the DCT is preferable over the discrete Fourier transform (DFT), because the ``ringing'' phenomenon (known also as ``aliasing'') is much less pronounced in the DCT; hence, we can avoid the use of ``windows''\footnote{The ringing results from discontinuities in the extended periodic function, represented by the truncated discrete Fourier series, or in its derivatives. This topic will be discussed further in Subsection~\ref{ssec:modif}. See also \cite{DCT}, for a more detailed explanation on the advantage of the DCT over the DFT\@.}.
\end{enumerate}

Now we return to our problem. The object of main interest for us is $\bar{\epsilon}(\omega)$, which represents the spectrum of the field. We want to find a way to affect its general shape. 

Note that the integration domain of the cosine transform does not have to be infinite in this case. The reason is that $\epsilon(t)$ is defined only in the time interval: $0\leq t\leq T$; hence, $\bar{\epsilon}(\omega)$ can be written as a finite time integral:
\begin{equation}\label{eq:epsw}
	\bar{\epsilon}(\omega) = \sqrt{\frac{2}{\pi}}\int_0^T \epsilon(t)\cos(\omega t)\,dt
\end{equation}

The same is true for the integration over $\omega$, in the inverse cosine transform: In practice, $\bar{\epsilon}(\omega)$ is negligible over some value of $\omega$. Moreover, when treating the problem numerically the maximal frequency possible is limited by the resolution of the time points by the Nyquist-Shannon theorem; according to this theorem, the maximal frequency cannot exceed $\pi/\Delta t$, where $\Delta t$ is the distance between neighbouring time points. Anyway, we can define the maximal relevant frequency in the system as $\Omega$. The inverse cosine transform can be written with $\Omega$ as the upper limit, instead of $\infty$:
\begin{equation}\label{eq:epst}
	\epsilon(t) =  \sqrt{\frac{2}{\pi}}\int_0^\Omega \bar{\epsilon}(\omega)\cos(\omega t)\,d\omega
\end{equation}

In order to restrict the spectrum of the forcing field, we introduce a new penalty functional, $J_{penal}$, which is formulated in the frequency domain instead of the time domain. We use a frequency dependent penalty factor:
\begin{equation}\label{eq:Jpenalw}
	J_{penal}[\bar{\epsilon}(\omega)] \equiv -\int_0^\Omega\alpha(\omega)\bar{\epsilon}^2(\omega)\,d\omega \qquad\qquad \alpha(\omega)>0
\end{equation}
The idea is to choose very large values for $\alpha(\omega)$ for the undesirable $\omega$ values, and small values for the desirable values. Later in this section, $\alpha(\omega)$ will attain a more precise meaning.

It is more convenient to replace the optimized function $\epsilon(t)$ by $\bar{\epsilon}(\omega)$. Accordingly, we replace the condition for an extremal in Eq.~\eqref{eq:dJdeps} by the following one:
\begin{equation}\label{eq:dJdepsw}
	\fnlderiv{J}{\bar{\epsilon}(\omega)} = 0
\end{equation}
When taking the functional derivative, we note that only $J_{penal}$ and $J_{con}$ are dependent on $\bar{\epsilon}(\omega)$ (when treated as independent of $\ket{\psi(t)}$ and $\ket{\chi(t)}$, according to the Lagrange-multiplier method). This makes the present treatment suitable for diverse kinds of problems (the same is true for the methods mentioned in the previous chapter). The dependence of $J_{penal}$ is simple. The dependence of $J_{con}$ is through $\epsilon(t)$ in the time-dependent Hamiltonian:
\begin{equation}\label{eq:Hepsw}
	\operator{H}(t) = \operator{H}_0 - \operator{\mu}\epsilon(t) = \operator{H}_0 - \operator{\mu}\left(\sqrt{\frac{2}{\pi}}\int_0^\Omega \bar{\epsilon}(\omega)\cos(\omega t)\,d\omega\right)
\end{equation}

The resulting Euler-Lagrange equation for $\bar{\epsilon}(\omega)$ is (the full derivation is given in App.~\ref{ap:derivation}):
\begin{equation}\label{eq:ELepsw}
	\epsw = -\sqrt{\frac{2}{\pi}}\frac{\int_0^T \Imag{\bracketsO{\chi(t)}{\operator{\mu}}{\psi(t)}}\cos(\omega t)\,dt}{\alpha(\omega)}
\end{equation}
This equation replaces Eq.~\eqref{eq:ELs2spsi} for $\epsilon(t)$ in regular control problems. In our problem, $\epsilon(t)$ can be easily computed from $\epsw$, by the inverse cosine transform.

The meaning of Eq.~\eqref{eq:ELepsw} will be more apparent if we define:
\begin{align}
	&\alpha(\omega) = \frac{\tilde{\alpha}}{\feps} \label{eq:deffw}\\
	&\tilde{\alpha} > 0 \label{eq:talphapos} \\
	&\int_0^\Omega \feps\, d\omega = 1 \label{eq:normf}
\end{align}
$\tilde{\alpha}$ is a positive constant. $\feps$ contains the dependence of $\alpha(\omega)$ on $\omega$. This function is always positive, due to the conditions in Eqs.~\eqref{eq:Jpenalw}, \eqref{eq:talphapos}. The normalization condition in \eqref{eq:normf} makes $\feps$ and $\tilde{\alpha}$ well defined.

We also define, for convenience, the following function of $t$:
\begin{equation}\label{eq:eta}
	\eta(t) \equiv -\frac{\Imag{\bracketsO{\chi(t)}{\operator{\mu}}{\psi(t)}}}{\tilde{\alpha}}
\end{equation}
Note that the RHS is just the expression for $\epsilon(t)$ in Eq.~\eqref{eq:ELs2seps}, for the regular control problems, where $\alpha$ from \eqref{eq:ELs2seps} is replaced by $\tilde{\alpha}$.

Now, let us write Eq.~\eqref{eq:ELepsw} as follows:
\begin{align}
	\epsw =& \feps\sqrt{\frac{2}{\pi}}\int_0^T \eta(t)\cos(\omega t)\,dt \nonumber \\
	=& \feps\mathcal{C}[\eta(t)] \label{eq:ELepswct}
\end{align}
The expression for $\epsilon(t)$ in our problem can be written as:
\begin{equation}\label{eq:ELepstct}
	\epsilon(t) = \mathcal{C}^{-1}\left\lbrace \feps\mathcal{C}[\eta(t)]\right\rbrace
\end{equation} 
Now, the meaning of the resulting Euler-Lagrange equation is obvious: We start from the field in the regular control problems, represented by $\eta(t)$; $\tilde{\alpha}$ has the role of a global, constant penalty factor. We transform this field to the frequency domain. Then, we multiply it by $\feps$, which has the meaning of a \emph{filter function}. The undesirable frequency components are filtered out, and an envelope function can be forced on the profile of the spectrum. Then, we transform the resulting spectral function back to the time domain, and get the $\epsilon(t)$ sequence.

The simplest choice for $\feps$ is the rectangular function. If the allowed frequencies are in the interval: $[\omega_{min},\;\omega_{max}]$, then the rectangular function is:
\begin{align}
	\feps =& 
	\begin{cases}
		0 & \qquad 0\leq\omega <\omega_{min} \\
		\frac{1}{\omega_{max} - \omega_{min}} & \qquad\omega_{min}\leq\omega \leq \omega_{max} \\
		0 & \qquad\omega_{max}<\omega
	\end{cases} \nonumber \\
	=& \frac{1}{\omega_{max} - \omega_{min}}u(\omega - \omega_{min})u(\omega_{max} - \omega) \label{eq:rect}
\end{align}
where $u(x)$ is the Heaviside step function, defined as:
\begin{equation}\label{eq:Heaviside}
	u(x) =
	\begin{cases}
		0 & \qquad x<0 \\
		1 & \qquad 0 \leq x
	\end{cases}
\end{equation}
With this choice for $\feps$, we get a complete filtration outside the interval. To be more strict, the value of $\feps$ cannot be absolutely $0$ --- otherwise, $\alpha(\omega)$ is undefined. It is more precise to refer to the \emph{limit}, in which outside the interval $\feps$ tends to $0$, and then, $\alpha(\omega)$ tends to $\infty$. Nevertheless, Eq.~\eqref{eq:rect} can be used as is for practical purposes, together with Eq.~\eqref{eq:ELepswct}.

\begin{figure}  
	\centering \includegraphics[width=3.0in]{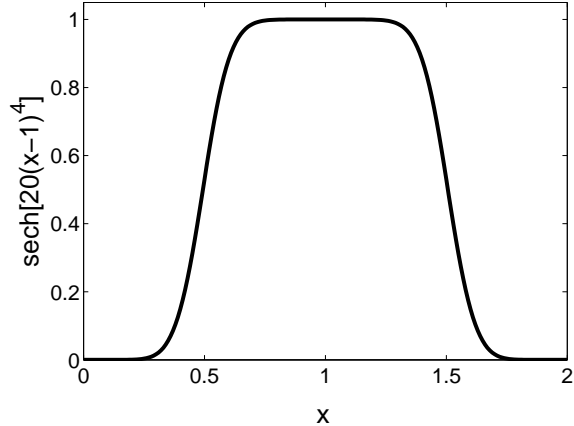}
	\caption{An example of a ``hat function'' --- the function: \text{$y(x)=\sech[20(x-1)^4]$}}\label{fig:hat}
\end{figure}

Other choices for $\feps$ are possible; for instance, we can choose a Gaussian function, or a ``hat function'' (see Fig.~\ref{fig:hat}, for an example of a hat function). These functions can be used when we want a ``smooth'' filtration. We can also choose $\feps$ with a special shape to enforce a desired envelope shape on the profile of $\epsw$.

The division of the quantity $1/\alpha(\omega)$ into the fraction of two quantities: $\tilde{\alpha}$ and $\feps$, is instructive conceptually; however, it has no meaning for practical use. The reason is that the exact value of $\tilde{\alpha}$ has no precise meaning --- it has to be determined by a trial and error process, and cannot be known in advance. Hence, there is no advantage in normalizing $\feps$. In practice, it is convenient to use the function $\tilde{f}_{\epsilon}(\omega)$ instead, defined as:
\begin{equation}\label{eq:tilf}
	\tilde{f}_{\epsilon}(\omega) = \frac{1}{\alpha(\omega)}
\end{equation}
For practical use, Eq.~\eqref{eq:ELepsw} is written conveniently as:
\begin{equation}\label{eq:ELepswuse}
	\epsw = \tilde{f}_{\epsilon}(\omega)\mathcal{C}\left[-\Imag{\bracketsO{\chi(t)}{\operator{\mu}}{\psi(t)}}\right]
\end{equation}

	\subsection{The relation between the new formulation and the existing methods}\label{ssec:relation}
Now we show the relation between the new formulation and the existing methods for restricting the spectrum of the forcing field. The formulation of Degani \etal is shown to be equivalent to the new formulation; that of Skinner and Gershenzon is shown to be a special case of it. There is also a close relation between the new formulation and the method employed by Werschnik and Gross.

We start with the method of Degani \etal . In our treatment of the new $J_{penal}$, we preferred to work in the frequency domain, with $\epsw$ as a variable; in this way, the frequency requirements are expressed more naturally. However, it is also possible to translate the whole problem to the time domain, with $\epsilon(t)$ as a variable. We express $\epsw$ in $J_{penal}$ as a cosine transform of $\epsilon(t)$ (Eq.~\eqref{eq:epsw}). The resulting $J_{penal}$ is expressed in the terms of $\epsilon(t)$:
\begin{equation}\label{eq:Jpenalwttp}
	J_{penal}[\epsilon(t)] = -\int_0^\Omega \alpha(\omega) \sqrt{\frac{2}{\pi}}\int_0^T \epsilon(t)\cos(\omega t)\,dt\, \sqrt{\frac{2}{\pi}}\int_0^T \epsilon(t')\cos(\omega t')\,dt'\,d\omega
\end{equation}
After rearrangement, Eq.~\eqref{eq:Jpenalwttp} can be written as follows:
\begin{align}
	& J_{penal}[\epsilon(t)] = - \int_0^T\negthickspace\int_0^T\epsilon(t)\beta(t, t')\epsilon(t')\,dt\,dt' \label{eq:penalwttp} \\
	& \beta(t, t') = \frac{2}{\pi}\int_0^\Omega\alpha(\omega)\cos(\omega t)\cos(\omega t')\,d\omega \label{eq:betaw} 
\end{align}
Note that Eq.~\eqref{eq:penalwttp} is the same as~\eqref{eq:penalttp} in the formulation of Degani \etal .

Alternatively, Eqs.~\eqref{eq:penalwttp}, \eqref{eq:betaw} can be written as:
\begin{align}
	& J_{penal}[\epsilon(t)] \equiv - \int_0^T\epsilon(t)\operator{B}\epsilon(t)\,dt \label{eq:penalwttpO}\\
	& \operator{B}\epsilon(t) = \mathcal{C}^{-1}\left\lbrace \alpha(\omega)\mathcal{C}[\epsilon(t)]\right\rbrace	\label{eq:Bw}
\end{align}
Eq.~\eqref{eq:penalwttpO} is the same as~\eqref{eq:penalttpO}. The meaning of the operator $\operator{B}$ in Eq.~\eqref{eq:Bw} is apparent: it transforms $\epsilon(t)$ to the frequency domain; then each frequency component is multiplied by its own penalty factor, $\alpha(\omega)$; the resulting function of $\omega$ is transformed back to the time domain.

The operator $\operator{B}$ proposed by Degani \etal (Eq.~\eqref{eq:DegB}) can be written in the form of the operator in Eq.~\eqref{eq:Bw}, with an appropriate $\alpha(\omega)$. Suppose that the ``good'' field functions are in the frequency domain:~$[0,\;\omega_{max}]$; then, the projection operators from~\eqref{eq:DegB} are defined by the following expressions:
\begin{align*}
	& \operator{P}_{good}\epsilon(t) = \mathcal{C}^{-1}\left\lbrace u(\omega_{max} - \omega)\mathcal{C}[\epsilon(t)]\right\rbrace \\
	& \operator{P}_{bad}\epsilon(t) = \mathcal{C}^{-1}\left\lbrace u(\omega - \omega_{max})\mathcal{C}[\epsilon(t)]\right\rbrace
\end{align*}
(Of course, another spectral transform can be chosen instead of the cosine transform.) The operator $\operator{B}$ from Eq.~\eqref{eq:DegB} can be written as in~\eqref{eq:Bw}, with the following $\alpha(\omega)$ (using the linearity property of the cosine transform):
\begin{align}
	\alpha(\omega) =& \alpha_{good}u(\omega_{max} - \omega) + \alpha_{bad}u(\omega - \omega_{max}) \nonumber \\
	 =&
	\begin{cases}
		\alpha_{good} &\qquad 0\leq\omega\leq\omega_{max} \\
		\alpha_{bad} &\qquad \omega_{max}<\omega
	\end{cases}	\label{eq:wdeg}
\end{align}
Note that Degani \etal use an extended definition of $\alpha(\omega)$, where it can attain negative values. Then, the interpretation of $\alpha(\omega)$ that was given here is not appropriate. We have decided to avoid using negative values to prevent the possibility of singularities in the hypersurface of $J$ (see Sec.~\ref{sec:OCTf} in the previous chapter).

We see that the formulation of Degani \etal is equivalent to ours. The difference between the two approaches is that we work in a basis in which $\operator{B}$ is diagonal --- the frequency basis, while Degani \etal work in a basis in which it is non-diagonal. It follows that our treatment is one-dimensional, while the treatment in the formulation of Degani \etal is two-dimensional. This is due to the fact that the requirements on the frequency are naturally formulated in the frequency domain, and the resulting expression for $J_{penal}$ is considerably simpler. The resulting Euler-Lagrange equation for the field is also more naturally expressed in the frequency domain --- compare the integral equation~\eqref{eq:ELDeg} for $\epsilon(t)$, with Eq.~\eqref{eq:ELepswuse} for $\epsw$.

One important disadvantage of the approach of Degani \etal is that a complete filtration, like the one achieved by the rectangular $\feps$, is impossible. The reason is that $\alpha(\omega)$ diverges, and there is no way to create the corresponding $\beta(t, t')$. The problem is not restricted only to a rectangular filter function --- any function that decays very rapidly (as in the case of an exponential decay, \eg in a Gaussian function, or in the hat function of Fig.~\ref{fig:hat}) is not suitable, because $\alpha(\omega)$ attains very large values that are not acceptable numerically. Another problem with rapidly decaying functions is the very sharp shape of the resulting $\alpha(\omega)$, which is not suitable for numerical treatment. For instance, in practice, the integral in Eq.~\eqref{eq:betaw} cannot be performed numerically, if $\alpha(\omega)$ is too sharp. In Sec.~\ref{sec:methnum} we will discuss another numerical problem which has a similar origin.

The two formulations result also in different numerical approaches for treating the problem; this will be discussed in Sec.~\ref{sec:methnum}.

Now we show that the formulation of Skinner and Gershenzon is a special case of the new formulation.

Consider the following filter function:
\begin{equation}\label{eq:fdeltas}
	\feps = \frac{2}{2N + 1}\sum_{n=0}^N \delta(\omega - n\,\Delta\omega)
\end{equation}
The factor $2/(2N+1)$ is a normalization constant (note that integration on $\delta(\omega)$ for the $n=0$ term gives only $1/2$, because the integration domain in Eq.~\eqref{eq:normf} starts from $0$). Let us derive the expression for $\epsilon(t)$, for this filter function. From Eq.~\eqref{eq:ELepswct} we have:
\begin{equation}\label{eq:epsweta}
	\epsw = \feps\etaw
\end{equation}
Using Eq.~\eqref{eq:epst}, we get the following expression for $\epsilon(t)$:
\begin{align}
	\epsilon(t) =& \sqrt{\frac{2}{\pi}}\int_0^\Omega \feps\etaw\cos(\omega t)\,d\omega \nonumber \\
	=& \sqrt{\frac{2}{\pi}}\,\frac{2}{2N + 1}\int_0^\Omega \sum_{n=0}^N \delta(\omega - n\,\Delta\omega) \etaw\cos(\omega t)\,d\omega \nonumber \\
	=& \sqrt{\frac{2}{\pi}}\,\frac{2}{2N + 1}\left[\frac{1}{2}\bar{\eta}(0)+\sum_{n=1}^N\bar{\eta}(n\,\Delta\omega)\cos(n\,\Delta\omega\,t)\right] \label{eq:epstsum}
\end{align}
Comparing Eq.~\eqref{eq:epstsum} with Eq.~\eqref{eq:epscosseries}, we can recognize that we got the same expression for $\epsilon(t)$ as that was used by Skinner and Gershenzon, with:
\begin{equation}\label{eq:coscoef}
	a_n \equiv \sqrt{\frac{2}{\pi}}\,\frac{2}{2N + 1}\times
	\begin{cases}
		\frac{1}{2}\bar{\eta}(0) & \qquad n=0 \\
		\bar{\eta}(n\,\Delta\omega) & \qquad n=1,2,\ldots, N
	\end{cases}
\end{equation}
Eq.~\eqref{eq:coscoef} is the Euler-Lagrange equation for the $a_n$'s. We see that the problem of optimizing a continuous function was reduced, with this choice of $\feps$, to the problem of optimizing a discrete set of parameters.

It is important to note, that in numerical computation the general continuous problem of the new formulation is approximated by a discretized version of the problem. This amounts to replacing the continuous cosine transform by a discrete cosine transform, which is a cosine series. If we also use a rectangular $\feps$, the problem solved numerically is essentially the same as in the formulation of Skinner and Gershenzon (see App.~\ref{ap:num}). The present formulation is still more general, because of the possibility of using other forms of $\feps$, for a smooth filtration, or in order to affect the shape of the spectrum. These advantages are applicable also for a  discrete version of the problem; starting from the general continuous problem of the new formulation, we may use a generalized version of Eq.~\eqref{eq:fdeltas}:
\begin{equation}\label{eq:fdeltasgen}
	\feps = \sum_{n=0}^N f_{\epsilon}^{(n)}\delta(\omega - n\,\Delta\omega)
\end{equation}
where the $f_{\epsilon}^{(n)}$ are a set of predefined constants which determine the desired shape of the spectrum in a similar manner to the continuous case. The derivation of the Euler-Lagrange equations is the same as above.

The brute force method of Werschnik and Gross is not equivalent to the new formulation. However, the new formulation provides some justification to this method.

Compare the field that is used to propagate $\ket{\psi(t)}$ in the brute force method (Eq.~\eqref{eq:Grossfil}, together with Eq.~\eqref{eq:Krotovteps}), to the field from Eq.~\eqref{eq:ELepstct}. The two equations are very similar. The only differences are:
\begin{enumerate}
	\item The Fourier transform from Eq.~\eqref{eq:Grossfil} is replaced by a cosine transform in Eq.~\eqref{eq:ELepstct}.
	\item $\alpha$ of the regular formulation is replaced by $\tilde{\alpha}$.
	\item $f(\omega)$ from Eq.~\eqref{eq:ELepstct} is replaced by $\feps$.
\end{enumerate}
According to the interpretation given here for $\tilde{\alpha}$ and $\feps$, the meaning of the two equations is essentially the same.

However, the field used to propagate $\ket{\chi(t)}$ in the brute force meth\-od (Eq.\ \eqref{eq:Krotovteps}), is not justified by the new formulation --- the function $\eta(t)$ does not have the interpretation of the field appropriate for this problem, before it is filtered, as in Eq.~\eqref{eq:ELepstct}.

\section{Numerical methods for the new formulation}\label{sec:methnum}
In this section, we discuss a few numerical methods that can be suggested for maximizing $J$, with the $J_{penal}$ presented in Sec.~\ref{sec:methlim}.

Usually, the Krotov method (or one of its variants) is considered the preferable method for solving QOCT problems. This is because of its monotonic convergence property and the relative ease of its application. However, the Krotov method is not applicable for the new formulation; the reason is, that $\epsilon(t)$ from Eq.~\eqref{eq:ELepstct} lacks the property of being dependent only on functions of the time $t$. This makes it necessary to look for other numerical methods, that are applicable and efficient when employed for this problem.

There are two possible approaches for numerical solution to our problem:
\begin{enumerate}
	\item We can use the Krotov-like approach, presented in the previous chapter (Subsection~\ref{ssec:Degani}). In this approach, we utilize the fact that the sequence of $\ket{\psi(t)}$ or $\ket{\chi(t)}$ does not have to be determined simultaneously in all times, for updating the field during the propagation. For applying the Krotov-like approach it is necessary to formulate the problem in the time domain, in the way that was presented in the previous chapter. Adopting this approach means, essentially, turning to the existing method presented by Degani \etal , with a somewhat different approach for the $\operator{B}$ operator.
	\item We can use the more natural formulation of the problem in the frequency domain, and employ general methods for nonlinear optimization problems (the time-domain formulation is not preferable for this approach, because the resulting update rules are more complicated, and more expansive numerically).
\end{enumerate}
Subsection~\ref{ssec:Degproblems} deals with the first approach; Subsections~\ref{ssec:BFGS}--\ref{ssec:relax} represent the second.  

	\subsection{The Krotov-like method of Degani \etal}\label{ssec:Degproblems}
When we tried to implement the Krotov-like algorithm, we encountered problems. The algorithm has been employed for the harmonic generation problem, with $J_{max}$ that has not been introduced yet, and will be introduced in Sec.~\ref{sec:methHG}. Anyway, the problems that we have encountered with seem to be general problems with this algorithm. Other groups also report (in private communications) on similar difficulties when applying the algorithm to the common control problem of Sec.~\ref{sec:OCTf}.

The algorithm was found to be causing a numerical instability during the propagation process. The origin of the instability was found to be the fact that $\beta(t, t')$ is very oscillatory in nature. The amplitudes of the oscillations become larger, when $\alpha(\omega)$ is chosen to attain larger values for undesirable frequency components. Larger oscillations increase the numerical instability.

It is necessary to mention here that the harmonic generation problem is a difficult problem. One reason is the typically small amplitudes of the higher frequency oscillations of the dipole that can be achieved by the low frequency field. Another reason is the difficulty in achieving a maximum, probably because of a complex structure of the hypersurface of the optimization functional $J$. It follows, that there is a need for very large values of $\alpha(\omega)$, to achieve a satisfactory filtration; otherwise, the optimization algorithm will prefer an easier path for increasing $J$, using high frequency fields. The values of $\alpha_{bad}$ that were used by Degani \etal in \cite{Degani} are several orders of magnitude smaller than the values that are required for the harmonic generation problem. When using $\alpha(\omega)$ values of the same order of magnitude  that was used in \cite{Degani} the numerical difficulty was found to be acceptable; however, such values do not satisfy the requirements of the harmonic generation problem.

The origin of this general problem with the algorithm can be considered to be the unnatural formulation of the problem in the time domain. This formulation makes it necessary to use the time-dependent, two-dimensional penalty function $\beta(t, t')$, instead of the much simpler $\alpha(\omega)$. The relatively simple frequency requirements of the problem are expressed in the time domain only by a complicated, non-smooth structure of $\beta(t, t')$, with large oscillations of very high frequencies.

Another problem that we encountered (even with relatively small values of $\alpha(\omega)$) is related to the convergence characteristics of the algorithm: the algorithm has shown monotonic convergence characteristics only for forcing fields with relatively low intensities. When the forcing field intensity became larger, the monotonic convergence characteristics were destroyed. We have not tried to find the origin of this problem. Possibly, the problem is just a result of inaccuracies in the propagation process due to the numerical instability mentioned above. On the other hand, the fact that this problem was reported also by other groups, together with the absence of available convergence analysis, allows to raise the suspicion that this Krotov-like algorithm is not always monotonically convergent (although it may show frequently monotonic convergence properties, as shown in \cite{Degani}, a fact that may be useful).

It is noteworthy that we tried to use another Krotov-like algorithm, based on updating the field in every time step of the propagation by Eq.~\eqref{eq:ELepstct}. This update rule is even less self-consistent than that used by Degani \etal . Although this algorithm shows sometimes monotonic convergence characteristics, usually it is even worse than the algorithm of the naive approach (Subsection~\ref{ssec:naive}).

The failure of the attempt to apply a Krotov-like algorithm for our problem, makes it necessary to turn to the second approach mentioned above.

	\subsection{The BFGS method}\label{ssec:BFGS}
A natural choice for a general optimization algorithm is a quasi-Newton algorithm, because of the good convergence characteristics near the maximum. In this subsection, we deal with the BFGS method, which is the most commonly used quasi-Newton method.

In the new formulation for imposing restrictions on the field spectrum, the update rule is more conveniently formulated for $\epsw$. Hence, we have to rewrite all the relevant equations for the quasi-Newton methods (Subsection~\ref{ssec:grad}) with $\epsw$ as the variable. The gradient with respect to $\epsw$ is\footnote{Eq.~\eqref{eq:gradw} raises a problem: when using a rectangular $\feps$ (Eq.~\eqref{eq:rect}), $\alpha(\omega)$ diverges for the undesired $\omega$ values. However, for these values, $\epsw$ is certainly $0$, so we can set in advance:
\[
	\epsw = 0, \qquad \text{for:\quad} \omega<\omega_{min},\; \omega>\omega_{max}
\]
without including these components of the field in the optimization process. This is equivalent to computing $\epsilon(t)$ only from frequency components in the domain:~$[\omega_{min},\;\omega_{max}]$, by changing the integration domain in Eq.~\eqref{eq:epst} to this domain. The integration domain in~\eqref{eq:Jpenalw} has to be changed accordingly. A similar problem exists for rapidly decaying $\feps$ functions, as was mentioned in another context in Subsection~\ref{ssec:relation}. The solution for this problem is similar: $\epsw$ is set to $0$ when the corresponding $\alpha(\omega)$ is large enough, so $\epsw$ can be assumed to be negligible (note that such a solution is impossible when dealing with the time domain formulation). The same problem exists for the computation of $J$ during the line-search, and the solution is similar.}:
\begin{equation}\label{eq:gradw}
	\nabla_{\epsw}J = \fnlderiv{J}{\epsw} = -2\left\lbrace\alpha(\omega)\epsw + \mathcal{C}\left[\Imag{\bracketsO{\chi(t)}{\operator{\mu}}{\psi(t)}}\right]\right\rbrace
\end{equation}
The Hessian is defined as:
\begin{equation}\label{eq:HessJw}
	s(\omega, \omega ') = \fnlderivdd{J}{\epsw}{\bar{\epsilon}(\omega ')}
\end{equation}
The operation of the Hessian on an arbitrary function of $\omega$, $g(\omega)$, is defined by:
\begin{equation}\label{eq:HessopJw}
	\operator{S}g(\omega) = \int_0^T s(\omega, \omega ')g(\omega ')\,d\omega '
\end{equation}
The update rule in Eq.~\eqref{eq:tryN} is replaced by the following equations:
\begin{align}
	& \bar{\epsilon}^{trial}(\omega) = \bar{\epsilon}^{(k)}(\omega) - K\,\operator{S}_{ap}^{(k)-1}[\nabla_{\epsw}J]^{(k)} \label{eq:tryNw} \\
	& \epsilon^{trial}(t) = \mathcal{C}^{-1}[\bar{\epsilon}^{trial}(\omega)] \label{eq:tryNwepst}
\end{align}
After the line-search was performed, and an optimal $K$ was found, $\epsw$ is updated; we add to Eq.~\eqref{eq:prgup} the additional update:
\begin{equation}\label{eq:upNw}
	\bar{\epsilon}^{(k+1)}(\omega) = \bar{\epsilon}^{trial}(\omega)
\end{equation}

The BFGS method was implemented by using the \verb"fminunc" function from the ``Optimization Toolbox'' of MATLAB (see~\cite{mathworks} for details). The method was applied successfully to simple problems of a two-level-system (TLS), with the $J_{max}$ of sections~\ref{sec:OCTf} and~\ref{sec:methHG}. Unfortunately, the method was found to be much time consuming in this case, so this option is not convenient for larger-scale problems. This raises the motivation to find an alternative method.

	\subsection{The relaxation method}\label{ssec:relax}
The ``relaxation method'' is a general method for ``helping'' an iterative process to converge. To our knowledge, it has not been employed yet for QOCT problems. In this subsection, we start from the description of the relaxation method, in the context of the simpler problems of the previous chapter, sections~\ref{sec:OCTf}, \ref{sec:OCTt}. A mathematical justification to the method is given. Numerical results for the simpler problems are presented . Then, the method is discussed in the context of the new formulation of Sec.~\ref{sec:methlim}. Numerical examples will be given for problems with the $J_{max}$ of Eq.~\eqref{eq:Jmaxs2s}.

The relaxation method, in the context of regular QOCT problems, is based on the following update rule:
\begin{align}
	& \epsilon^{new}(t) = K\epsilon^{EL}(t) + (1-K)\epsilon^{old}(t) \qquad 0<K\leq 1 \label{eq:relaxt} \\
	& \epsilon^{EL}(t) \equiv -\frac{\Imag{\bracketsO{\chi(t)}{\operator{\mu}}{\psi(t)}}}{\alpha}\biggm|_{\epsilon(t)=\epsilon^{old}(t)} \label{eq:epsEL}
\end{align}
$\epsilon^{EL}(t)$ is the field from the Euler-Lagrange equation~\eqref{eq:ELs2seps}, using the $\ket{\psi(t)}$ and $\ket{\chi(t)}$ sequences, propagated by $\epsilon^{old}(t)$. Actually, this is the new field, according to the update rule in the naive approach (see Subsection~\ref{ssec:naive}). The idea is to ``mix'' the old solution with the new solution of the naive approach. $K$ is a parameter, which determines the weights of the old and new solutions. Its value has to be determined by a trial and error process. Note that if we take: $K=1$, we return to the naive approach update rule. Although the naive approach seldom converges, the relaxation method may be convergent, by an appropriate choice of $K$. Usually, the value of $K$ should be decreased during the optimization process.

We propose the following scheme for the implementation of the relaxation method, for the standard QOCT problems:
\begin{enumerate}
	\item Guess a field sequence: $\epsilon^{(0)}(t)$. \label{pr:eps0}
	\item Guess an initial value for $K$. \label{pr:upK}
	\item Propagate $\ket{\psi^{(0)}(t)}$ forward from $t=0$ to $t=T$, with $\epsilon^{(0)}(t)$.
	\item Calculate $J^{(0)}$ with $\ket{\psi^{(0)}(t)}$ and $\epsilon^{(0)}(t)$.
	\item (k = 0)
	\item Repeat the following steps, until convergence:\label{pr:rdoconv}
	\begin{enumerate}
		\item Set $\ket{\chi^{(k)}(T)}$ according to the suitable boundary condition, using  $\ket{\psi^{(k)}(T)}$.
		\item Propagate $\ket{\chi^{(k)}(t)}$ backward from $t=T$ to $t=0$, with $\epsilon^{(k)}(t)$.
		\item Do the following steps, and repeat while $J^{trial}\leq J^{(k)}$:
		\begin{enumerate}
			\item Set a new field, using Eq.~\eqref{eq:relaxt}:
			\begin{equation}\label{eq:tryr}
				\epsilon^{trial}(t) = K\left[-\frac{\Imag{\bracketsObiggm{\chi^{(k)}(t)}{\operator{\mu}}{\psi^{(k)}(t)}}}{\alpha}\right] + (1-K)\epsilon^{(k)}(t)
			\end{equation}
			\item Propagate $\ket{\psi^{trial}(t)}$ forward from $t=0$ to $t=T$, with $\epsilon^{trial}(t)$.
			\item Calculate $J^{trial}$ with $\ket{\psi^{trial}(t)}$ and $\epsilon^{trial}(t)$.
			\item If $J^{trial}\leq J^{(k)}$, then set: $K = K/2$
		\end{enumerate}
		\item Update all the variables:
		\begin{equation}
			\epsilon^{(k+1)}(t) = \epsilon^{trial}(t) \qquad \ket{\psi^{(k+1)}(t)} = \ket{\psi^{trial}(t)} \qquad J^{(k+1)} = J^{trial} \label{eq:pruprlx}		
		\end{equation}
		\item (k = k + 1)
	\end{enumerate}
\end{enumerate}

During this procedure, the value of $K$ decays exponentially with the number of its updates. Most frequently, it decays in the former iterations, and remains constant during the rest of the process.

We denote the initial guess for $K$ in step~\ref{pr:upK} as: $K_i$. Usually, $K_i=1$ is a good choice. However, often it may be advantageous to choose a smaller value, for the following reasons:
\begin{enumerate}
	\item When $\alpha$ is very small, a numerical instability in the propagation process might appear, unless $K$ is small enough (this numerical problem exists also in the Krotov method).
	\item Large $K$ may be advantageous in the former iteration; however, after many iterations, we may get a better result with a smaller $K_i$.
\end{enumerate}

Now we are going to give a mathematical justification for the relaxation method. We show, that the relaxation method is actually a \emph{quasi-Newton method}. As we mentioned in the previous chapter (Subsection~\ref{ssec:grad}), in practice, the Hessian (Eq.~\eqref{eq:HessJ}) cannot be computed in every iteration, because of the complex dependence of $\ket{\psi(t)}$ and $\ket{\chi(t)}$ on $\epsilon(t)$ (which are present in the expression for the gradient, Eq.~\eqref{eq:gradt}). Now we claim, that we can get a useful approximation to the Hessian by ignoring this dependence, and treating $\ket{\psi(t)}$ and $\ket{\chi(t)}$ as independent of $\epsilon(t)$. This is equivalent to the approximation of the Hessian by the Hessian of $J_{penal}$. To be more general, we include the possibility of a time-dependent penalty factor, $\alpha(t)$ (Eq.~\eqref{eq:penalt}). The approximation for the Hessian is defined by the following equations:
\begin{align}
	& s(t, t') \approx s_{ap}(t, t') \label{eq:aproxs}\\
	& s_{ap}(t, t') \equiv \fnlderivdd{J_{penal}}{\epsilon(t)}{\epsilon(t')} = -2\alpha(t')\delta(t'-t) \label{eq:saprlx}
\end{align}
The operation of the Hessian on an arbitrary function of $t$, $g(t)$, is defined by:
\begin{equation} \label{eq:Sapgrlx}
	\operator{S}_{ap}g(t) = -2\int_0^T\alpha(t')\delta(t'-t)g(t')\,dt' = -2\alpha(t)g(t)
\end{equation}
We see that $\operator{S}_{ap}$ is \emph{diagonal} in the time basis. Now, it is a trivial matter to get the inverse of $\operator{S}_{ap}$:
\begin{equation} \label{eq:invSap}
	\operator{S}_{ap}^{-1}g(t)=-\frac{1}{2\alpha(t)}g(t)
\end{equation}
Let us write the expression to the update rule in the quasi-Newton method (Eq.~\eqref{eq:qNtneps}), with the new $\operator{S}_{ap}$, using~\eqref{eq:gradt} and~\eqref{eq:invSap}:
\begin{align*}
	\epsilon^{new}(t) =& \epsilon^{old}(t) - K\,\operator{S}_{ap}^{-1}\nabla_{\epsilon(t)}J\biggm|_{\epsilon(t)=\epsilon^{old}(t)} \\
	=& \epsilon^{old}(t) - K\frac{1}{2\alpha(t)}2\left[\alpha(t)\epsilon^{old}(t) +\Imag{\bracketsO{\chi(t)}{\operator{\mu}}{\psi(t)}}\biggm|_{\epsilon(t)=\epsilon^{old}(t)}\right] \\
	=& K\epsilon^{EL}(t) + (1-K)\epsilon^{old}(t)
\end{align*}
which is the same as~\eqref{eq:relaxt}. Note, however, that $K$ has a slightly different definition in a quasi-Newton method, because it can attain, in principle, values greater than $1$.

In the regular Newton method, with the exact Hessian, $K$ is absent, which is equivalent to setting: $K=1$ (see Eq.~\eqref{eq:Newtoneps}). The fact that the naive approach, with $K=1$, is unsuccessful, indicates that the approximation of Eq.~\eqref{eq:aproxs} is not a very good one. Anyway, this approximation proves itself to be useful for determining the direction of search, in the context of a quasi-Newton method.

Note that when $\alpha$ is a constant, the operation of $\operator{S}_{ap}^{-1}$ on the gradient vector is just a multiplication by a negative constant. The resulting direction of search is the direction of the gradient, as in the first order gradient method (see Eq.~\eqref{eq:updateg}). However, the gradient vector, which determines the direction of search, is multiplied here by the constant $1/(2\alpha)$. This may have an effect on the numerical search procedure.

This new interpretation of Eq.~\eqref{eq:relaxt} raises the possibility of using the common scheme for the quasi-Newton methods, with a line-search in every iteration, instead of the decay of $K$ proposed above. To distinguish between the two possibilities, we will call the scheme that was proposed above: ``the decaying $K$ scheme'', and the one that employs a line-search: ``the line-search scheme''.

First, we discuss the numerical results of the decaying $K$ scheme.

To demonstrate the efficiency of the relaxation method, we compare the convergence curve of this method with that of the Krotov method. In the convergence curve, we plot $J$ vs.\ a variable which is a measure of the numerical effort; here it is the number of propagations. We have chosen two simple state-to-state problems, with $J_{max}$ of the form of Eq.~\eqref{eq:Jmaxs2s}, where the projection operator from Eq.~\eqref{eq:POf} stands for $\operator{O}$. 
 
\begin{figure}
	\centering \includegraphics[width=3.0in]{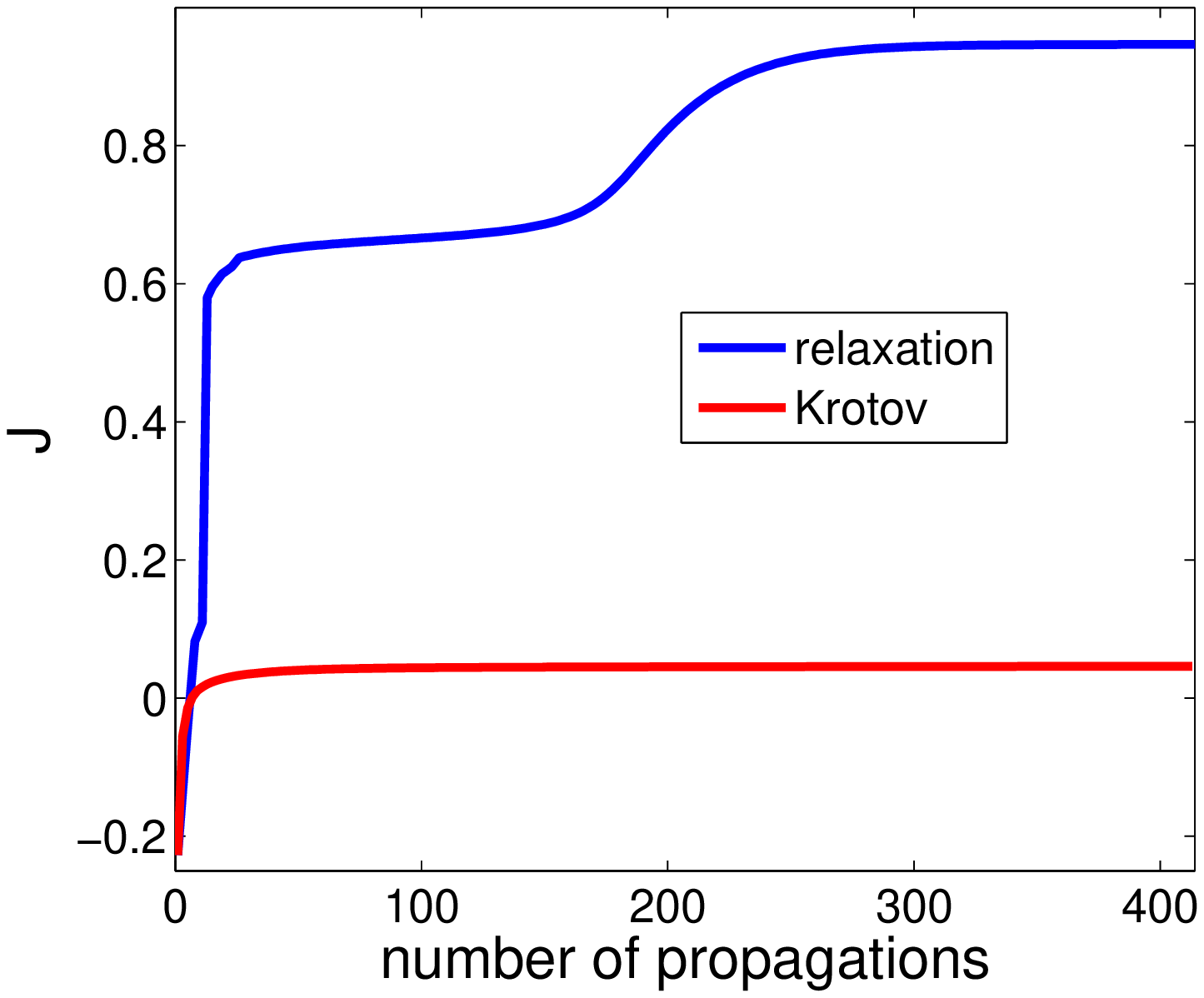} 
	\input{TLSKrcomparison}
	\caption{The convergence curves of the functional $J$, vs. the number of propagations, for a state-to-state problem of a TLS. The results for the Krotov and the relaxation methods are presented. The number of propagations represents the amount of numerical effort.}\label{fig:TLSs2s}
\end{figure}

The first problem is a TLS problem. The problem is to find a field for a transition from the initial state --- the ground-state, to the target state --- the excited state. In our TLS examples, we take the dipole moment operator to be the $x$ Pauli matrix:
\begin{equation}\label{eq:TLSmu}
	\operator{\mu}= \sigma_x =
	\begin{bmatrix}
		0 & 1 \\
		1 & 0
	\end{bmatrix}
\end{equation}
The rest of the details of this problem, and the resulting curve, are presented in Fig.~\ref{fig:TLSs2s}.

\begin{figure}
	\centering \includegraphics[width=3.0in]{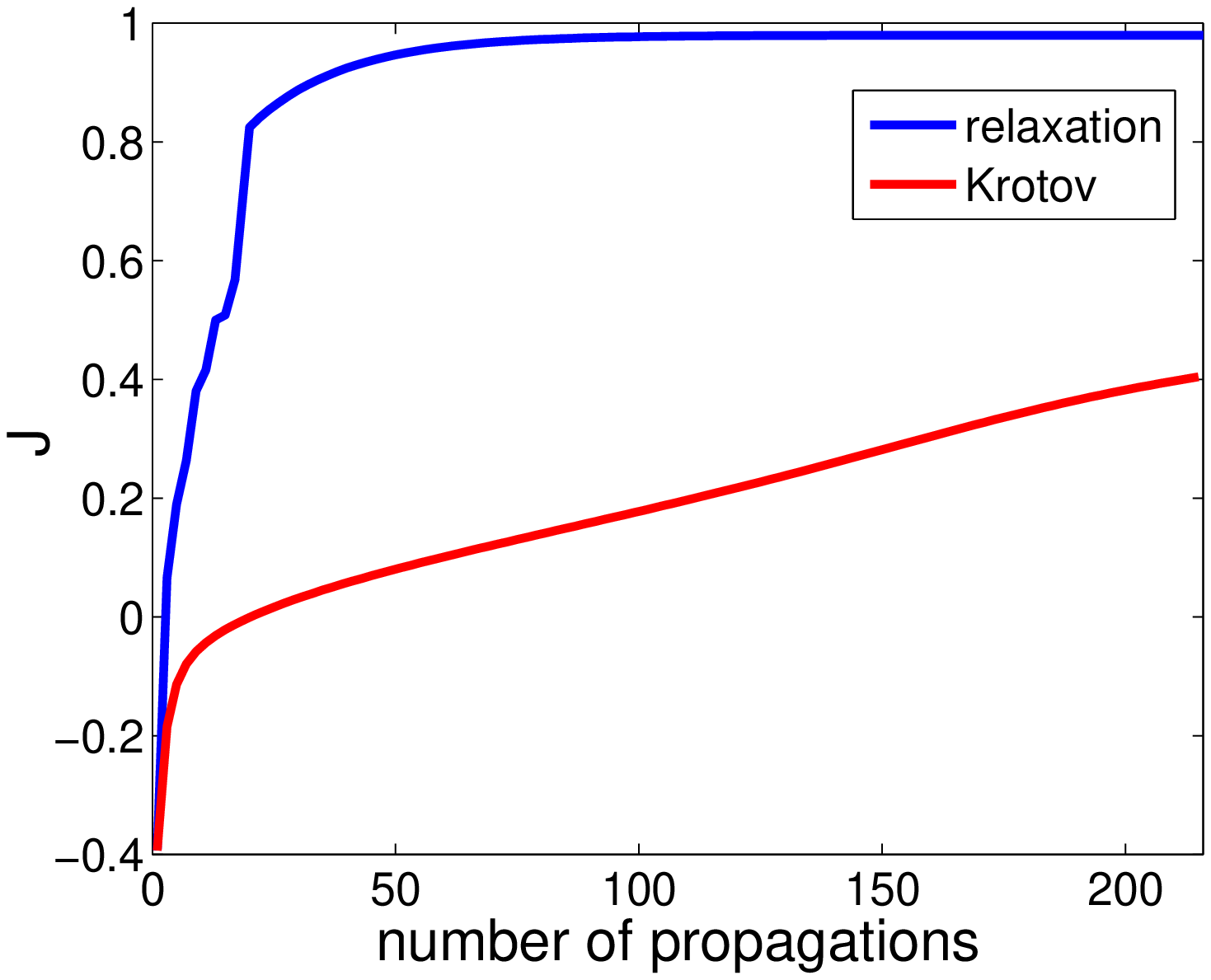}
	\input{HOKrcomparison}
	\caption{The convergence curves of the functional $J$, vs. the number of propagations, for a state-to-state problem of a harmonic oscillator. The initial state is the ground state, and the excited state is the ground state shifted one unit to the positive direction. The results for the Krotov and the relaxation methods are presented.}\label{fig:HOs2s}
\end{figure}

The system of the second problem is a one-dimensional harmonic oscillator. The initial state is the ground-state. The target state is a ``shifted'' ground-state. The details and the resulting curve are presented in Fig.~\ref{fig:HOs2s}.

The results show, that in these simple cases, the relaxation method is definitely advantageous over the Krotov method. An additional research is required to determine the efficiency of the relaxation method, compared with the Krotov method, in more complex problems. This is beyond the scope of our research. 

We have also tried the line-search scheme. It has been implemented using \verb"fminunc". We found that this method is much less expansive than the BFGS method, and it converges to better solutions. However, it is considerably more expansive than the decaying $K$ scheme. Moreover, somewhat surprisingly, the resulting solutions, after convergence, are of smaller $J$ values, compared with the decaying $K$ scheme solutions. Hence,  the decaying $K$ scheme was adopted in the present work, and is used in all the examples throughout this text.

It is noteworthy, that after the decaying $K$ scheme has converged to some solution, it is possible to get some improvement using the line-search scheme, with this solution as a first guess. We observed also, that a further  improvement can be achieved, by using the BFGS method after this stage.

To employ the relaxation method to the new formulation of restricting the forcing field spectrum we have to reformulate the update rule in the frequency domain. Eqs.~\eqref{eq:relaxt}, \eqref{eq:epsEL} are replaced by the following equations (compare Eq.~\eqref{eq:ELepswuse}):
\begin{align}
	& \bar{\epsilon}^{new}(\omega) = K\bar{\epsilon}^{EL}(\omega) + (1-K)\bar{\epsilon}^{old}(\omega) \qquad 0<K\leq 1 \label{eq:relaxw} \\
	& \bar\epsilon^{EL}(\omega) \equiv \tilde{f}_{\epsilon}(\omega)\mathcal{C}\left[-\Imag{\bracketsO{\chi(t)}{\operator{\mu}}{\psi(t)}}\biggm|_{\epsw=\bar\epsilon^{old}(\omega)}\right] \label{eq:epsELw}
\end{align}
Accordingly, Eq.~\eqref{eq:tryr} is replaced by the following equations:
\begin{align}
	&\bar\epsilon^{trial}(\omega) = K\tilde{f}_{\epsilon}(\omega) \mathcal{C} \left[-\Imag\bracketsObiggm{\chi^{(k)}(t)}{\operator{\mu}}{\psi^{(k)}(t)}\right] + (1-K)\bar\epsilon^{(k)}(\omega) \label{eq:tryrw}\\
	& \epsilon^{trial}(t) = \mathcal{C}^{-1}[\bar{\epsilon}^{trial}(\omega)] \label{eq:tryrlxepst}
\end{align}
The additional update of Eq.~\eqref{eq:upNw} is added to Eq.~\eqref{eq:pruprlx}.

A practical remark: the first guess of the field (stage~\ref{pr:eps0} in the procedure) should also be formulated in the frequency domain; a reasonable choice will be of the same form as $\feps$.

Now, the new method of restricting the forcing field spectrum is ready for application. We have chosen two examples of state-to-state problems. In both cases, the field is restricted to be around the frequency: $\omega=1_{a.u.}$. We use $\tfeps$ of the form of the hat function from Fig.~\ref{fig:hat}:
\begin{equation}\label{eq:hat}
	\tfeps = A\,\sech[20(\omega - 1)^4] \qquad A>0
\end{equation}
where $A$ is a positive parameter, adjusted to our needs.

The first example is of a TLS system, with the following unperturbed Hamiltonian:
\begin{equation}\label{eq:TLSH14}
	\operator{H}_0 =
	\begin{bmatrix}
		1 & 0 \\
		0 & 4
	\end{bmatrix}
\end{equation}
The initial state is the ground state, and the target state is the excited state. Without a restriction on the frequency, the preferable field for this transition is of the main resonance frequency, with the Bohr frequency: $\omega=3_{a.u.}$. However, there are other, much smaller resonances, in the odd fractions of the Bohr frequency. The most pronounced of them is at $\omega=1_{a.u.}$. This makes this problem a test for the ability of the method to restrict the field spectrum and to find a satisfactory restricted field, when it is known to be possible.

\begin{figure} 
	\begin{center}
		\begin{tabular}{cc}
			\includegraphics[scale=.4]{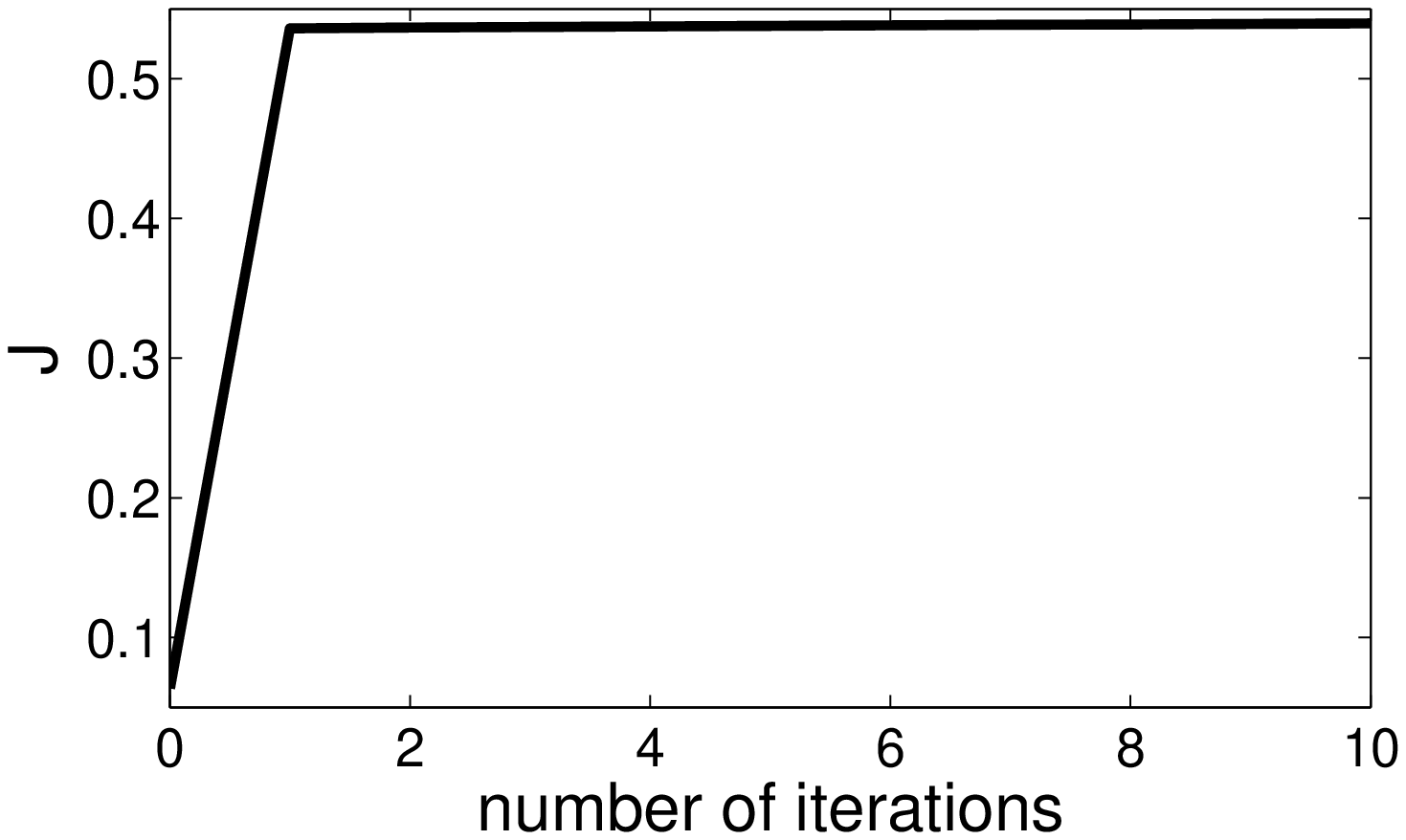} & \includegraphics[scale=.4]{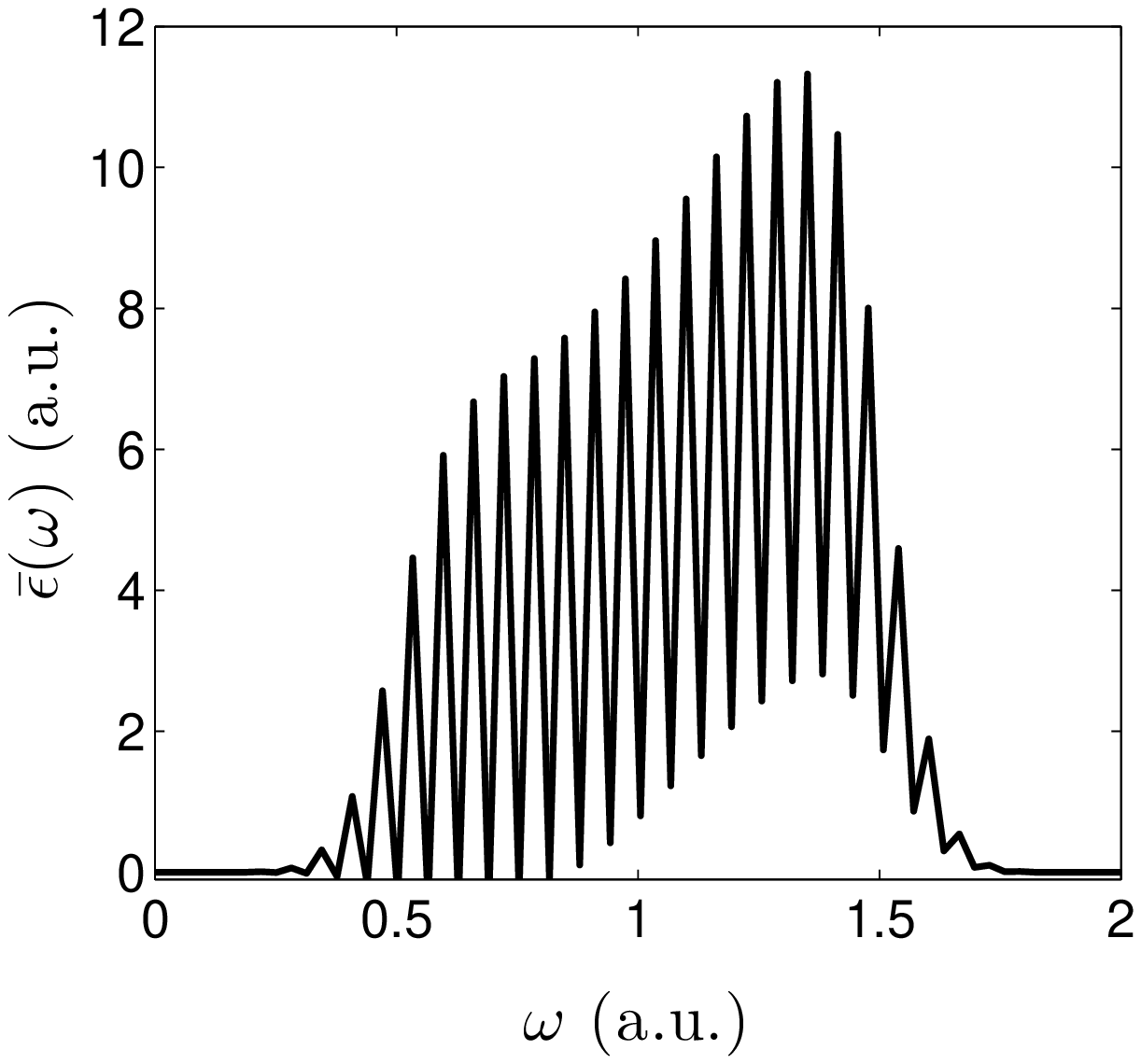}
		\end{tabular}	
	\end{center}
	\input{TLSlimf}
	\caption{The results of the state-to-state problem of the TLS, with a restricted spectrum field. The convergence curve is shown in the left; the resulting spectrum is shown in the right.}\label{fig:TLSs2sres}
\end{figure}

The results are shown in Fig.~\ref{fig:TLSs2sres}. The convergence curve is shown, where this time, $J$ is plotted vs.\ the number of iterations. We see that the first guess, of the form of $\feps$, gave poor results. The algorithm converges very fast to a solution, and after only one iteration, $J$ attains almost its maximal value.

The resulting $\epsw$ is also shown in Fig.~\ref{fig:TLSs2sres}. The spectrum has a discrete character, due to the use of the DCT for the finite time integral in Eq.~\eqref{eq:epsw}. The resolution of $\epsw$ is determined by $T$, and a larger $T$ is required for a smoother curve. The envelope shape of the spectrum, in the form of the hat function (Eq.~\eqref{eq:hat}), is apparent. 

The value of $J_{max}$ is of interest --- it has the physical significance of the yield in the process: 
\[
	J_{max}=\left|\brackets{\phi}{\psi(T)}\right|^2
\]	
Its value at the end of the optimization process was very close to $1$: \text{$J_{max}=0.99894$}.

\begin{figure} 
	\centering \includegraphics[width=3in]{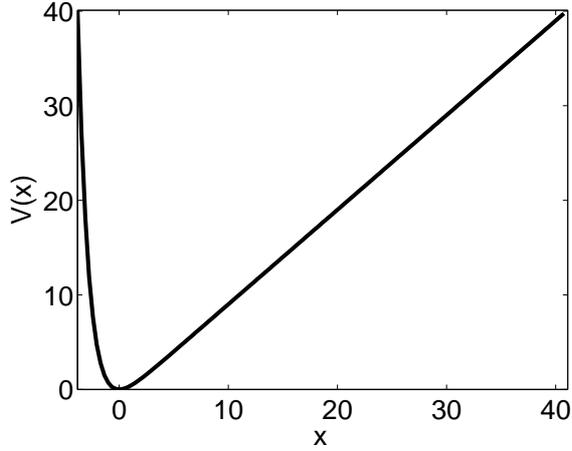} 
	\caption{The Toda potential, with a simple choice of parameters; \text{$V(x)=\exp(-x)+x-1$}}\label{fig:Toda}
\end{figure}

The second problem is a more challenging one --- the system is a one-dimensional anharmonic oscillator, with the potential:
\begin{equation}\label{eq:Toda}
	V(x) = \exp(-x) + x - 1
\end{equation}
This potential is the so called: ``Toda potential'' (with a simple choice of its adjustable parameters; see~\cite{Toda}). The potential is plotted in Fig.~\ref{fig:Toda}. Its general form is close to that of the potential of a chemical bond; the difference between the forms is that the Toda potential is not dissociative, \ie at $x\longrightarrow\infty$, also $V(x)\longrightarrow\infty$. This simplifies the problem, because the possibility of dissociation is prevented. This potential will be used also in Ch.~\ref{ch:RD} (in Subsection~\ref{ssec:modif}, we present a solution for the possibility of dissociation in dissociative potentials, like the Morse potential).

\begin{figure}
	\begin{center}
		\begin{tabular}{cc}
			\includegraphics[scale=.4]{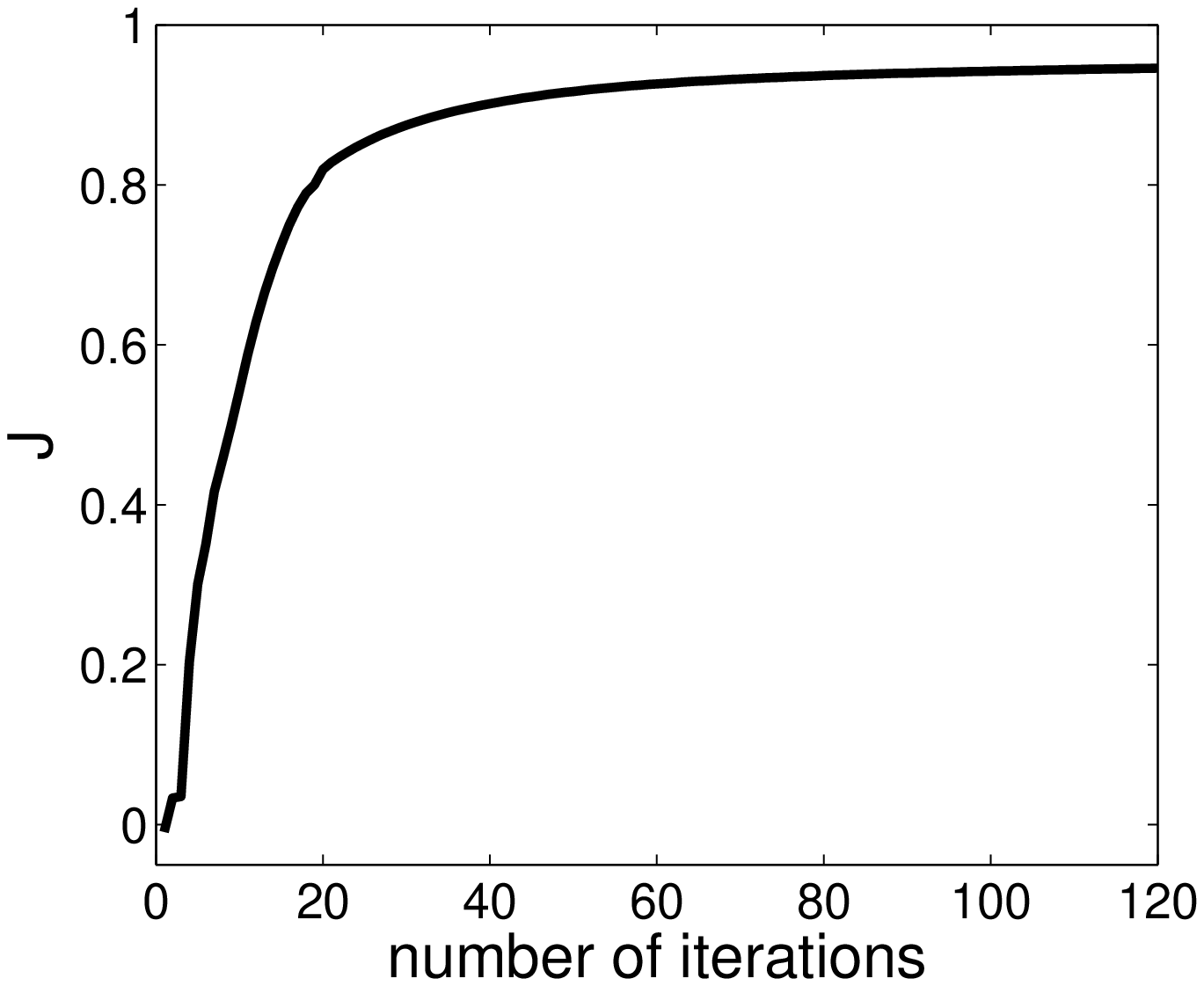} & \includegraphics[scale=.4]{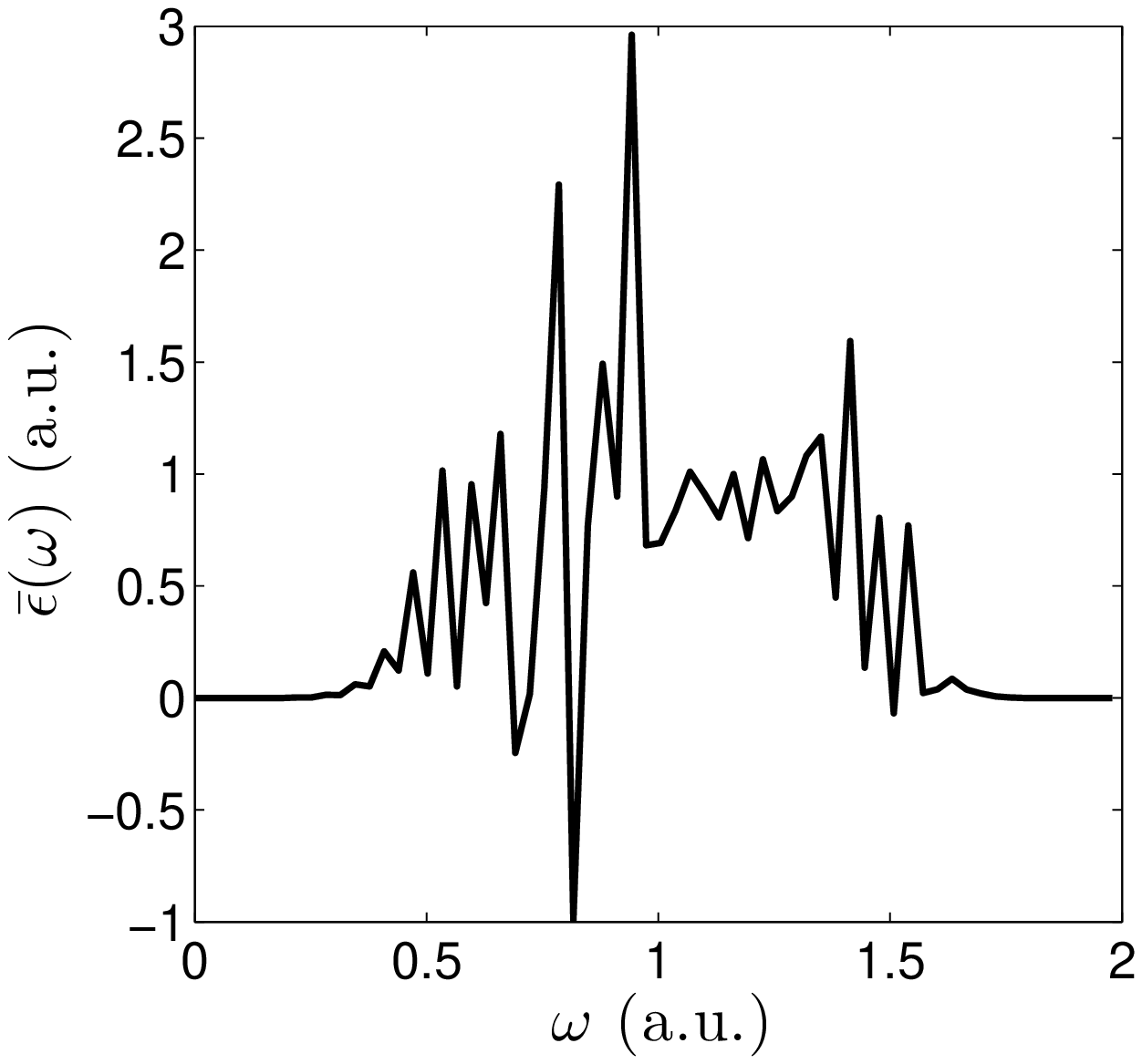}
		\end{tabular}	
	\end{center}
	\input{Todalimf}
	\caption{The results of the state-to-state problem of the Toda potential anharmonic oscillator, with a restricted spectrum field. The convergence curve is shown in the left; the resulting spectrum is shown in the right.}\label{fig:Todas2s}
\end{figure}

The initial state of the Toda potential problem is the ground-state: $\ket{\varphi_0}$ and the target state is: $\ket{\varphi_4}$. The results are shown in Fig.~\ref{fig:Todas2s}. $\epsw$ is very different from the one of the TLS problem, but the envelope shape is still apparent. The yield was: \text{$J_{max} = 0.95919$}. A better yield could be achieved if we used a smaller tolerance parameter for the convergence of the field.

These examples show that the new method for quantum control with a restricted forcing field spectrum is applicable. In the next section, we use this method for the problem of harmonic generation.

\section{New formulation for harmonic generation}\label{sec:methHG}
The second part of our task is the control of the spectrum of the oscillating dipole-moment, which determines the spectrum of the emitted field. It is treated in a similar way to the first part of our task.

In Subsection~\ref{ssec:muw}, we introduce a new $J_{max}$, which represents the requirements on the dipole. In Subsection~\ref{ssec:full}, we present the full maximization problem, and the resulting Euler-Lagrange equations. Subsection~\ref{ssec:modif} deals with a generalization and optional modifications for the new functional.

	\subsection{New maximization functional for controll\-ing the sp\-e\-ctrum of the dipole moment}\label{ssec:muw}
The requirements for the response of the system to the forcing field belong to $J_{max}$. Hence, we introduce a new $J_{max}$ that represents the special requirements of the harmonic generation problem.

The quantity that we want to control is the dipole-moment expectation value:
\begin{equation}\label{eq:evmu}
	\exval{\operator{\mu}}\!(t) = \bracketsO{\psi(t)}{\operator{\mu}}{\psi(t)}
\end{equation}
We want that the oscillations of this quantity, in the desired spectrum, will be of maximal amplitudes. The function of interest is a spectral representation of $\exval{\operator{\mu}}\!(t)$. We use the cosine transform of $\exval{\operator{\mu}}\!(t)$:
\begin{equation}\label{eq:muw}
	\muw = \mathcal{C}[\evmu]=\sqrt{\frac{2}{\pi}}\int_0^T \evmu\cos(\omega t)\,dt
\end{equation}
Here, again, we take $T$ to be the upper limit of integration, because the problem is undefined for $t>T$. 

$\muw$ can attain negative values; we want to maximize the \emph{absolute value} of $\muw$ in the desired spectrum. It is more convenient to handle with the \emph{square} of this quantity. Hence, we define the new $J_{max}$ in the following way:
\begin{equation}\label{eq:Jmaxw}
	J_{max} \equiv \frac{1}{2}\int_0^\Omega \fmu\overline{\exval{\operator{\mu}}}^2(\omega)\,d\omega \qquad\qquad \fmu\geq 0
\end{equation}
where $\fmu$ satisfies also a normalization condition:
\begin{equation}\label{eq:normfmu}
	\int_0^\Omega \fmu\, d\omega = 1 
\end{equation}
$\fmu$ has the role of a \emph{filter function}. As in the case of $\feps$, we can either choose a rectangular function, for complete filtration, or another function, for smooth filtration. Note that the role of the filtration in the present case, is different from that of Sec.~\ref{sec:methlim}: we do not require that there will be no response to the forcing field outside the desired spectrum, represented by $\fmu$; we simply do not encourage this response.

The conditions for an extremal in Eqs.~\eqref{eq:dJdpsitk}, \eqref{eq:dJdpsitb}, result in the following equation for $\ket{\chi(t)}$ (instead of~\eqref{eq:ELs2schi} or~\eqref{eq:ELTDchi}; the derivation is given in App.~\ref{ap:derivation}):
\begin{equation}\label{eq:ELchiw}
	\pderiv{\ket{\chi(t)}}{t} = -i\operator{H}(t)\ket{\chi(t)} - \mathcal{C}^{-1}\left[\fmu\muw\right]\operator{\mu}\ket{\psi(t)}, \qquad\qquad 	\ket{\chi(T)} = 0
\end{equation}
The condition on $\ket{\chi(T)}$ represents the natural boundary conditions of the problem, as in Sec.~\ref{sec:OCTt}.

The expression:
\[
	\mathcal{C}^{-1}\left[\fmu\muw\right]
\]	
is a function of $t$. Its meaning will be more apparent if we  write it in the following way:
\begin{equation}\label{eq:mufil}
	\mathcal{C}^{-1}\left\lbrace\fmu\mathcal{C}\left[\evmu\right]\right\rbrace
\end{equation}
The meaning of this expression can be described as follows: We transform $\evmu$ to the frequency domain; then, we apply a spectral filtration to the resulting function of $\omega$, by multiplying by the filter function $\fmu$; the result is transformed back to the time domain. We can give this expression the interpretation of the filtered, time dependent dipole moment expectation value.

Note the similarity of Eq.~\eqref{eq:ELchiw} to Eq.~\eqref{eq:ELTDchi}: Here, again, we have the form of the inhomogeneous Schr\"odinger equation; The time dependent operator $w(t)\operator{O}(t)$ is replaced by the time dependent operator:
\begin{equation}\label{eq:muwop}
	\mathcal{C}^{-1}\left\lbrace\fmu\mathcal{C}\left[\evmu\right]\right\rbrace\operator{\mu}
\end{equation}
We can give a similar interpretation to the two operators: The operator $w(t)\operator{O}(t)$ is interpreted as the operator $\operator{O}(t)$, weighted by $w(t)$ in the time domain; likewise, \eqref{eq:muwop} may be interpreted as the time dependent operator $\evmu\operator{\mu}$, weighted by $\fmu$ in the frequency domain. 

$J_{penal}$ is not involved in the derivation of Eq.~\eqref{eq:ELchiw}, because it does not have an explicit dependence on $\ket{\psi(t)}$. It follows, that the new $J_{max}$ may be used with various kinds of $J_{penal}$, with the same Euler-Lagrange equation, \eqref{eq:ELchiw}. For the harmonic generation problem, we use the $J_{penal}$ from Eq.~\eqref{eq:Jpenalw}.

	\subsection{The full maximization problem}\label{ssec:full}
We present here, for convenience, the full maximization problem for harmonic generation.

The maximization functional is defined as:
\begin{align}
	& J \equiv J_{max} + J_{penal} + J_{con} \label{eq:Jw}\\ 
	& J_{max} \equiv \frac{1}{2}\int_0^\Omega \fmu\overline{\exval{\operator{\mu}}}^2(\omega)\,d\omega & \fmu\geq 0 \label{eq:Jmaxw2} \\
	& J_{penal} \equiv -\int_0^\Omega\frac{1}{\tfeps}\bar{\epsilon}^2(\omega)\,d\omega & \tfeps>0 \label{eq:Jpenalw2}\\
	& J_{con} \equiv -2\Real{\int_0^T\bracketsO{\chi(t)}{\pderiv{}{t}+i\operator H(t)}{\psi(t)}\,dt}
\end{align}

The variables are subject to the Schr\"odinger equation constraints in Eqs.~\eqref{eq:Schrodinger}, \eqref{eq:conjSchrodinger}.

The conditions for an extremal in Eqs.~\eqref{eq:dJdeps}-\eqref{eq:dJdpsitb}, together with the constraints and the natural boundary conditions, result in the following set of Euler-Lagrange equations:
\begin{align}
	&\pderiv{\ket{\psi(t)}}{t} = -i\operator{H}(t)\ket{\psi(t)}, & 	&\ket{\psi(0)} = \ket{\psi_0} \label{eq:ELpsiw} \\ 
	&\pderiv{\ket{\chi(t)}}{t} = -i\operator{H}(t)\ket{\chi(t)} - \mathcal{C}^{-1}\left[\fmu\muw\right]\operator{\mu}\ket{\psi(t)}, & &\ket{\chi(T)} = 0 \label{eq:ELchiw2} \\ 
	&\operator{H}(t) = \operator{H}_0 - \operator{\mu}\mathcal{C}^{-1}[\epsw] \nonumber \\
	& \epsw = \tilde{f}_{\epsilon}(\omega)\mathcal{C}\left[-\Imag{\bracketsO{\chi(t)}{\operator{\mu}}{\psi(t)}}\right] \label{eq:ELepswuse2}
\end{align}
	
	\subsection{Modifications to the maximization problem}\label{ssec:modif}
The skeleton of the method was summarised in Subsection~\ref{ssec:full}. In this subsection we present several modifications to the optimization problem that are occasionally necessary.

\subsubsection*{Generalization of $J_{max}$ to an arbitrary Hermitian operator}
We start with a generalization of the problem. In the formulation that was presented, the dipole moment operator plays two distinct roles:
\begin{enumerate}
	\item The dipole moment operator serves as the operator adjacent to the control field.
	\item The dipole moment operator expectation value is the subject of the control problem.
\end{enumerate}  
It is not necessary that the same operator play both roles; for example, we can use $x$ polarised laser field and require the maximal response of the dipole in the $y$ direction in the desired spectrum. The operator $\operator{\mu}_x$ will play the first role, and $\operator{\mu}_y$ will play the second.

The generalization of the formulation to include such a case is trivial; the operator $\operator{\mu}$ represents the operator adjacent to the control field, and plays the first role. The second role is played by an arbitrary Hermitian operator $\operator{O}$\@. Eq.~\eqref{eq:Jmaxw2} is replaced by:
\begin{equation}\label{eq:JmaxwO}
	J_{max} \equiv \frac{1}{2}\int_0^\Omega f_O(\omega)\overline{\exval{\operator{O}}}^2(\omega)\,d\omega \qquad\qquad  \fO\geq 0
\end{equation}
where $\fO$ has the same role as $\fmu$, and is subject to the same normalization condition as in~\eqref{eq:normfmu}. Accordingly, Eq.~\eqref{eq:ELchiw2} is replaced by:
\begin{equation}\label{eq:ELchiwO}
	\pderiv{\ket{\chi(t)}}{t} = -i\operator{H}(t)\ket{\chi(t)} - \mathcal{C}^{-1}\left[\fO\Ow\right]\operator{O}\ket{\psi(t)}, \qquad\qquad \ket{\chi(T)} = 0 
\end{equation}

In the present work, only the case of $\operator{O}=\operator{\mu}$ was implemented.

\subsubsection*{Including an adjustable coefficient in $J_{max}$}
The \emph{relative weight} of $J_{penal}$ in $J$ determines the ``cost'' of large fields. The relative weight can be controlled, in principle, by only one adjustable coefficient --- the coefficient $\tilde{\alpha}$ of $J_{penal}$. In practice, it is often recommended to include also an adjustable positive coefficient in $J_{max}$:
\begin{equation}\label{eq:lambmax}
	J_{max} \equiv \frac{1}{2}\lambda\int_0^\Omega \fO\overline{\exval{\operator{O}}}^2(\omega)\,d\omega \qquad\qquad \lambda>0
\end{equation}
The reason is, that increasing $\tfeps$ (which means decreasing the relative weight of $J_{penal}$) sometimes results in a numerical instability. In such a case, it is recommended to increase $\lambda$ instead.

It is convenient to define:
\begin{equation}\label{eq:tilfO}
	\tfO = \lambda\fO
\end{equation}
as we did in Subsection~\ref{ssec:epsw} in the context of the forcing field filter function.

Eq.~\eqref{eq:ELchiwO} is modified in the following way:
\begin{equation} \label{eq:ELchitil}
	\pderiv{\ket{\chi(t)}}{t} = -i\operator{H}(t)\ket{\chi(t)} - \mathcal{C}^{-1}\left[\tfO\Ow\right]\operator{O}\ket{\psi(t)}
\end{equation}

\subsubsection*{Preventing the system from occupying undesirable states}
When dealing with the problem of harmonic generation for a realistic chemical-bond potential (like the Morse potential), it is important to prevent the possibility of dissociation. Occupation of states with eigenenergies above the depth of the well means a partial dissociation; hence, the occupation of these states should be prohibited. Moreover, the states with energy above the depth of the well have no physical meaning in the numerical method of representing the spatial variable, used in the present work (see App.~\ref{ap:num}). For these reasons, it is important to include in our formulation the possibility of restricting the allowed state-space of the quantum system.

To deal with the problem, we should insert two changes in the optimization problem:
\begin{enumerate}
	\item The forbidden components of the state vector, that may increase $J_{max}$, are encouraged by the recent version of $J_{max}$. First, we should avoid this encouragement.
	\item A penalty should be put on the forbidden states. 
\end{enumerate}

The first change is easily achieved, by using only the allowed components of $\ket{\psi(t)}$ in $J_{max}$. Let us define the projection operator on the allowed subspace:
\begin{equation}\label{eq:projal}
	\operator{P}_{a} \equiv \sum_{n=0}^L\ket{\varphi_n}\bra{\varphi_n}
\end{equation}
where $E_L$ is the maximal allowed eigenenergy. The ``refined'' expectation value of $\operator{O}$, computed only from the allowed components, is defined as:
\begin{equation}\label{eq:refevO}
	\exval{\operator O}_{a}(t) \equiv \bracketsO{\operator{P}_{a}\psi(t)}{\operator O}{\operator{P}_{a}\psi(t)} = \bracketsO{\psi(t)}{\operator{P}_{a}\operator O\operator{P}_{a}}{\psi(t)}
\end{equation}
Let us define the following operator:
\begin{equation}\label{eq:defOa}
	\operator{O}_a = \operator{P}_{a}\operator O\operator{P}_{a}
\end{equation}
The modified $J_{max}$ is:
\begin{equation}
	J_{max} \equiv \frac{1}{2}\int_0^\Omega \tfO\overline{\exval{\operator{O}_a}}^2(\omega)\,d\omega 
\end{equation}
Now it is apparent, that Eq.~\eqref{eq:ELchitil} can be used also in this case, just by replacing $\operator O$ with $\operator{O}_a$.

The idea of penalizing the functional for forbidden states has already been suggested in~\cite{Jose}. We insert into $J$ an additional penalty term, defined as:
\begin{equation}\label{eq:Jforb}
	J_{forb} \equiv -\gamma\int_0^T \bracketsO{\psi(t)}{\operator{P}_f}{\psi(t)}\,dt \qquad\qquad \gamma>0
\end{equation}
$\gamma$ is an adjustable penalty factor; $\operator{P}_f$ is the projection operator on the subspace of forbidden functions, where: $\operator{P}_f = \operator{I} - \operator{P}_a$.

Our experience shows that including $J_{forb}$ in its present form increases greatly the difficulty in the optimization process. If it is practical to compute the forbidden eigenstates we can offer a modified version to $J_{forb}$, in which this difficulty is decreased significantly:
\begin{align}
	& J_{forb} \equiv -\int_0^T \bracketsO{\psi(t)}{\operator{P}_f^\gamma}{\psi(t)}\,dt \label{eq:Jforb2} \\
	& \operator{P}_f^\gamma \equiv \sum_{n=L+1}^{N-1}\gamma_n\ket{\varphi_n}\bra{\varphi_n} & \gamma_n>0 \label{eq:Pfg}
\end{align}
where $N$ is the dimension of the eigenvector. In this way, we can choose $\gamma_n$ to increase gradually with $n$, and to achieve a smoother filtration of undesirable states. Note that when $\gamma_n$ is constant in $n$, we return to \eqref{eq:Jforb}, so \eqref{eq:Jforb2} is a generalization of~\eqref{eq:Jforb}.

This method of penalizing $J$ for undesirable states should not be used with too large $\gamma_n$ values --- this might cause the maximum of $J$ to be $0$, with $\epsilon(t) \equiv 0$. Large $\gamma_n$ values might also result in a numerical instability in the propagation process. This limits the efficiency of the method.

After inserting the two modifications into the functional, the Euler-Lagrange equation~\eqref{eq:ELchitil} is replaced by the following one:
\begin{equation} \label{eq:ELchiforb}
	\pderiv{\ket{\chi(t)}}{t} = -i\operator{H}(t)\ket{\chi(t)} - \left\lbrace\mathcal{C}^{-1}\left[\tfO\overline{\exval{\operator{O}_a}}(\omega)\right]\operator{O}_a - \operator{P}_f^\gamma\right\rbrace\ket{\psi(t)}
\end{equation}

\subsubsection*{Prevention of the ``ringing'' phenomenon}
$J_{max}$ presented in~\eqref{eq:Jw} do not include a boundary term in $t=T$, like the one that appears in \eqref{eq:JTD2}. This results in the natural boundary condition: $\ket{\chi(T)}=0$. Using a natural boundary condition is, in some sense, a ``waste'' of an opportunity for an additional control requirement. In this section, we introduce a boundary term, that results in a non-natural boundary condition for $\ket{\chi(t)}$.

The necessity of this boundary term is numerical, and not physical --- it helps in preventing the ``ringing'' phenomenon, that may appear when using the discrete-cosine-transform (DCT) for the finite time integral:
\begin{equation}\label{eq:Ow}
	\Ow = \sqrt{\frac{2}{\pi}}\int_0^T \exval{\operator O}(t)\cos(\omega t)\,dt
\end{equation}

\begin{figure}
	\centering \includegraphics[width=3in]{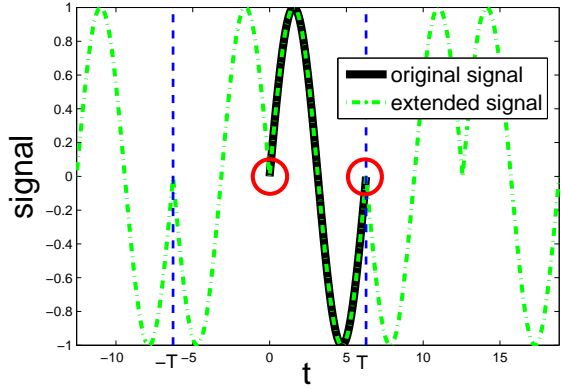} 
	\caption{The figure demonstrates the source of the ringing phenomenon in DCT\@. The original signal, defined in the time interval:~$[0,\;T]$, is plotted in black; the DCT of this signal represents the extended periodical function (in green). This function is given by ``folding'' the signal, and treating the resulting pattern, defined in~$[-T,\;T]$, as a single period of the periodical function. Discontinuities in the first derivative (marked by red circles) are present in the extended function (unless the first derivative at the boundaries of the original signal is $0$). These discontinuities contain very high frequencies, which result in the ringing phenomenon.}\label{fig:extsignal}
\end{figure}

The extended periodical function, represented by the DCT of the signal, has discontinuities in the first derivative, unless the first derivative of the signal is $0$ at the boundaries of the signal (see Fig.~\ref{fig:extsignal}). This discontinuity includes very high frequencies, which result in the ringing phenomenon throughout the transformed vector. The effect is much smaller than the one that appears when using a discrete-Fourier-transform (DFT), where there is a discontinuity in the periodical function itself (see~\cite{DCT}); however, the small effect becomes a trouble, when there is an interest in small amplitude zones in the spectrum.

Usually, in our case, there is no problem with the boundary $t=0$; the reason is that the initial state is typically chosen to be the ground state, and:
\[ 
	\deriv{\exval{\operator{O}}(0)}{t}=0
\]
However, there is a problem for the $t=T$ boundary. Frequently, the harmonic generation effect is very small, and the noise resulting by the ringing is a serious trouble. The situation is even worse: The algorithm tends to increase the ringing, in order to increase $J_{max}$, instead of using a physical mechanism; the intensity of the resulting fields is greatly increased close to $T$; as a result, the dipole oscillates in a wild manner, in a way that the first derivative at $T$ is maximized, and the discontinuity in the extended periodic function is increased. In this situation, the method completely fails.

We propose to solve this problem by trying to force:
\[
	\deriv{\exval{\operator{O}}(T)}{t}=0
\]
This is achieved by the insertion of the following boundary term into the functional:
\begin{equation}\label{eq:Jbring}
	J_{bound} \equiv -\frac{1}{2}\kappa\left[\deriv{\exval{\operator{O}}(T)}{t}\right]^2 \qquad\qquad \kappa>0
\end{equation}
$\kappa$ is a positive parameter, that determines the relative importance of $J_{bound}$ in the functional. This boundary term is maximized when the magnitude of the derivative at the boundary is minimized. Actually, it is a penalty term, and $\kappa$ has the role of a penalty factor.

Taking the expectation value of both sides of the Heisenberg equation, we have:
\begin{equation}\label{eq:Heis}
	\deriv{\exval{\operator{O}}(T)}{t} = i\exval{\left[\operator{H}(T), \operator{O}\right]}(T)
\end{equation}
In the special case that \text{$\commut{\operator{\mu}}{\operator{O}}=\operator{0}$}, we have:
\begin{equation}\label{eq:Heis0}
	\deriv{\exval{\operator{O}}(T)}{t} = i\exval{\left[\operator{H}_0, \operator{O}\right]}(T)
\end{equation}
In this case, the insertion of $J_{bound}$ results in a relatively simple modification of the Euler-Lagrange equations. Using Eq.~\eqref{eq:Heis0}, we are able to derive the following boundary condition (see App.~\ref{ap:derivation}):
\begin{equation}\label{eq:chiTring}
	\ket{\chi(T)} = \kappa\exval{\left[\operator{H}_0, \operator{O}\right]}(T) \left[\operator{H}_0, \operator{O}\right]\ket{\psi(T)}
\end{equation}
In the case that \text{$\commut{\operator{\mu}}{\operator{O}} \neq \operator{0}$}, the insertion of $J_{bound}$ results also in a modification of the expression for $\epsw$, which becomes an integral equation. This case requires a special treatment, and will not be discussed here. 

The insertion of this $J_{bound}$ into the functional might increase the difficulty in the optimization process. Hence, it should be used only when necessary.

\subsubsection*{The full general maximization problem}
For convenience, we summarise all the mentioned modifications, by presenting the most general version of the optimization problem (with the exception of the case that \text{$\commut{\operator{\mu}}{\operator{O}} \neq \operator{0}$} and \text{$\kappa>0$}).

The maximization functional is defined as:
\begin{align}
	& J \equiv J_{max} + J_{bound} + J_{forb} + J_{penal} + J_{con} \label{eq:Jw2}\\ 
	& J_{max} \equiv \frac{1}{2}\int_0^\Omega \tfO\overline{\exval{\operator{O}_a}}^2(\omega)\,d\omega  & \tfO\geq 0 \label{eq:Jmaxgen} \\
	& J_{bound} \equiv -\frac{1}{2}\kappa\left[\deriv{\exval{\operator{O}}(T)}{t}\right]^2 & \kappa \geq 0 \label{eq:Jbringal}\\
	& J_{forb} \equiv -\int_0^T \bracketsO{\psi(t)}{\operator{P}_f^\gamma}{\psi(t)}\,dt \\
	& J_{penal} \equiv -\int_0^\Omega\frac{1}{\tfeps}\bar{\epsilon}^2(\omega)\,d\omega & \tfeps>0 \\
	& J_{con} \equiv -2\Real{\int_0^T\bracketsO{\chi(t)}{\pderiv{}{t}+i\operator H(t)}{\psi(t)}\,dt}
\end{align}

The resulting Euler-Lagrange equations are Eqs.~\eqref{eq:ELpsiw}, \eqref{eq:ELepswuse2}, together with the following equation for $\ket{\chi(t)}$:
\begin{align} \label{eq:chigen}
	&\pderiv{\ket{\chi(t)}}{t} = -i\operator{H}(t)\ket{\chi(t)} - \left\lbrace\mathcal{C}^{-1}\left[\tfO\overline{\exval{\operator{O}_a}}(\omega)\right]\operator{O}_a - \operator{P}_f^\gamma\right\rbrace\ket{\psi(t)} \nonumber \\
	&\ket{\chi(T)} = \kappa\exval{\left[\operator{H}_0, \operator{O}\right]}(T) \left[\operator{H}_0, \operator{O}\right]\ket{\psi(T)}
\end{align}
	
%

	\chapter{Results and discussion}\label{ch:RD}
%
%
%
%
%
%
In the present chapter, the new method is applied to several simple problems. Two classes of problems are discussed:
\begin{enumerate}
	\item Many level system problems, formulated in the basis of the $\operator{H}_0$ eigenstates;
	\item Anharmonic oscillator problems.
\end{enumerate}
The results are analysed and discussed in order to obtain more general conclusions on possible mechanisms of harmonic generation.

When discussing mechanisms of harmonic generation, we should distinguish between two elements of the mechanism; in general, the mechanism is characterized by a steady state, in which the dipole spectrum contains the desired frequency. The steady state is characterized by a periodic pattern of changes in the system and the forcing field. The mechanism consists of the following elements:
\begin{enumerate}
	\item The mechanism of achieving the steady state;
	\item The mechanism of the process in the steady state.
\end{enumerate}
These two elements cannot always be assigned to two distinct stages in the process: using the new method, we frequently observe that the system passes during the process through many such ``steady states''; any of these has its own character during a portion of the process, and produces the desired frequency; then, it is replaced by another, improved ``steady state''.

In order to be able to choose appropriate problems to test the new method, we need some previous insight into  possible mechanisms of harmonic generation. We start from the second element of the mechanism, mentioned above. The characteristic frequencies of the system are the \emph{Bohr frequencies}; the Bohr frequency of two eigenstates, $\eigs{m}$ and $\eigs{n}$, is defined as:
\begin{equation}\label{eq:Bohr}
	\omega_{mn} = E_m - E_n
\end{equation}
It is certain that the system is able to emit radiation in the Bohr frequency of two states, when they are coupled by the dipole moment operator, and the higher energy level is occupied. Hence, the problems are constructed in a way that the target frequency is one of the Bohr frequencies, $\omega_{mn}$, and the two states are coupled by $\operator{\mu}$. The expected mechanism is based on the occupation of the higher energy level, $E_m$. 

The first element of the mechanism must be based on fields with lower frequencies than $\omega_{mn}$. Hence, we cannot use the main resonance frequency for the transition from $\eigs n$ to $\eigs m$, which is equal to $\omega_{mn}$. The occupation transfer may be achieved by utilizing other resonance frequencies of the system, in two ways:
\begin{enumerate}
	\item Using a secondary resonance frequency for the transition from $\eigs{n}$ to $\eigs{m}$, \ie using one of the odd fractions of $\omega_{mn}$:
	\begin{equation}
		\omega = \frac{\omega_{mn}}{l} \qquad\qquad l=3,5,7,\ldots
	\end{equation}
	\item Using intermediate states for the occupation transfer; this will be demonstrated in Subsection~\ref{ssec:3LS}.
\end{enumerate}

Our discussion on possible mechanisms will concentrate on the second element of the mechanisms.
  
In Sec.~\ref{sec:MLS}, the results for many level systems are presented. In Sec.~\ref{sec:analysis}, the results of Sec.~\ref{sec:MLS} are analysed. The analysis leads to general conclusions on the second element of harmonic generation mechanisms, mentioned above. In Sec.~\ref{sec:anharmonic}, the results for the more complicated anharmonic oscillator problems are presented. The conclusions from Sec.~\ref{sec:analysis} are used to analyse the results. In Sec.~\ref{sec:problems}, the deficiencies and the problems in the new method are discussed.

\emph{A general remark for all problems}: the initial state is always chosen to be the ground state of the system.

\section{Many level systems}\label{sec:MLS}

\subsection{Two level system (TLS)}\label{ssec:TLS}
We start from the simplest problem --- harmonic generation in a TLS\@. We try to see if the new method will follow a simple mechanism of harmonic generation; it consists of utilizing one of the secondary resonances of the system for transition to the excited state, and emission at the Bohr frequency, $\omega_{1,0}$.

\begin{figure}
	\centering \includegraphics[width=3in]{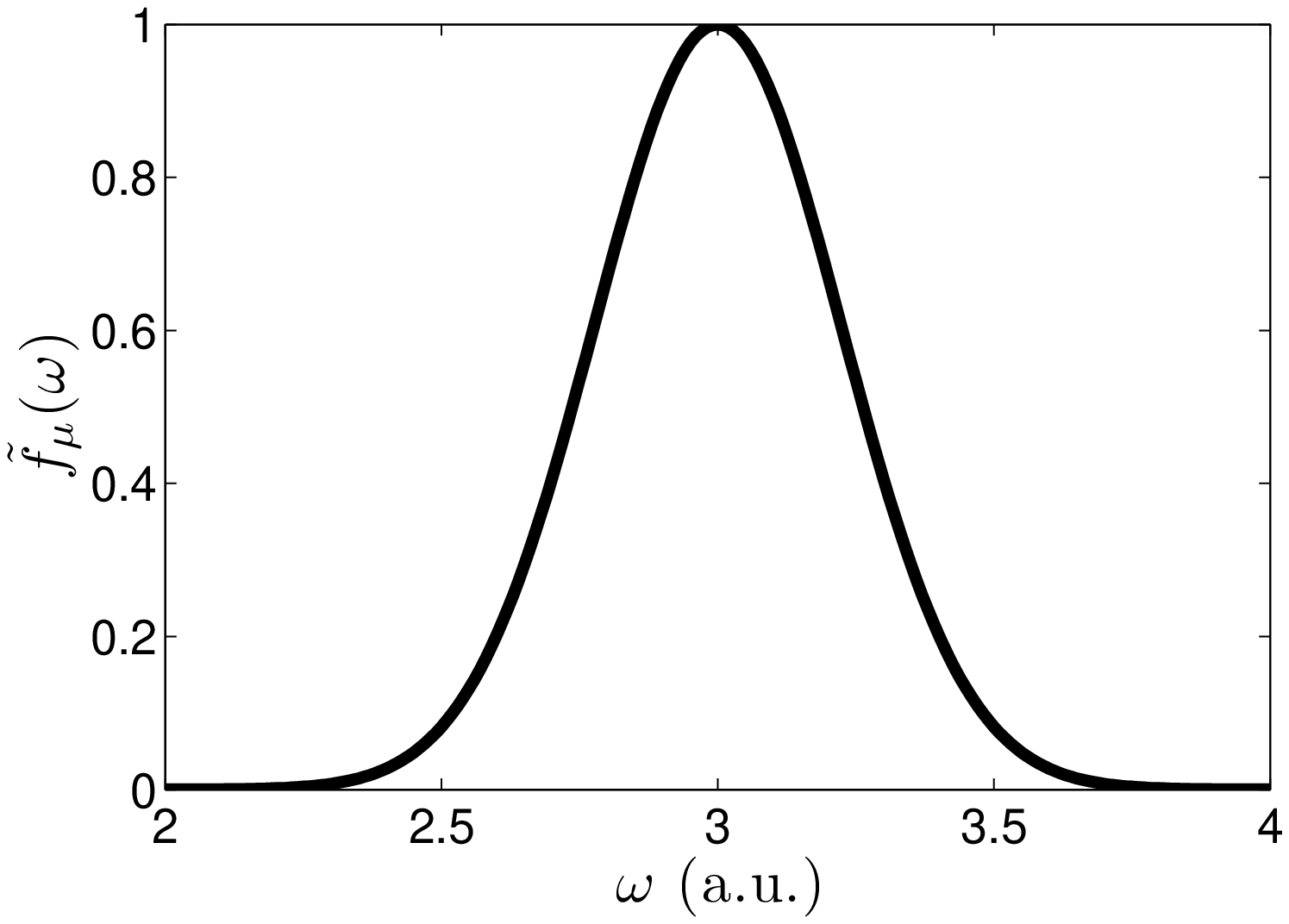}
	\caption{$\tfmu$ of the TLS problem}\label{fig:gauss10}	
\end{figure}

The unperturbed Hamiltonian and the dipole moment operator are the same as in Eqs.~\eqref{eq:TLSH14}, \eqref{eq:TLSmu}, respectively (see also Table~\ref{tab:TLS}). We want to maximize the emission at \text{$\omega_{1,0}=3_{a.u.}$}. As in Subsection~\ref{ssec:relax}, the forcing field is restricted to be around \text{$\omega=1_{a.u.}$}, the first odd fraction ($1/3$) of $\omega_{1,0}$. $\tfeps$ is a hat function (see Fig.~\ref{fig:hat} for the general shape), and $\tfmu$ is a Gaussian function (see Fig.~\ref{fig:gauss10}).

The details of the problem are summarised in Table~\ref{tab:TLS}.

\begin{table}
	\begin{equation*}
	\renewcommand{\arraystretch}{1.25}
	\begin{array}{|c||c|}
		\hline
		\operator{H}_0 & 
		\begin{bsmatrix}
			1 & 0 \\
			0 & 4
		\end{bsmatrix} \\ \hline
		\operator{\mu} &
		\begin{bsmatrix}
			0 & 1 \\
			1 & 0
		\end{bsmatrix} \\ \hline
		\ket{\psi_0} & 
		\begin{bsmatrix}
			1  \\
			0	
		\end{bsmatrix} \\ \hline
		T & 100 \\ \hline
		\tfeps & 20\,\sech[20(\omega - 1)^4] \\ \hline
		\tfmu & \exp[-10(\omega -3)^2] \\ \hline
		\bar\epsilon^{0}(\omega) & \sech[20(\omega - 1)^4] \\ \hline
		K_i & 0.5 \\ \hline
		\text{tolerance} & 10^{-3} \\ \hline
	\end{array}
\end{equation*}
	\caption{The details of the TLS problem}\label{tab:TLS}
\end{table}

\begin{figure}
	\centering \includegraphics[width=3in]{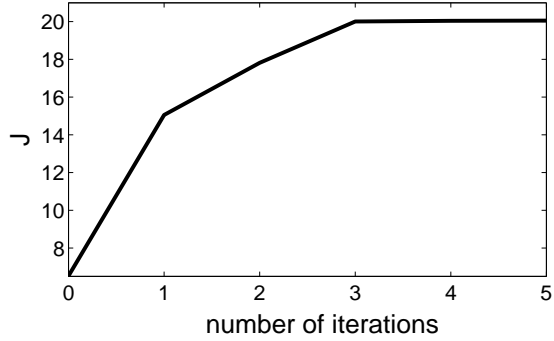}
	\caption{The convergence curve of the TLS problem}\label{fig:TLSHGconv}
\end{figure}

\begin{figure}
	\centering \includegraphics[width=3in]{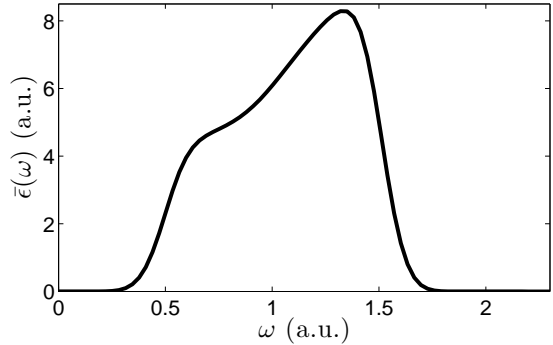}
	\caption{The $\epsw$ curve of the TLS problem; the spectrum is characterized by a large distribution around \text{$\omega=1_{a.u.}$}. The hat function envelope is apparent.}\label{fig:TLSHGepsw}	
\end{figure}

\begin{figure}
	\centering \includegraphics[width=3in]{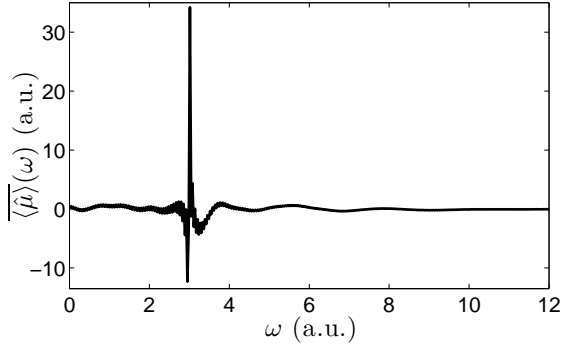}
	\caption{The $\muw$ curve of the TLS problem; the dipole oscillates mainly at the Bohr frequency: \text{$\omega_{1,0}=3_{a.u.}$}. A small response exists also in the neighbourhood of $\omega_{1,0}$, in accordance with the Gaussian shape of $\tfmu$ (see Fig.~\ref{fig:gauss10})}\label{fig:TLSHGmuw}	
\end{figure}

\begin{figure}
	\centering \includegraphics[width=3in]{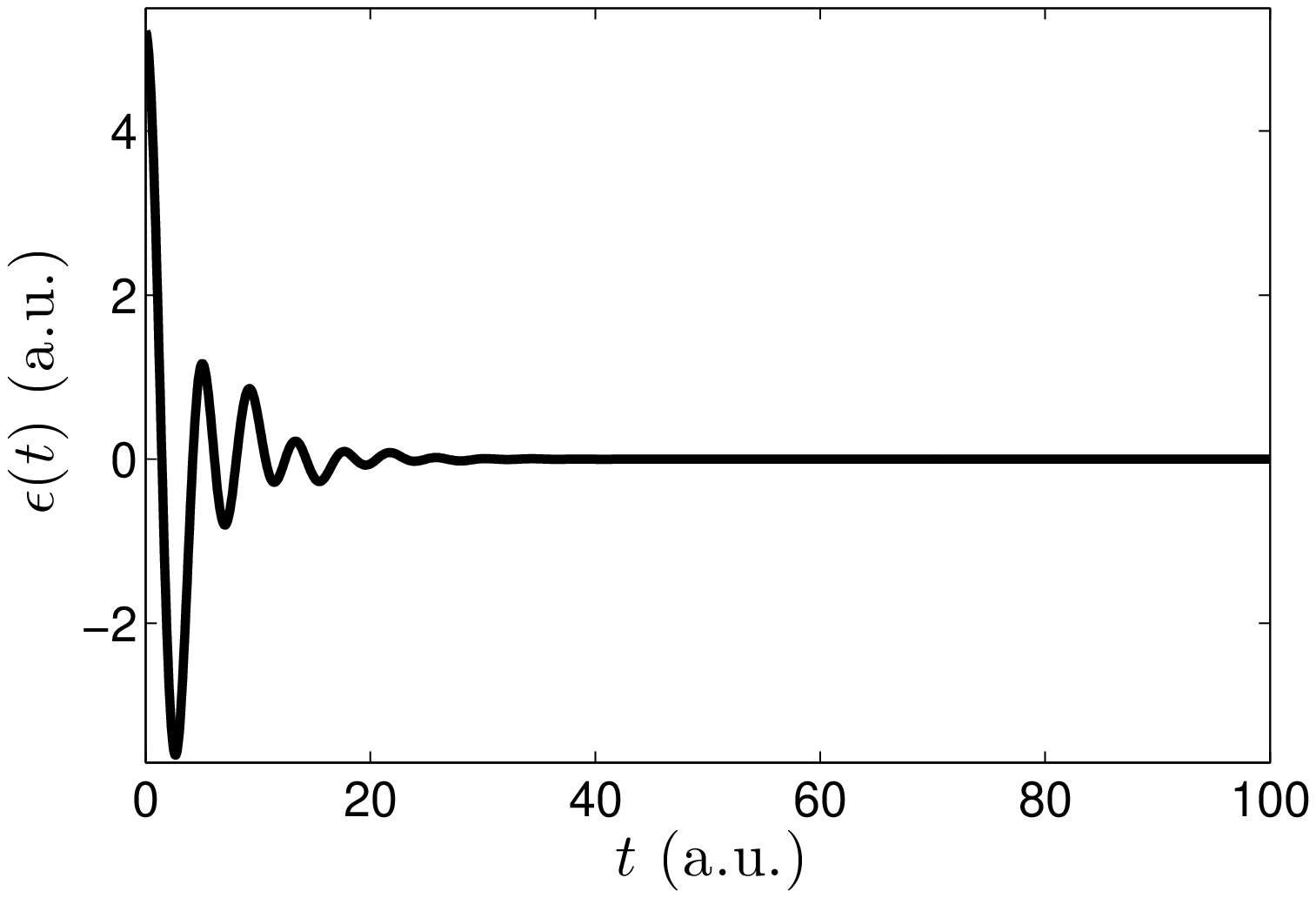} \\	
	\centering \includegraphics[width=3.3in]{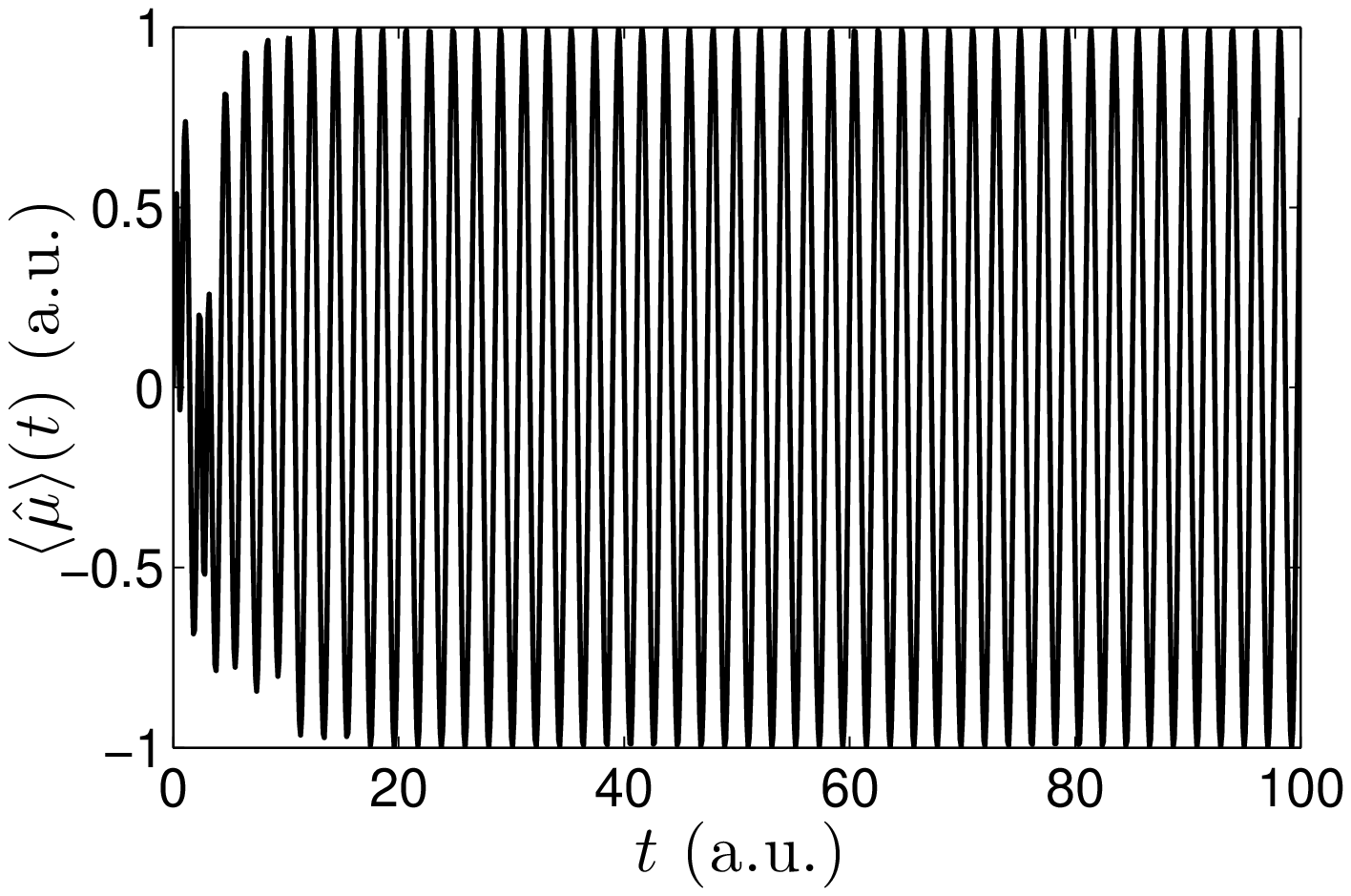} \\
	\centering \includegraphics[width=3in]{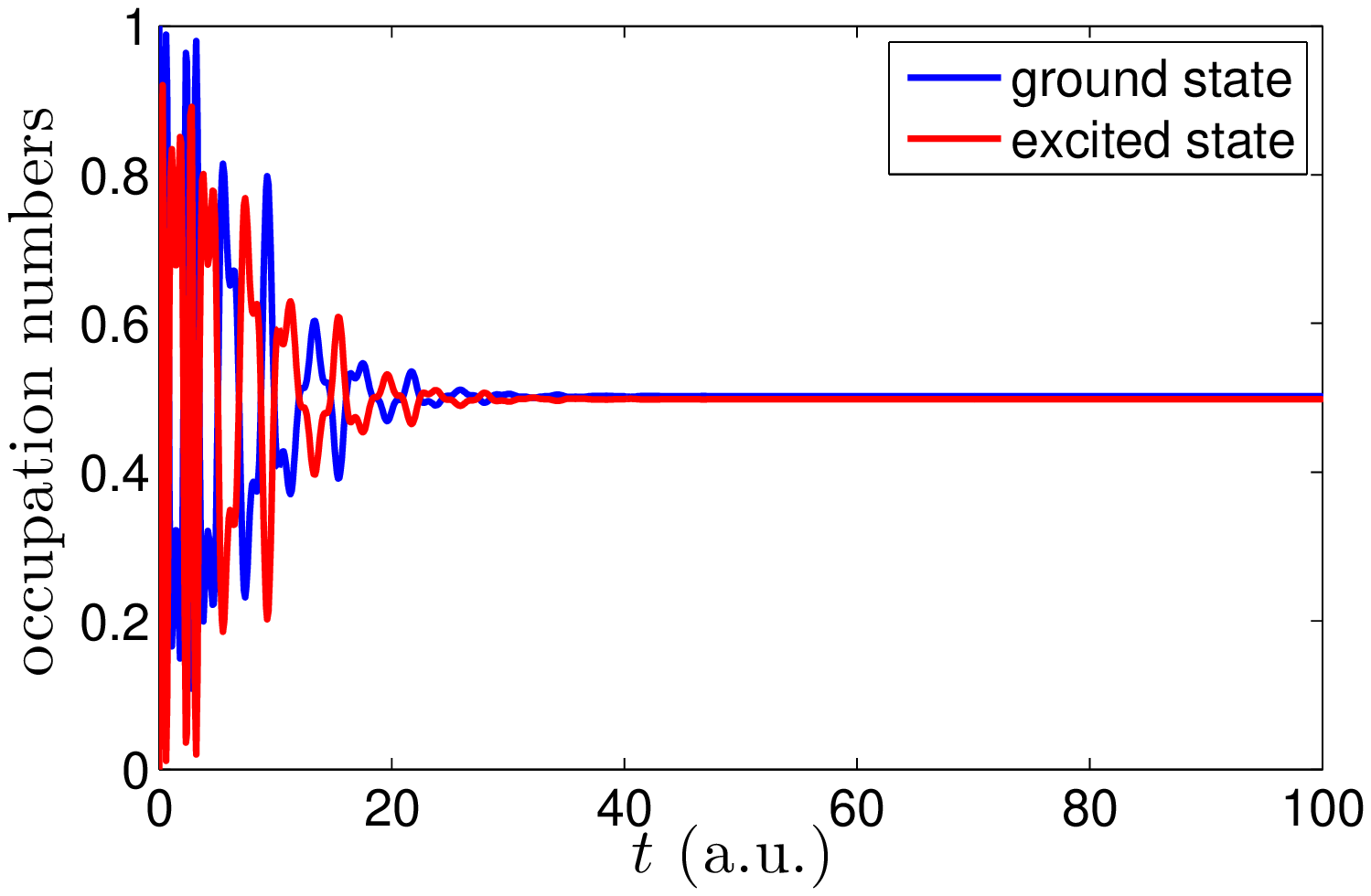}
	\caption{The time picture of the system, for the TLS problem; at the top: The $\epsilon(t)$ curve; in the middle: The $\evmu$ curve; at the bottom: The occupation vs.\ time curves; once the two levels are equally occupied, the dipole-moment oscillations reach their maximal amplitude: $1_{a.u.}$, and the forcing field is ``turned off''. }\label{fig:TLSt}	
\end{figure}

The optimization process converges rapidly to a solution (Fig.~\ref{fig:TLSHGconv}). The results are presented in Figs.~\ref{fig:TLSHGepsw}-\ref{fig:TLSt}.

The resulting forcing field (Fig.~\ref{fig:TLSHGepsw}) is rather different from our expectations: instead of being consisted of a single frequency, \text{$\omega=1_{a.u.}$}, $\epsw$ has a large distribution of frequencies around \text{$\omega=1_{a.u.}$}. The hat function envelope is apparent.

The resulting $\muw$ (Fig.~\ref{fig:TLSHGmuw}) mainly consists of a large peak at $\omega_{1,0}$, as could be expected. There is also a small response in the neighbourhood of $\omega_{1,0}$. This is probably a sequence of the chosen shape of $\tfmu$, which is not completely localized.

Examining the time-picture of the system (Fig.~\ref{fig:TLSt}) is very edifying. We observe that the system reaches rapidly an equal occupation of the two eigenstates. Once the equal occupation is achieved, the $\evmu$ oscillations reach their maximal amplitude: $1_{a.u.}$, and the forcing field is ``turned off''. This apparently means that the system achieved its optimal state for the emission at $\omega_{1,0}$. This is the steady state of the harmonic generation process.

\subsection{Three level system (3LS)}\label{ssec:3LS}
The next example is of a 3 level system (3LS). The unperturbed Hamiltonian is:
\begin{equation}\label{eq:thLSH0}
	\operator{H}_0 =
	\begin{bmatrix}
		1 & 0 & 0 \\
		0 & 1.9 & 0 \\
		0 & 0 & 3
	\end{bmatrix}
\end{equation}
The energy levels are almost equidistant, with an energy difference of around $1_{a.u.}$\footnote{The levels in this example, and in the example of Subsection~\ref{ssec:11LS}, are not chosen to be exactly equidistant; this is because a system with equidistant levels possesses symmetry properties; these may limit the \emph{controllability} of the system. The topic of controllability is discussed in \cite{David, tutorial}.}. The dipole moment operator is:
\begin{equation}\label{eq:thLSmu}
	\operator{\mu} =
	\begin{bmatrix}
		0 & 1 & 1 \\
		1 & 0 & 1 \\
		1 & 1 & 0
	\end{bmatrix}
\end{equation}
This operator couples between all levels.

\begin{figure}
	\centering \includegraphics[width=3in]{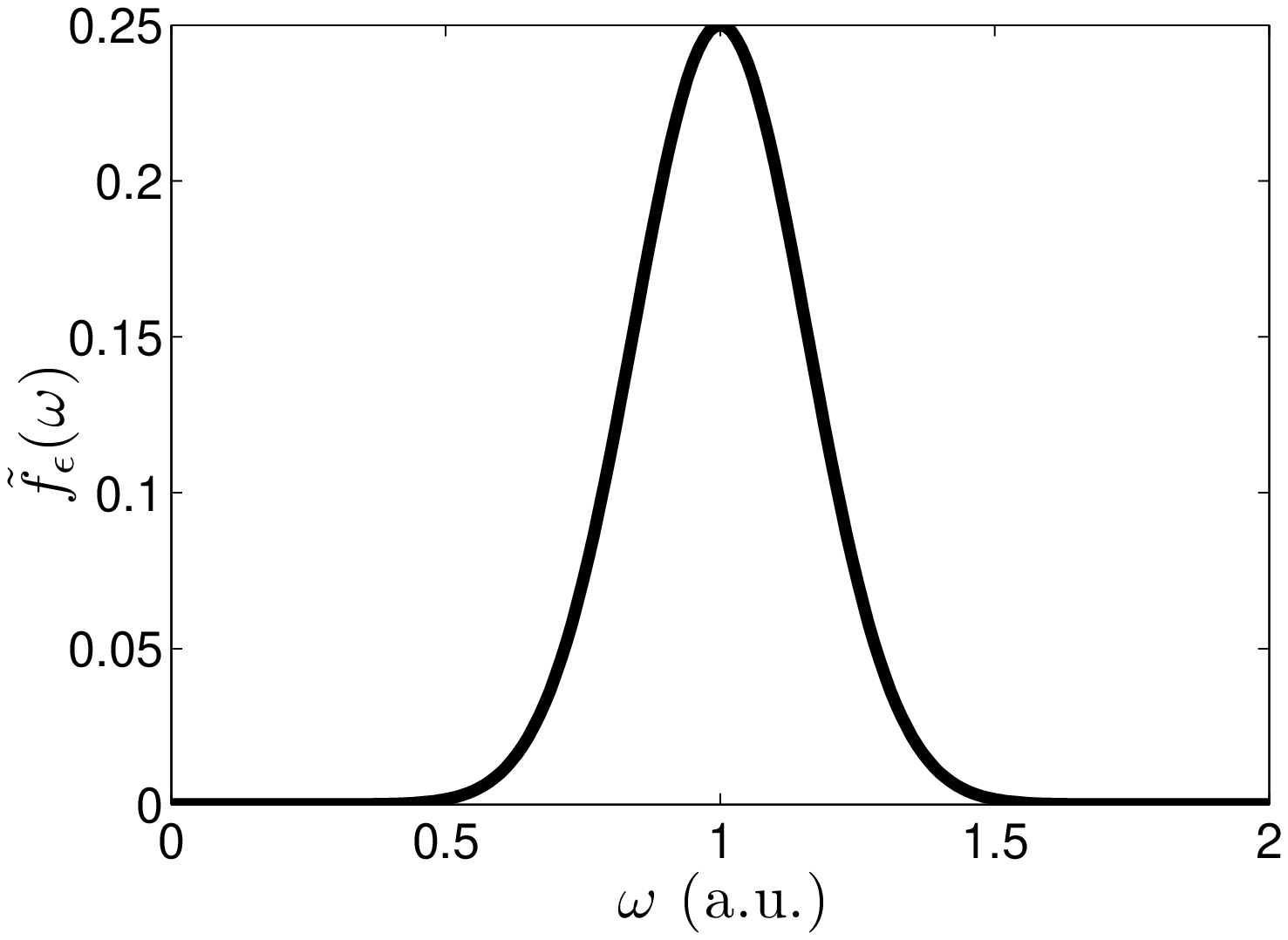}
	\caption{$\tfeps$ of the 3LS problem}\label{fig:gauss20}	
\end{figure}

In this problem, we examine the ability of the method to follow another mechanism for achieving the steady state, based on an intermediate state for occupation transfer. The field is restricted to be with frequencies around \text{$\omega=1_{a.u.}$}. This is the region of the resonance frequencies of the neighbouring levels; the resonance frequency of the levels $\ket{\varphi_0}$ and $\ket{\varphi_2}$, \ie \text{$\omega_{2,0}=2_{a.u.}$}, is out of the allowed frequency region. $\tfeps$ is a narrow Gaussian (see Fig.~\ref{fig:gauss20}). We require that the dipole moment will oscillate at the Bohr frequency \text{$\omega_{2,0}$}. $\tfmu$ has the same form as $\tfeps$.

According to our requirements on the forcing field spectrum, the forcing field is at resonance for the transition between neighbouring levels, but is off resonant for the transition between the non-neighbouring levels. In the expected mechanism, $\ket{\varphi_1}$ serves as an intermediate state for transferring occupation from $\ket{\varphi_0}$ to $\ket{\varphi_2}$.

The details of the problem are summarised in Table~\ref{tab:3LS}. The parameter $\tilde{\alpha}$ is chosen to be relatively large. After discussing the results of this problem, we will repeat the problem with another choice of $\tilde{\alpha}$.

\begin{table}
	\begin{equation*}
	\renewcommand{\arraystretch}{1.5}
	\begin{array}{|c||c|}
		\hline
		\operator{H}_0 & 
		\begin{bsmatrix}
			1 & 0 & 0 \\
			0 & 1.9 & 0 \\
			0 & 0 & 3
		\end{bsmatrix} \\ \hline
		\operator{\mu} &
		\begin{bsmatrix}
			0 & 1 & 1 \\
			1 & 0 & 1 \\
			1 & 1 & 0
		\end{bsmatrix} \\ \hline
		\ket{\psi_0} & \ket{\varphi_0} \\ \hline
		T & 100 \\ \hline
		\tfeps & 0.25\,\exp[-20(\omega- 1)^2] \\ \hline
		\tfmu & \exp[-20(\omega -2)^2] \\ \hline
		\bar\epsilon^{0}(\omega) & \exp[-20(\omega- 1)^2] \\ \hline
		K_i & 1 \\ \hline
		\text{tolerance} & 10^{-2} \\ \hline
	\end{array}
\end{equation*}
	\caption{The details of the 3LS problem, with large $\tilde{\alpha}$}\label{tab:3LS}
\end{table}

\begin{figure}
	\centering \includegraphics[width=3in]{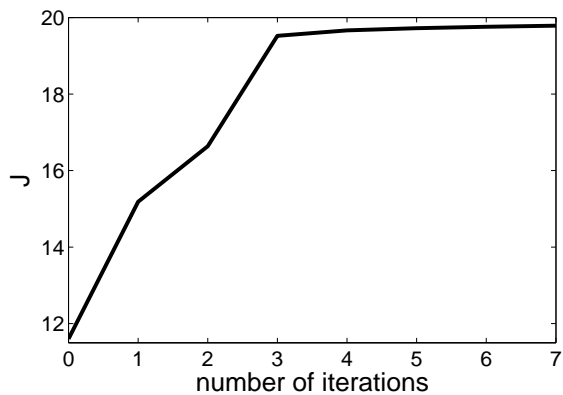}
	\caption{The convergence curve of the first 3LS problem}\label{fig:3LSconv}
\end{figure}

\begin{figure}
	\centering \includegraphics[width=3in]{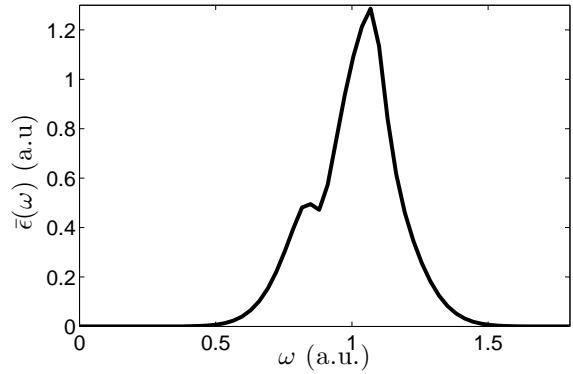}
	\caption{The $\epsw$ curve of the first 3LS problem; the frequencies are distributed around the Bohr frequencies of the neighbouring energy levels. The Gaussian envelope is apparent.}\label{fig:3LSepsw}	
\end{figure}

\begin{figure}
	\centering \includegraphics[width=3in]{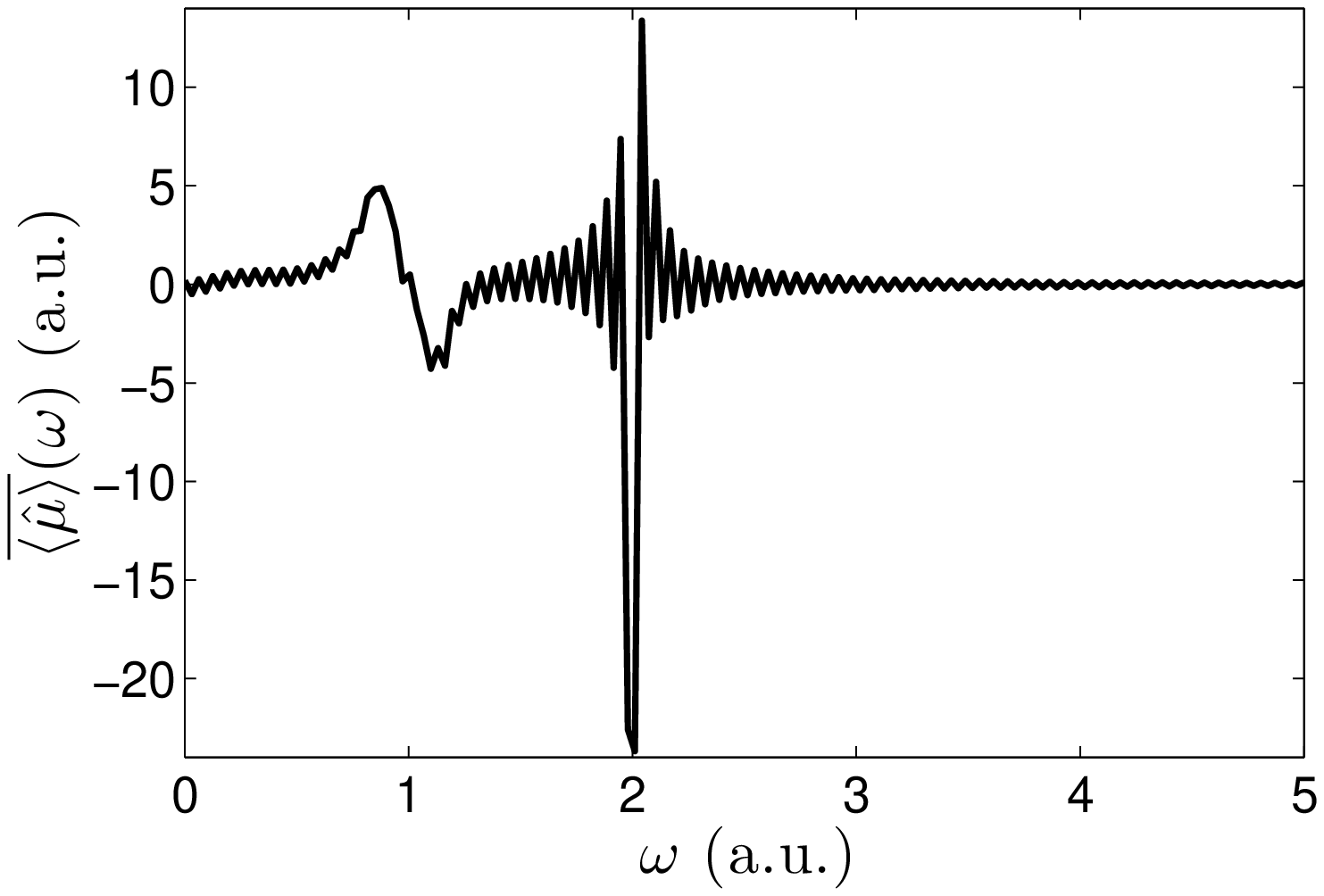}
	\caption{The $\muw$ curve of the first 3LS problem; the spectrum mainly consists of the Bohr frequencies of the system. The largest component of the spectrum is at the Bohr frequency of the non-neighbouring levels: \text{$\omega_{2,0}=2_{a.u.}$}. Smaller extrema exist at the Bohr frequencies of the neighbouring levels, near  \text{$\omega=1_{a.u.}$}.}\label{fig:3LSmuw}	
\end{figure}

\begin{figure}
	\centering \includegraphics[width=3in]{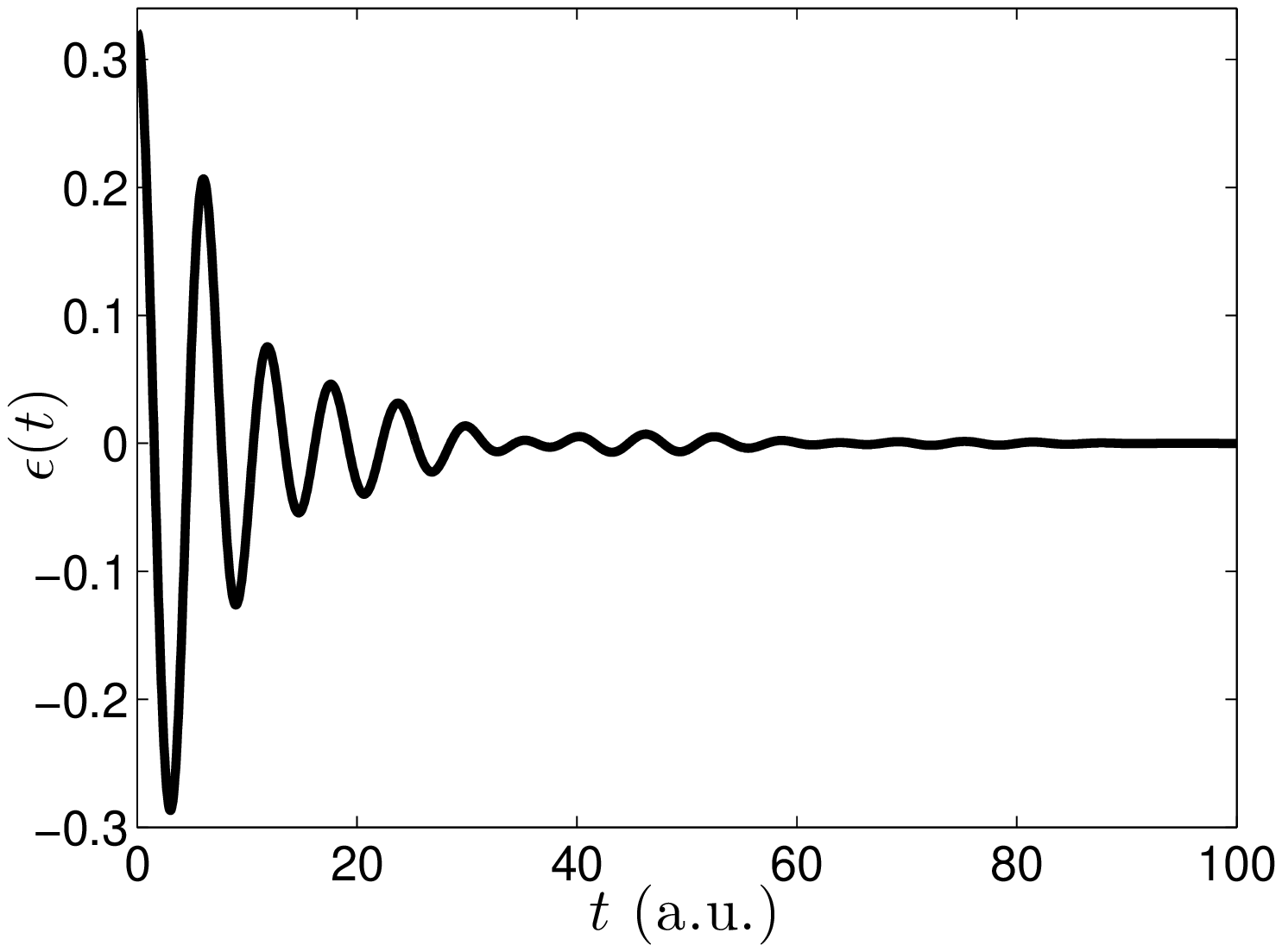} \\	
	\centering \includegraphics[width=3in]{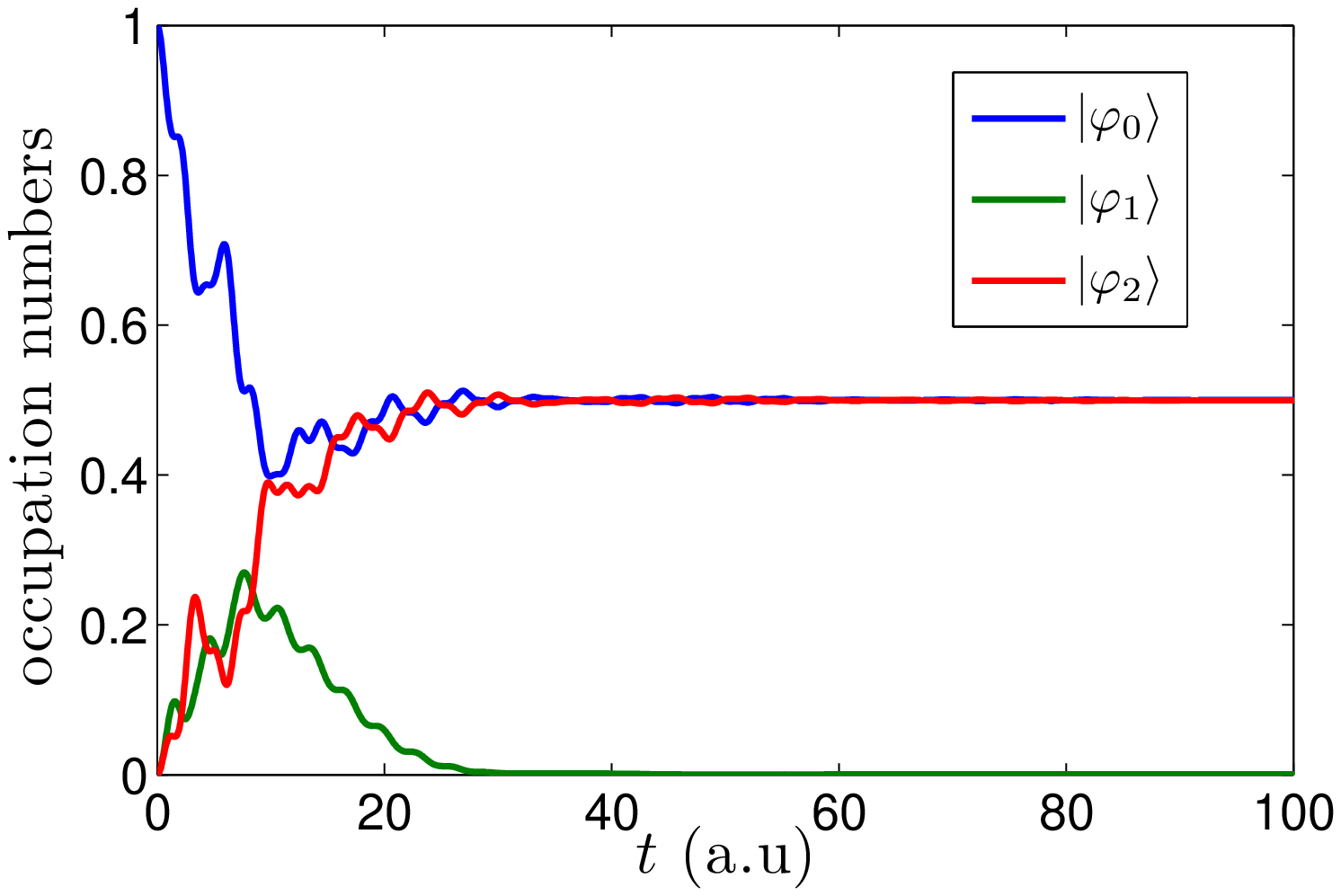}
	\caption{The time picture of the system for the first 3LS problem; at the top: The $\epsilon(t)$ curve; at the bottom: The occupation vs.\ time curves; the states $\ket{\varphi_0}$ and $\ket{\varphi_2}$ reach an equal occupation of $0.5$, and the forcing field is ``turned off''.}\label{fig:3LSt}	
\end{figure}

The convergence of the optimization process is very fast (Fig.~\ref{fig:3LSconv}). The results are shown in Figs.~\ref{fig:3LSepsw}-\ref{fig:3LSt}.

The resulting $\epsw$ (Fig.~\ref{fig:3LSepsw}) has a very apparent Gaussian envelope, in accordance with $\tfeps$. The frequencies are distributed around the region of the resonance frequencies of the neighbouring levels.

The $\muw$ spectrum (Fig.~\ref{fig:3LSmuw}) mainly consists of the Bohr frequencies of the system. In accordance with our requirements, there is a large response at $\omega_{2,0}$. Smaller response exists at the forcing field frequency region, with extrema at the Bohr frequencies of the neighbouring levels.

The time-picture of the system is presented in Fig.~\ref{fig:3LSt}. It seems from the occupation curve that the mechanism of harmonic generation is similar to the expected one; at the beginning of the process, there is a gradual increase in the occupation of $\ket{\varphi_1}$, along with an increase in the occupation of $\ket{\varphi_2}$; then, the occupation of $\ket{\varphi_1}$ falls off to $0$, and is transferred to $\ket{\varphi_2}$. This implies that $\ket{\varphi_1}$ serves as an intermediate state for occupation transfer. However, it is not absolutely certain that there is no direct transfer between $\ket{\varphi_0}$ and $\ket{\varphi_2}$.

Despite the difference in the mechanism between the TLS and 3LS problems, similar features are observed in the time-pictures of the two systems. The 3LS system reaches an equal occupation of $0.5$ in the states $\ket{\varphi_0}$ and $\ket{\varphi_2}$, which their Bohr frequency is equal to the desired target frequency; then, the forcing field is ``turned off'' again.

Now, we solve the same problem with a smaller $\tilde{\alpha}$. The details of the problem are summarised in Table~\ref{tab:3LS1}.

\begin{table}
	\begin{equation*}
	\renewcommand{\arraystretch}{1.5}
	\begin{array}{|c||c|}
		\hline
		\operator{H}_0 & 
		\begin{bsmatrix}
			1 & 0 & 0 \\
			0 & 1.9 & 0 \\
			0 & 0 & 3
		\end{bsmatrix} \\ \hline
		\operator{\mu} &
		\begin{bsmatrix}
			0 & 1 & 1 \\
			1 & 0 & 1 \\
			1 & 1 & 0
		\end{bsmatrix} \\ \hline
		\ket{\psi_0} & \ket{\varphi_0} \\ \hline
		T & 100 \\ \hline
		\tfeps & 50\,\exp[-20(\omega- 1)^2] \\ \hline
		\tfmu & \exp[-20(\omega -2)^2] \\ \hline
		\bar\epsilon^{0}(\omega) & \exp[-20(\omega- 1)^2] \\ \hline
		K_i & 0.1 \\ \hline
		\text{tolerance} & 10^{-3} \\ \hline
	\end{array}
\end{equation*}

	\caption{The details of the second 3LS problem, with a smaller $\tilde{\alpha}$}\label{tab:3LS1}
\end{table}

\begin{figure}
	\centering \includegraphics[width=3in]{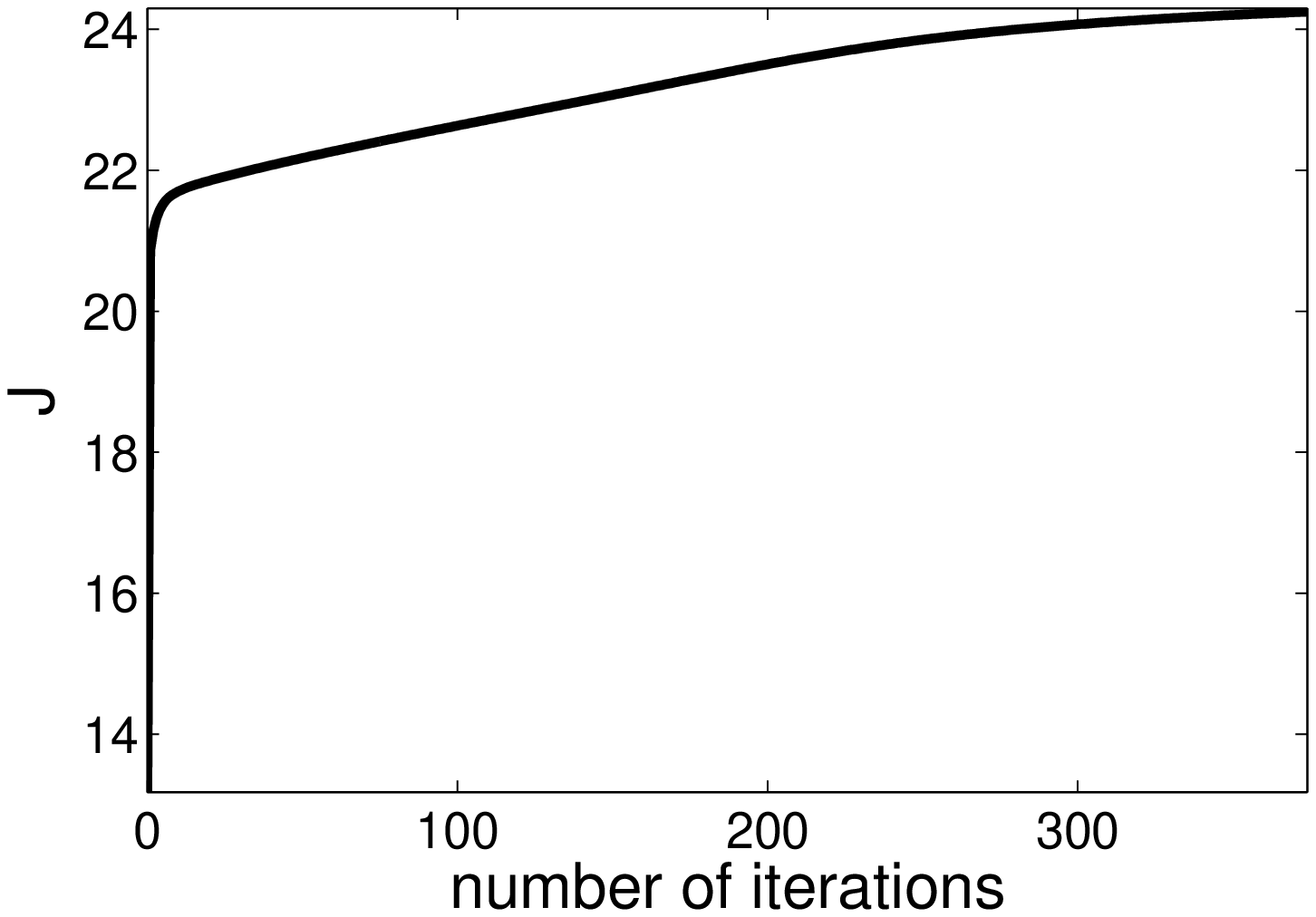}
	\caption{The convergence curve for the second 3LS problem}\label{fig:3LS1conv}
\end{figure}

\begin{figure}
	\centering \includegraphics[width=3in]{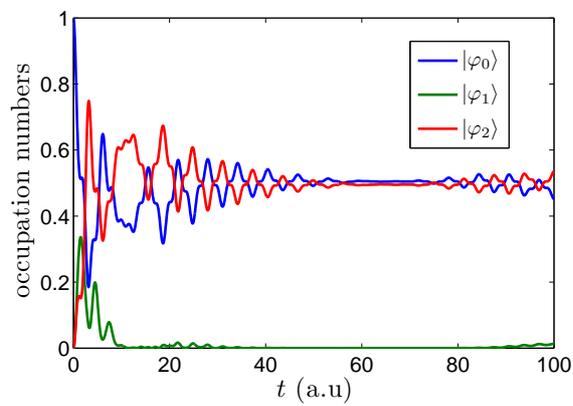}
	\caption{The occupation vs.\ time curves of the system, for the second 3LS problem; the states $\ket{\varphi_0}$ and $\ket{\varphi_2}$ reach an equal occupation of $0.5$, as in the previous problems; however, in this case, the occupation transfers between the levels continue, and the occupation of $\ket{\varphi_1}$ is not completely $0$ during the rest of the propagation.}\label{fig:3LS1oc}	
\end{figure}

The convergence curve is shown in Fig.~\ref{fig:3LS1conv}. The resulting spectra of $\epsw$ and $\muw$ do not seem to possess any new interesting features. However, the time-picture of the system is different from that of the previous problems: in this case, the forcing field is not completely turned off. The occupation time picture (Fig.~\ref{fig:3LS1oc}) has a general similarity to that of the previous problem: the states $\ket{\varphi_0}$ and $\ket{\varphi_2}$ reach an equal occupation of $0.5$, and the general picture of the system do not change much at the rest of the propagation. However, small occupation transfers between the levels continue, and we observe small ``jumps'' in the occupation of $\ket{\varphi_1}$. It seems that this state has a role also in the second element of the harmonic generation process. 

It is interesting to compare the value of $J_{max}$ of the two 3LS problems, with that computed using the following $\ket{\psi(t)}$ sequence:
\begin{equation}\label{eq:halfoc1}
	\ket{\psi(t)} = \exp\left(-i\operator{H}_0 t\right)\left\lbrace\frac{1}{\sqrt{2}}\left[\ket{\varphi_0} +\ket{\varphi_2}\right]\right\rbrace
\end{equation}
In this $\ket{\psi(t)}$ sequence, the eigenstates $\ket{\varphi_0}$ and $\ket{\varphi_2}$ have a constant occupation of $0.5$. Using this $\ket{\psi(t)}$ sequence, we have: \text{$J_{max}=24.0$}. In the first 3LS problem, we have a smaller value: \text{$J_{max}=21.5$}. This is not surprising, because the mechanism of harmonic generation seems to be based on achieving the steady state of an equally occupied quantum state; this is not achieved immediately at the beginning of the process. In the second 3LS problem, we have: \text{$J_{max}=24.3$}, which is even larger than that of the sequence of Eq.~\eqref{eq:halfoc1}. This means that the steady state of equal occupation of $0.5$ is not necessarily the one that yields the maximal emission, for a system with more than two states.

\subsection{Eleven level system (11LS)}\label{ssec:11LS}
The next problem is a more challenging one: We want to observe the 10'th harmonic, using an 11 level system (11LS). This problem is based on the same principle as that of the 3LS problem. The purpose of this example is to test the ability of the new method in generating higher harmonics.

The unperturbed Hamiltonian is:
\begin{equation}\label{eq:elLSH}
	\setcounter{MaxMatrixCols}{11}
	\operator{H}_0 = 
	\begin{bmatrix}
		1   &   &   &   &   &   &   &   &   &   & \\
		    &2.1&   &   &   &   &   &   &   &   & \\
		    &   &  3&   &   &   &   &   &   &   & \\
		    &   &   &3.9&   &   &   &   &\BigZero &   & \\
		    &   &   &   &  5&   &   &   &   &   & \\
		    &   &   &   &   &6.1&   &   &   &   & \\
		    &   &   &   &   &   &  7&   &   &   & \\
		    &   &\BigZero &   &   &   &   &8.1&   &   & \\
   		    &   &   &   &   &   &   &   &  9&   & \\
		    &   &   &   &   &   &   &   &   &9.9& \\
		    &   &   &   &   &   &   &   &   &   &11
	\end{bmatrix}
\end{equation}
The dipole moment operator is:
\begin{equation}\label{eq:elLSmu}
	\setcounter{MaxMatrixCols}{11}
	\operator{\mu} =
	\begin{bmatrix}
		   0&  1&   &   &   &   &   &   &   &   & 1\\
		   1&  0&  1&   &   &   &   &   &   &   & \\
		    &  1&  0&  1&   &   &   &   &   &   & \\
		    &   &  1&  0&  1&   &   &   &   &   & \\
		    &   &   &  1&  0&  1&   &   &\BigZero&   & \\
		    &   &   &   &  1&  0&  1&   &   &   & \\
		    &   &\BigZero&   &   &  1&  0&  1&   &   & \\
		    &   &   &   &   &   &  1&  0&  1&   & \\
		    &   &   &   &   &   &   &  1&  0&  1& \\
		    &   &   &   &   &   &   &   &  1&  0& 1\\
		   1&   &   &   &   &   &   &   &   &  1& 0
	\end{bmatrix}
\end{equation}
This $\operator{\mu}$ couples between neighbouring eigenstates, and between the outer eigenstates: $\ket{\varphi_0}$ and $\ket{\varphi_{10}}$. The forcing field is restricted not to exceed the region of the resonance frequencies of the neighbouring levels: \text{$\omega_{n+1,n}=1_{a.u.}$}. We require an emission in the neighbourhood of the Bohr frequency of the outer energy levels: \text{$\omega_{10,0}=10_{a.u.}$}. We want to see if the new method succeeds to find the route of ``climbing up'' to $\ket{\varphi_{10}}$, in order to generate oscillations in the requested frequency. This time, we use rectangular functions for $\tfeps$ and $\tfmu$.

The details of the problem are summarized in Table~\ref{tab:11LS}.

\begin{table}
	\begin{equation*}
	\renewcommand{\arraystretch}{1.5}
	\begin{array}{|c||c|}
		\hline
		\operator{H}_0 & \text{eq.~\eqref{eq:elLSH}} \\ \hline
		\operator{\mu} & \text{eq.~\eqref{eq:elLSmu}} \\ \hline
		\ket{\psi_0} & \ket{\varphi_0} \\ \hline
		T & 100 \\ \hline
		\tfeps & 50\,u(1.3-\omega) \\ \hline
		\tfmu & u(\omega - 9.9)\,u(10.1-\omega) \\ \hline
		\bar\epsilon^{0}(\omega) & u(1.3-\omega) \\ \hline
		K_i & 1 \\ \hline
		\text{tolerance} & 10^{-3} \\ \hline
	\end{array}
\end{equation*}
	\caption{The details of the 11LS problem}\label{tab:11LS}
\end{table}

\begin{figure}
	\centering \includegraphics[width=3in]{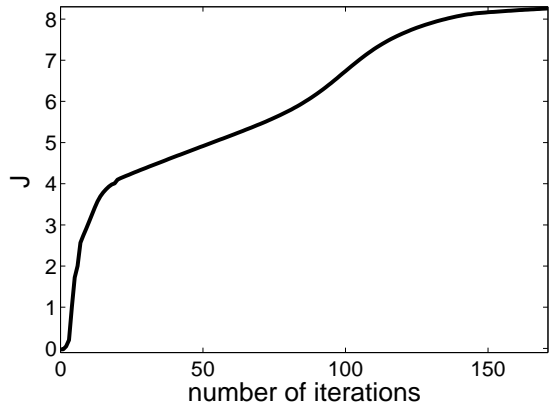}
	\caption{The convergence curve of the 11LS problem}\label{fig:11LSconv}	
\end{figure}

The convergence curve is shown in Fig.~\ref{fig:11LSconv}. This time, the starting point of the first guess is much less successful than in the previous examples. Nevertheless, the new method succeeds in the task of finding a satisfactory solution for the problem.

\begin{figure}
	\centering \includegraphics[width=3in]{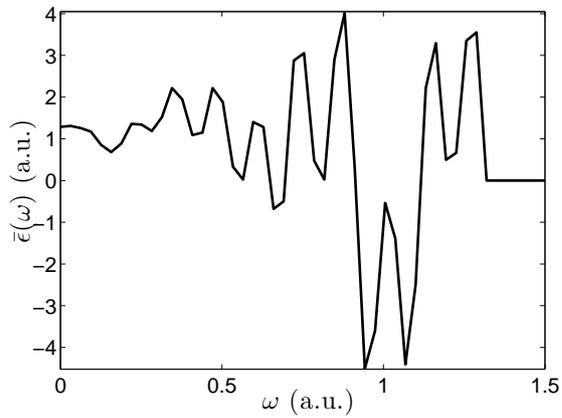}
	\caption{The $\epsw$ curve of the 11LS problem; the main components of the spectrum are in the region of the Bohr frequencies of the neighbouring energy levels, near \text{$\omega=1_{a.u.}$}. There are also important components in other frequencies.}\label{fig:11LSepsw}	
\end{figure}

The $\epsw$ curve is shown in Fig.~\ref{fig:11LSepsw}. The main components of the spectrum are in the region of the Bohr frequencies of the neighbouring levels. However, there are also important components in other frequencies. The spectrum is cut sharply at the maximal allowed frequency, due to the rectangular $\tfeps$.

\begin{figure}
	\centering \includegraphics[width=3in]{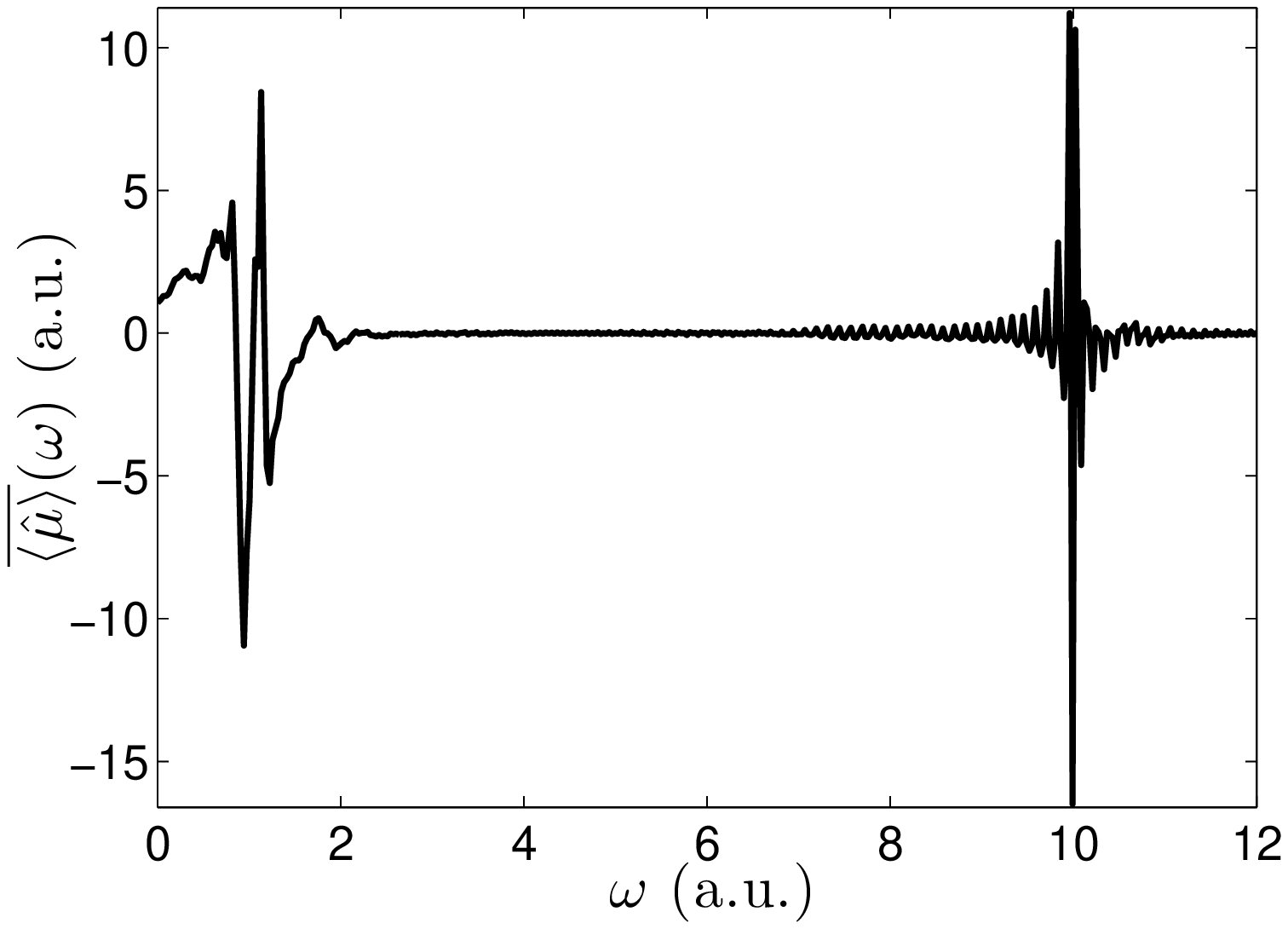}
	\caption{The $\muw$ curve of the 11LS problem; the largest component of the spectrum is at \text{$\omega_{10,0}=10_{a.u.}$}. Smaller components exist at the region Bohr frequencies of the neighbouring levels, near \text{$\omega=1_{a.u.}$}.}\label{fig:11LSmuw}	
\end{figure}

The resulting $\muw$ curve is shown in Fig.~\ref{fig:11LSmuw}. It does not seem to contain any new interesting features.

\begin{figure}
	\centering \includegraphics[scale=0.6]{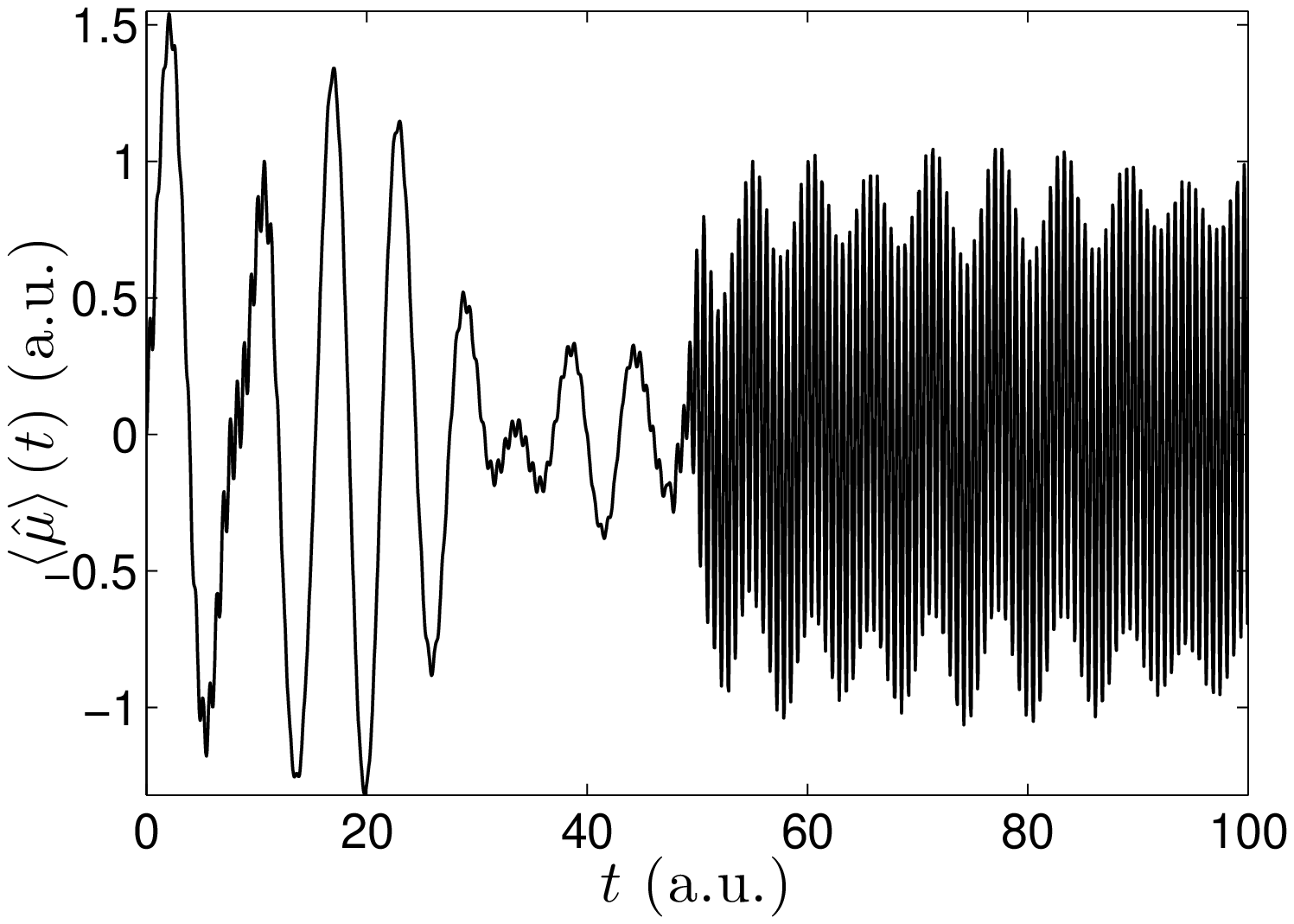} \\
	\centering \includegraphics[scale=0.6]{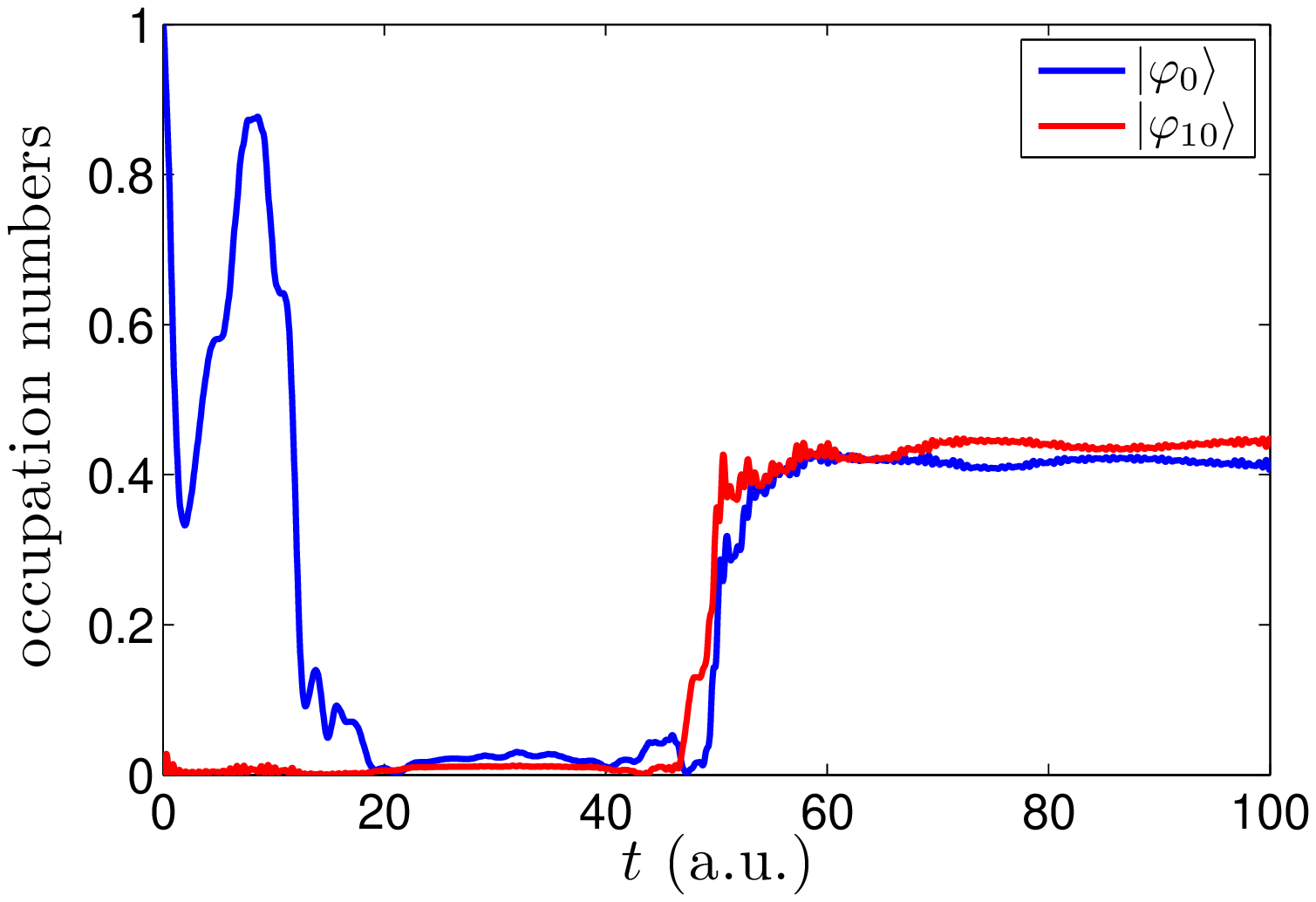}
	\caption{The time picture of the system, for the 11LS problem; At the top: The $\evmu$ curve; at the bottom:  The occupation vs.\ time curves of $\eigs 0$ and $\eigs{10}$; this time, the system only gets close to $0.5$ occupation.}\label{fig:11LSt}
\end{figure}

The time picture of the system is presented in Fig.~\ref{fig:11LSt}. The system does not achieve the $0.5$ occupation steady state, but gets close to it. We observed that it gets closer to this steady state after a larger number of iterations. It seems that the equally occupied state is more difficult to be achieved in this more complex system.

\section{Mathematical analysis of the results}\label{sec:analysis}
We want to get more insight into the mechanism that produces the requested frequencies at the steady states that were observed in Sec.~\ref{sec:MLS}. For this purpose, we analyse the spectrum of the expectation value of an arbitrary Hermitian operator. Then, we will be able to explain the observations of the previous section. The analysis will be very helpful for understanding the much more complex results of Sec.~\ref{sec:anharmonic}.

Consider a quantum system of dimension $N$; first, as a matter of convenience, let us extend the definition of $\eigs n$, to include $n<0$ or $n>N-1$:
\begin{equation}\label{eq:defphi}
	\eigs{n}\equiv
	\begin{cases}
		\text{the $n$'th eigenstate of } \operator{H}_0 & \qquad 0\leq n \leq N-1 \\
		\vec{0} & \qquad\text{otherwise}
	\end{cases}
\end{equation}

An arbitrary operator (not necessarily Hermitian) $\operator{o}$ that operates in the $N$ dimensional space is defined by the following set of equations:
\begin{equation}\label{eq:defo}
	\operator{o}\eigs{j} = \sum_{i=0}^{N-1}o_{ij}\eigs{i} \qquad \qquad 0\leq j \leq N-1
\end{equation}
(Note that using this definition, the index of the matrix elements $o_{ij}$ starts from $0$.)

Now we are going to decompose $\operator{o}$ into a sum of operators; this decomposition will be helpful for the analysis of the spectrum of a Hermitian operator expectation value.

Let us introduce a set of $2N-1$ operators: 
\[
	\operator{q}^{(n)}, \qquad \qquad -(N-1)\leq n \leq N-1
\]
The operator $\operator{q}^{(i)}$ is defined by the following set of $N$ equations:
\begin{equation}\label{eq:defq}
	\operator{q}^{(i)}\eigs{j} = \eigs{j+i} \qquad \qquad 0\leq j \leq N-1
\end{equation}
These operators will be found to be very useful for our analysis. The matrix elements of $\operator{q}^{(i)}$ are:
\begin{equation}\label{eq:qmatel}
	\left[\operator{q}^{(i)}\right]_{kj}=\bracketsO{\varphi_k}{\opq{i}}{\varphi_j} = \delta_{k,i+j}
\end{equation}
For instance, the matrix representation of $\opq{1}$ is:
\begin{equation}\label{eq:q1mat}
	\begin{bmatrix}
		0 &  &  & &\BigZero \\
		1 & 0& 	& & \\
		  & 1& 0& & \\
		  &  & 1&0& \\
		\BigZero & & &\ddots & \ddots
	\end{bmatrix}		
\end{equation}

Using Eq.~\eqref{eq:defq}, Eq.~\eqref{eq:defo} may be written in the following way:
\begin{equation}\label{eq:defoq}
	\operator{o}\eigs{j} = \sum_{i=0}^{N-1}o_{ij}\opq{i-j}\eigs{j} = \sum_{i=-j}^{N-j-1}o_{i+j,j}\opq{i}\eigs{j}
\end{equation}

Let us define the following set of constants, characteristic to the operator $\operator{o}$:
\begin{equation}\label{eq:defd}
	d_{ij}^o \equiv
	\begin{cases}
		o_{i+j,j} & \qquad (-i\leq j \leq N-i-1) \bigcap (0 \leq j \leq N-1) \\
		0 & \qquad \text{otherwise}
	\end{cases}
\end{equation}
Let us introduce the following set of $2N-1$ operators, characteristic to the operator $\operator{o}$:
\[
	\opdo{n}, \qquad \qquad -(N-1)\leq n \leq N-1
\]
The operator $\opdo{i}$ is defined by the following set of $N$ equations:
\begin{equation}\label{eq:defdop}
	\opdo{i}\eigs{j} = d_{ij}^o\eigs{j} \qquad \qquad 0\leq j \leq N-1
\end{equation}
The $\opdo{n}$ are diagonal operators; The main diagonal of $\opdo{i}$ contains the $i$'th diagonal of $\operator{o}$.


Using the operators $\opdo{n}$, we can write Eq.~\eqref{eq:defoq} in the following way:
\begin{equation}\label{eq:defoqd}
	\operator{o}\eigs{j} = \sum_{i=-(N-1)}^{N-1}d_{ij}^o\opq{i}\eigs{j} = \sum_{i=-(N-1)}^{N-1}\opq{i}\opdo{i}\eigs{j}
\end{equation}
Hence, we can write:
\begin{equation}\label{eq:osumqd}
	\operator{o} = \sum_{i=-(N-1)}^{N-1}\opq{i}\opdo{i}
\end{equation}
The $\opq{n}\opdo{n}$ operators have an important significance; The matrix elements of the operator $\opq{i}\opdo{i}$ are:
\begin{equation}\label{eq:qdmatel}
	\left[\opq{i}\opdo{i}\right]_{kj}=\bracketsO{\varphi_k}{\opq{i}\opdo{i}}{\varphi_j} = d_{ij}^o\delta_{k,i+j} = o_{i+j,j}\delta_{k,i+j} = o_{kj}\delta_{k,i+j}
\end{equation}
For instance, the matrix representation of $\opq{1}\opdo{1}$ is:
\begin{equation}\label{eq:qd1mat}
	\begin{bmatrix}
		0      &  		&  		& &\BigZero \\
		o_{10} & 0		& 		& & \\
		   	   & o_{21}	& 0		& & \\
		  	   &  		& o_{32}&0& \\
		\BigZero & 		& 		&\ddots & \ddots
	\end{bmatrix}		
\end{equation}
A well known example of an operator that may be written in the form of $\opq{i}\opdo{i}$ is $\operator{a}$, used in the treatment of the quantum harmonic oscillator system. $\operator{a}$ is defined by the following equation:
\begin{equation}\label{eq:defa}
	\operator{a}\eigs{n} = \sqrt{n}\eigs{n-1}
\end{equation}
If we define $\operator{o}$ as:
\begin{equation}
	\operator{o} \equiv \sqrt{2m\omega_0}\operator{X}
\end{equation}
($m$ is the mass, and $\omega_0$ is the characteristic frequency of the oscillator) we have:
\begin{equation}
	o_{kj} = \sqrt{n}(\delta_{k,j+1} + \delta_{k,j-1})
\end{equation}
Hence, we can write:
\begin{equation}
	\operator{a} = \opq{-1}\opdo{-1}
\end{equation}
Other well known examples are the operators $\operator{J}_+$ and $\operator{J}_-$, used in the treatment of  angular momentum in quantum mechanics.

Now, consider the case when $\operator{o}\equiv\operator{O}$, where $\operator{O}$ is an arbitrary \emph{Hermitian operator}. It is easy to show that $\opq{-i}\opdO{-i}$ is the adjoint operator of $\opq{i}\opdO{i}$:
\begin{equation}\label{eq:mdadj}
	\opq{-i}\opdO{-i} = \left(\opq{i}\opdO{i}\right)^+
\end{equation}
Hence, Eq.~\eqref{eq:osumqd} may be written in the following way:
\begin{equation}\label{eq:Osumqd}
	\operator{O} = \sum_{i=0}^{N-1}\frac{1}{l_i}\left[\opq{i}\opdO{i} + \left(\opq{i}\opdO{i}\right)^+ \right]
\end{equation}
where:
\begin{equation}\label{eq:defl}
	l_n \equiv 
	\begin{cases}
		2 & \qquad n=0 \\
		1 & \qquad n>0
	\end{cases}
\end{equation}
Let us define the following set of $N$ Hermitian operators:
\begin{equation}\label{eq:defOn}
	\opO{n} \equiv \frac{1}{l_n}\left[\opq{n}\opdO{n} + \left(\opq{n}\opdO{n}\right)^+ \right] \qquad\qquad 0 \leq n \leq N-1
\end{equation}
Now, Eq.~\eqref{eq:Osumqd} may be written as a sum of the $\opO{n}$:
\begin{equation}\label{eq:OsumOn}
	\operator{O} = \sum_{i=0}^{N-1}\opO{i}
\end{equation}

The $\opO{n}$ operators have an important physical significance: the operator $\opO{i}$ couples between the $i$'th nearest neighbours of the $\operator{H}_0$ eigenstates. These operators are useful when the energy levels are nearly equally spaced.

The matrix representation of $\opO{1}$, for instance, is:
\begin{equation}\label{eq:O1mat}
	\begin{bmatrix}
		0      &O_{10}^*&  		& &\BigZero \\
		O_{10} & 0		&O_{21}^*& & \\
		   	   & O_{21}	& 0		&O_{32}^*& \\
		  	   &  		& O_{32}&0& \\
		\BigZero & 		& 		&\ddots & \ddots
	\end{bmatrix}		
\end{equation}
There are well known examples of Hermitian operators, that contain only this component ($i=1$) in the sum of Eq.~\eqref{eq:OsumOn}: $\operator{X}$ and $\operator{P}$ of the harmonic oscillator system, and the angular momentum operators, $\operator{J}_x$ and $\operator{J}_y$, for a system that can be represented by an irreducible representation of the full rotation group.

Using \eqref{eq:OsumOn}, we can write:
\begin{equation}\label{eq:evOsum}
	\exval{\operator{O}} = \sum_{i=0}^{N-1}\exval{\opO{i}}
\end{equation}
Let us define:
\begin{equation}\label{eq:defc}
	c_n \equiv \brackets{\varphi_n}{\psi}
\end{equation}
$\ket{\psi}$ may be written as:
\begin{equation}
	\ket{\psi} = \sum_{j=0}^{N-1}c_j\eigs j
\end{equation}
Let us find the expression for $\exval{\opO{i}}$, in the terms of the $c_n$. Using \eqref{eq:qdmatel}, we have:
\begin{equation}\label{eq:evqd}
	\bracketsO{\psi}{\opq{i}\opdO{i}}{\psi} = \sum_{k=0}^{N-1}\sum_{j=0}^{N-1}c_k^*c_j\bracketsO{\varphi_k}{\opq{i}\opdO{i}}{\varphi_j} = \sum_{j=0}^{N-i-1}c_{i+j}^*c_j d_{ij}^O
\end{equation}
From \eqref{eq:defOn} and \eqref{eq:evqd}, we have:
\begin{equation}\label{eq:evOi}
	\exval{\opO{i}} = \frac{2}{l_i}\Real\left[\sum_{j=0}^{N-i-1}c_{i+j}^*c_j d_{ij}^O\right]
\end{equation}

Now, we have the necessary tools for the analysis of the spectrum of $\exval{\operator{O}}(t)$. The spectrum of $\exval{\operator{O}}(t)$ is the sum of the spectra of the $\exval{\opO{n}}(t)$. It will be instructive to express the $\exval{\opO{n}}(t)$ in the terms of the components of the state in the \emph{interaction picture}. The state in the interaction picture is:
\begin{equation}\label{eq:psiI}
	\ket{\psi_I(t)} \equiv \exp\left(i\operator{H}_0 t\right)\ket{\psi(t)}
\end{equation}
We define the components of $\ket{\psi_I(t)}$:
\begin{equation}\label{eq:defb}
	b_n(t) \equiv \brackets{\varphi_n}{\psi_I(t)}
\end{equation}
We have:
\begin{equation}\label{eq:cb}
	c_n(t) = b_n(t)\exp(-iE_n t)
\end{equation}

First, we consider an important special case, when \text{$\operator{H}(t)=\operator{H}_0$}. In this case, the occupation of the $\eigs {n}$ is constant, and so are the $b_n$. Using Eqs.~\eqref{eq:evOi}, \eqref{eq:cb}, we can write:
\begin{align}
	\exval{\opO{i}}(t) &= \frac{2}{l_i}\Real\left[\sum_{j=0}^{N-i-1}b_{i+j}^*b_j d_{ij}^O\exp(i\omega_{i+j,j}t)\right] \nonumber \\
	& = \frac{2}{l_i}\sum_{j=0}^{N-i-1}\left|b_{i+j}\right|\,\left|b_j\right|\,\left|d_{ij}^O\right|\cos(\omega_{i+j,j}t+\phi_{ij}) \nonumber \\
	\phi_{ij} &\equiv \arg\left(b_{i+j}^*b_j d_{ij}^O\right) \label{eq:evOit}
\end{align} 
According to this expression, the spectrum of $\exval{\opO{i}}(t)$ consists of the Bohr frequencies of the $i$'th nearest neighbouring eigenstates. The amplitude of the term with the Bohr frequency: $\omega_{i+j,j}$, is determined by the occupation of the eigenstates: $\eigs{i+j}$, $\eigs{j}$, and the magnitude of $d_{ij}^O$, which represents the coupling between these eigenstates. Of course, the spectrum of $\exval{\operator{O}}(t)$ consists of all the Bohr frequencies which have non-zero amplitudes. When the energy levels are roughly equally spaced, it is convenient, both conceptually and practically, to group together the Bohr frequencies that belong to the spectrum of any of the $\exval{\opO{n}}(t)$.

Consider the case when we want to maximize the amplitude of only one of the Bohr frequencies of the system: $\omega_{mn}$; in addition, $\eigs m$ and $\eigs n$ are the only pair of eigenstates with this Bohr frequency. We ask: what is the optimal steady state for maximizing the amplitude, when the occupation remains constant? It is clear from Eq.~\eqref{eq:evOit}, that all the occupation should be at the states $\eigs m$, $\eigs n$, \ie:
\begin{equation}
	\left|b_m\right|^2 + \left|b_n\right|^2 = 1
\end{equation}
Let us define:
\begin{align*}
	& \left|b_m\right| = \cos\theta &\nonumber \\
	& \left|b_n\right| = \sin\theta & 0\leq\theta\leq \frac{\pi}{2}
\end{align*}
We have:
\begin{equation*}
	2\left|b_m\right|\,\left|b_n\right| = \sin(2\theta)
\end{equation*}
It follows that the maximal amplitude in \eqref{eq:evOit} is achieved when $\theta=\pi/4$, and:
\begin{equation}\label{eq:equal}
	\left|b_m\right| = \frac{1}{\sqrt{2}} \qquad \qquad \left|b_n\right| = \frac{1}{\sqrt{2}}
\end{equation}
\ie the optimal steady state is the equally occupied state.

When the occupation is not constant, $b_n=b_n(t)$. In this case, we should represent $b_n(t)$ by its spectral components. The Fourier transform of the $b_n(t)$ is:
\begin{equation}\label{eq:Ftb}
	\widetilde{b}_n(\omega) = \frac{1}{\sqrt{2\pi}}\int_{-\infty}^{\infty}b_n(t)\exp(i\omega t)\,dt
\end{equation}
$b_n(t)$ may be written as the inverse Fourier transform of $\widetilde{b}_n(\omega)$:
\begin{equation}\label{eq:binvFt}
	b_n(t) = \frac{1}{\sqrt{2\pi}}\int_{-\infty}^{\infty}\widetilde{b}_n(\omega)\exp(-i\omega t)\,d\omega
\end{equation}
Using Eqs.~\eqref{eq:evOi}, \eqref{eq:cb}, \eqref{eq:binvFt}, we can write:
\begin{align}
	& \exval{\opO{i}}\!(t) = \frac{1}{\pi l_i}\Real\left\lbrace\!\!\sum_{j=0}^{N-i-1}\!\!\! \iint \! \widetilde{b}_{i+j}^*(\omega)\widetilde{b}_j(\omega') d_{ij}^O\exp\left[i\left(\omega_{i+j,j}+\omega-\omega'\right)t\right] \,d\omega\,d\omega' \right\rbrace \nonumber \\
	& = \frac{1}{\pi l_i}\!\!\sum_{j=0}^{N-i-1}\!\!\! \iint \left|\widetilde{b}_{i+j}(\omega)\right|\,\left|\widetilde{b}_j(\omega')\right|\,\left|d_{ij}^O\right|\cos\left[\left(\omega_{i+j,j}+\omega-\omega'\right)t+\phi_{ij}(\omega, \omega')\right]\,d\omega\,d\omega' \nonumber\\
&\phi_{ij}(\omega, \omega') \equiv \arg\left[\widetilde{b}_{i+j}^*(\omega)\widetilde{b}_j(\omega') d_{ij}^O\right] \label{eq:evOitbt}
\end{align}
We see that the spectrum of $\exval{\opO{i}}\!(t)$ consists of the frequencies: \text{$\omega_{i+j,j}+\omega-\omega'$} in which $\left|\widetilde{b}_{i+j}(\omega)\right|\,\left|\widetilde{b}_j(\omega')\right|\,\left|d_{ij}^O\right|$ is significant. This allows the appearance of  frequencies other than the Bohr frequencies. Usually, we observe that the main contribution to the spectrum is from \text{$\omega$, $\omega'$} values in which the difference: \text{$\omega-\omega'$} is rather small, compared with $\omega_{i+j,j}$. Hence, most frequently, the spectrum will be mainly distributed around the Bohr frequencies. If the energy levels are roughly equally spaced, the main contribution to the spectrum of $\exval{\opO{i}}(t)$ will be concentrated in a single region of the spectrum.

We return to the analysis of the results of Sec.~\ref{sec:MLS}. The most pronounced observation of Sec.~\ref{sec:MLS} was the steady state of the equal occupation of the two levels with the appropriate Bohr frequency. This is explained by the discussion above on the ideal steady state with constant occupation. In the TLS example, where there is a contribution only from one pair of levels, it is certain that this constantly occupied state is the ideal one in general.

This is not necessarily true for the 3LS problem; $\operator{\mu}$ of this problem (see Eq.~\eqref{eq:thLSmu}) may be decomposed into two Hermitian components:
\begin{equation}\label{eq:3LSmusum}
	\operator{\mu} = \opmu{1} + \opmu{2} = 
	\begin{bmatrix}
		0 & 1 & 0 \\
		1 & 0 & 1 \\
		0 & 1 & 0
	\end{bmatrix}
	+
	\begin{bmatrix}
		0 & 0 & 1 \\
		0 & 0 & 0 \\
		1 & 0 & 0
	\end{bmatrix}
\end{equation}
The above statement for the TLS problem is true if the only contribution to $\muw$ is from $\muwn{2}$; however, a contribution from $\muwn{1}$ is also possible. It is certain that in order to get contribution from $\muwn{1}$, the state $\eigs 1$ must be occupied. Moreover, the Bohr frequencies that characterise $\muwn{1}$ are around \text{$\omega=1_{a.u.}$}; in order to get response in the neighbourhood of \text{$\omega=2_{a.u.}$}, the occupation should not be constant (see Eq.~\eqref{eq:evOitbt}). These requirements for contribution from $\muwn{1}$ contradicts the constantly equally occupied picture.

The first 3LS example, with a relatively large $\tilde{\alpha}$, demonstrates a more energetically economical use of the forcing field; the altering of the occupation, required for a contribution from $\muwn{1}$, requires energy, and it is preferable to use the equal occupation mechanism, with $0$ field. In this case, there is  contribution only from $\muwn{2}$. In the second 3LS example, the smaller $\tilde{\alpha}$ allows the use of a slightly different mechanism, in which there is also a small contribution from $\muwn{1}$.

\begin{figure}
	\centering \includegraphics[width=3in]{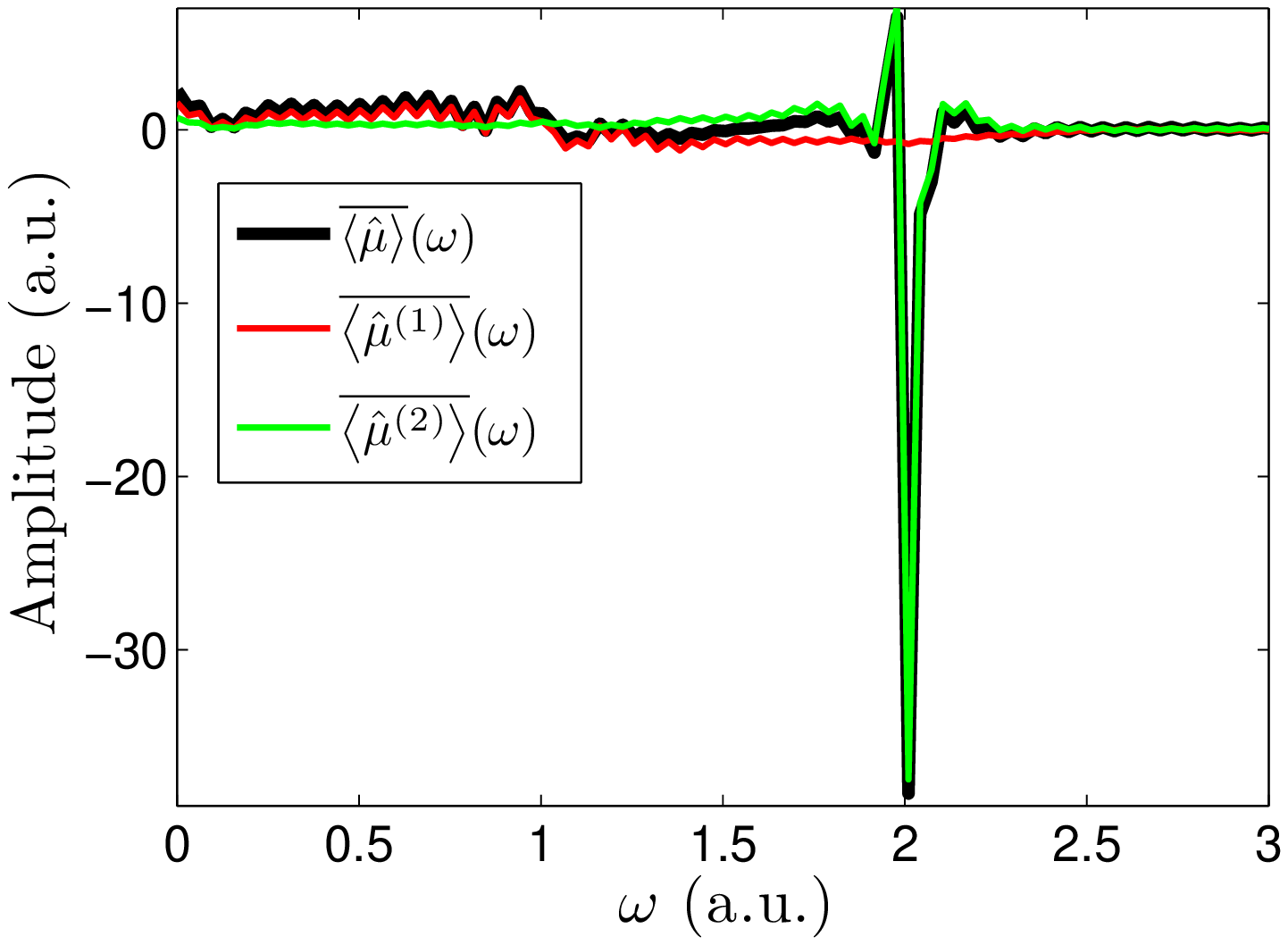} \\
	\centering \includegraphics[width=3in]{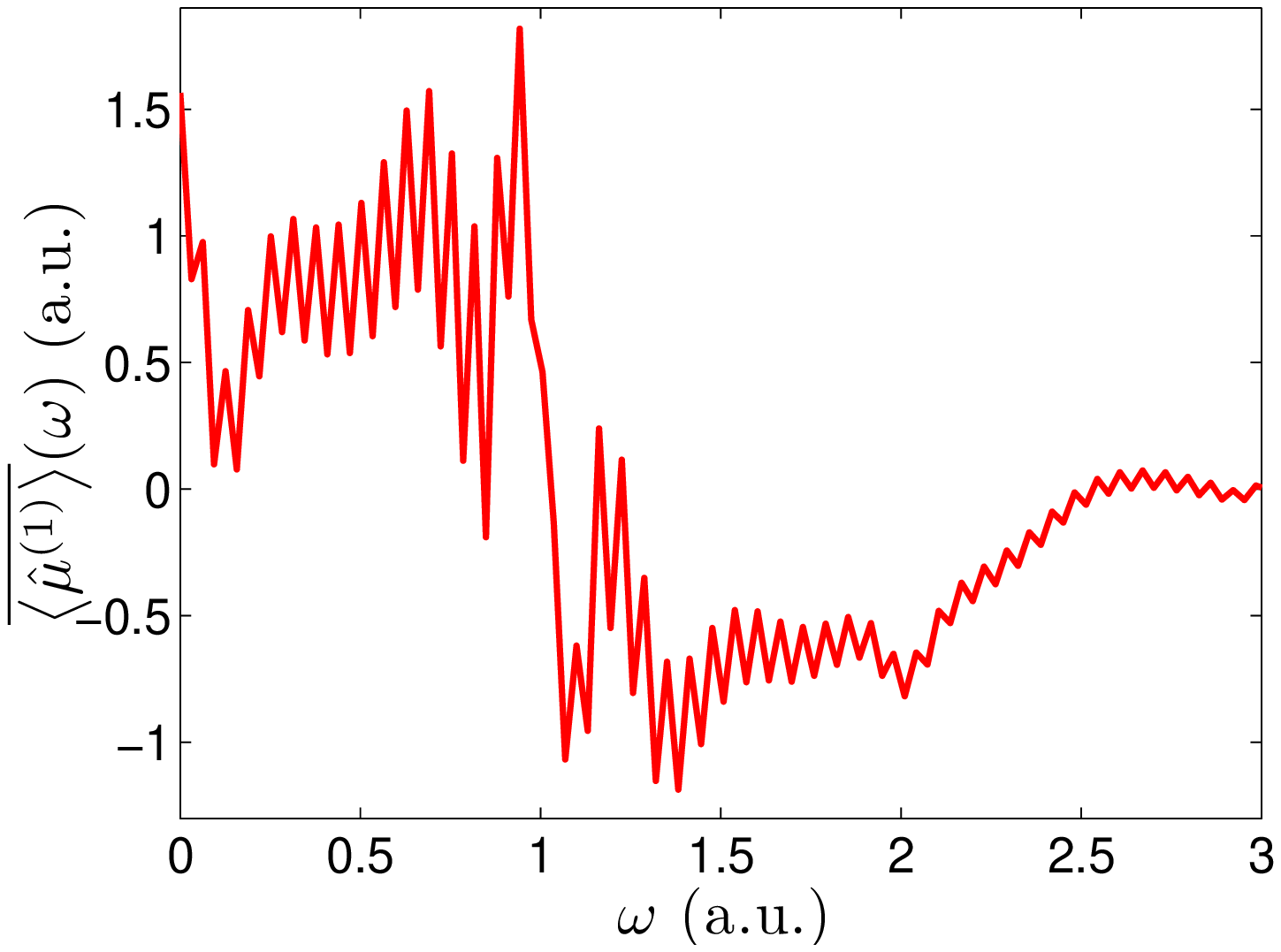}
	\caption{The composition of $\muw$, in the terms of the $\muwn{n}$, for the second 3LS problem; at the top: The curves of $\muw$ and its Hermitian components; at the bottom: A close up view of $\muwn{1}$; the main contribution to $\muw$ is from $\muwn{2}$. However, there is also a small contribution from $\muwn{1}$.}\label{fig:3LS1muw}
\end{figure}

In Fig.~\ref{fig:3LS1muw}, $\muw$ for the second 3LS is shown, along with the spectrum of the Hermitian components. The main contribution is from $\muwn{2}$, as could be expected; however, we observe that there is also a small contribution from $\muwn{1}$. This contribution makes the $J_{max}$ value larger than the maximal possible with the equal occupation mechanism.

This simple example illustrates the effect of the generation of other frequencies than the Bohr frequencies characteristic to $\muwn{n}$. In this case, the effect gives only a small contribution to $J_{max}$. In more complex systems we may observe that the main contribution to $J_{max}$ involves such effects. We also frequently observe that an important contribution to $J_{max}$ comes from more than one Hermitian component. Typically, the most important contribution comes from the $\muwn{n}$ that their characteristic Bohr frequencies are close to the non-negligible part of $\fmu$.

We introduce a useful tool for finding the relative importance of the contributions of the Hermitian components; let us define the following set of $N$ functionals:
\begin{equation}\label{eq:Jmaxn}
	J_{max}^{(n)} \equiv \frac{1}{2}\int_0^\Omega \tfmu\overline{\exval{\operator{\mu}^{(n)}}}^2(\omega)\,d\omega \qquad \qquad 0\leq n \leq N-1
\end{equation}
Of course:
\[
	J_{max} \neq \sum_{n=0}^{N-1}J_{max}^{(n)}
\]
because of the cross terms; this set of functionals has to be considered just a useful tool for qualitative view.

\section{Anharmonic oscillators}\label{sec:anharmonic}
In this section, we test the capability of the new method when dealing with the more complex systems of anharmonic oscillators. We deal with anharmonic potentials with the general form of a chemical bond potential (the Morse potential, Subsections.~\ref{ssec:HCl}, \ref{ssec:HClfic}), or a similar form (the Toda potential, Subsection.~\ref{ssec:Toda}).

The energy levels of an anharmonic oscillator are roughly equally spaced. Hence, the treatment of Sec.~\ref{sec:analysis} is appropriate here.

We begin with a description of the structure of the dipole operator of an anharmonic oscillator, with the general form of a chemical bond potential. This description is necessary for the discussion on the results.

First, we describe the structure of a dipole operator of the form: \text{$\operator{\mu}\propto\operator{X}$}.

The dipole operator of the \emph{harmonic} oscillator system has a simple structure: Only neighbouring states are coupled by $\operator{\mu}$, and \text{$\operator{\mu}=\opmu{1}$}. The only active Bohr frequency in $\operator{\mu}$ is: $\omega_{n+1,n}=\omega_0$, where $\omega_0$ denotes the characteristic frequency of the oscillator. This structure determines the selection rules of the harmonic oscillator.

The dipole operators of anharmonic oscillator systems have a more complex structure; the largest couplings are the $d_{1n}^\mu$, because of the similarity to the harmonic oscillator system. We may also have relatively large $d_{0n}^\mu$ values, particularly for larger $n$ values; they represent the deviation of the $\bracketsO{\varphi_n}{\operator{X}}{\varphi_n}$ from the bottom of the well, where \text{$x=0$}. The $d_{mn}^\mu$ values for $i>1$ are usually much smaller in magnitude. We also observe the following trends for the $d_{mn}^\mu$:
\begin{enumerate}
	\item $d_{mn}^\mu$ decays rapidly with $m$, for $m>1$.
	\item $d_{mn}^\mu$ increases with $n$.
\end{enumerate}

The explanation for the first trend is simple: the selection rules of the harmonic oscillator originates from its symmetry properties. In anharmonic oscillator systems, there are deviations from this symmetry. A pronounced value for $d_{mn}^\mu$ with larger $m$, means a larger deviation from the harmonic oscillator selection rules, and requires greater break of symmetry.

The second trend characterizes also the $d_{1n}^\mu$ values of the harmonic oscillator system, which increase with the square-root of $n$. This trend is explained by the fact that higher energy eigenstates are characterized by a high density of $\varphi_n(x)$ at larger $x$ values. Hence, the coupling between higher energy states represents larger amplitude phenomena, and is larger in magnitude. In the context of a chemical bond potential, an additional explanation may be given to the second trend: The deviation of the chemical bond potential from the harmonic oscillator symmetry is greater for larger deviations from $x=0$.

The dipole operator of the form: \text{$\operator{\mu}\propto\operator{X}$}, is a good approximation for a chemical bond system, in the region of the equilibrium state. However, this description is inadequate for larger deviations from equilibrium, where changes in the charge separation have to be taken into account. When dealing with larger deviations, the dependence of the dipole on $x$ has to be described by a more complicated dipole function, $\mu(x)$. The description of the $\operator{\mu}$ structure given above for the linear form of $\mu(x)$, might be inadequate for other forms. Nevertheless, the main features are the same, at least for the lower energy levels.

When using a non-linear functional form of $\mu(x)$, the deviations from the selection rules of the harmonic oscillator for the linear functional form, are larger in magnitude.

\subsection{The HCl molecule}\label{ssec:HCl}
The first example is the H$\,^{35}$Cl molecule. We have chosen this problem, to test the method for a realistic choice of parameters. However, there is no intention of giving accurate predictions in this example. Hence, we make use of approximations that may be somewhat crude.

The coordinate of the one-dimensional oscillator is the deviation of the inter-nuclei distance, $r_{H-Cl}$, from the distance at the bottom of the well ($r^*$):
\begin{equation}\label{eq:HClx}
	x = r_{H-Cl} - r^*
\end{equation}

The potential of the bond was obtained by adjusting the parameters of the Morse potential:
\begin{equation}\label{eq:Morse}
	V(x) = D_0\left[\exp(-ax)-1\right]^2
\end{equation}
to experimental data on HCl: The atomization energy of HCl, and the frequency of vibration, using the IR absorption frequency for the transition to the fundamental state. We made a few reasonable approximations. The resulting potential is brought in Table~\ref{tab:HCl}. The characteristic frequency of the bottom of the well is:
\begin{equation}\label{eq:HClw0}
	\omega_0 = {1.35\cdot 10^{-2}}_{a.u.}
\end{equation}

\begin{table}
	\begin{equation*}
	\renewcommand{\arraystretch}{2}
	\begin{array}{|c||c|}
		\hline
		\operator{H}_0 & \frac{\operator{P}^2}{2\cdot 1785} + 0.171\left[\exp\left(-0.975\,\operator{X}\right)-\operator{I}\right]^2 \\ \hline
		\operator{\mu} &  \left(0.19309\,\operator{X}\right)\;\times \\
		& \left\lbrace \operator{I}-\Real\left[\tanh\left((0.17069 +0.056854\,i)\left(\operator{X} - 0.10630\,\operator{I}\right)^{1.8977}\right)\right]\right\rbrace\\ \hline
		\ket{\psi_0} & \ket{\varphi_0} \\ \hline
		T & 10^4 \\ \hline
		\tfeps & 2500\,u(0.015-\omega) \\ \hline
		\tfmu & 100\,u(\omega - 0.025)\,u(0.027-\omega) \\ \hline
		L & 19 \\ \hline
		\gamma_n & (n-19)^2 \\ \hline
		\bar\epsilon^{0}(\omega) & u(0.015-\omega) \\ \hline
		K_i & 1 \\ \hline
		x \text{ domain} & [-0.69407,\;3.51178) \\ \hline
		N_{grid} & 32 \\ \hline
		\text{tolerance} & 10^{-3} \\ \hline
	\end{array}
\end{equation*}
	\caption{The details of the HCl problem}\label{tab:HCl}
\end{table}
 

The dipole function was obtained by adjusting experimental data to a reasonable functional form. The data is the first 4 derivatives of $\mu(x)$ at equilibrium \cite{HCl}:
\[
	\left(\nderiv{\mu}{x}{n}\right)_{eq} \qquad\qquad n=1,2,3,4
\]
The functional form is:
\begin{equation}\label{eq:mutanh}
	\mu(x) = a_1x\left\lbrace 1-\tanh\left[a_2(x-a_3)^{a_4}\right]\right\rbrace
\end{equation}
This functional form is intended to represent a nearly linear form at $x=0$, that decays to $0$ at the dissociation region of the potential. This behaviour is typical to a homolitic bond cleavage. We made the approximation:
\[
	\left(\nderiv{\mu}{x}{n}\right)_{x=0}\approx\left(\nderiv{\mu}{x}{n}\right)_{eq}
\]
The resulting system of equations was solved using the Symbolic Math Toolbox of MATLAB\@. The resulting function is complex; we take its real part (see Table~\ref{tab:HCl}).

\begin{figure}
	\centering \includegraphics[width=3in]{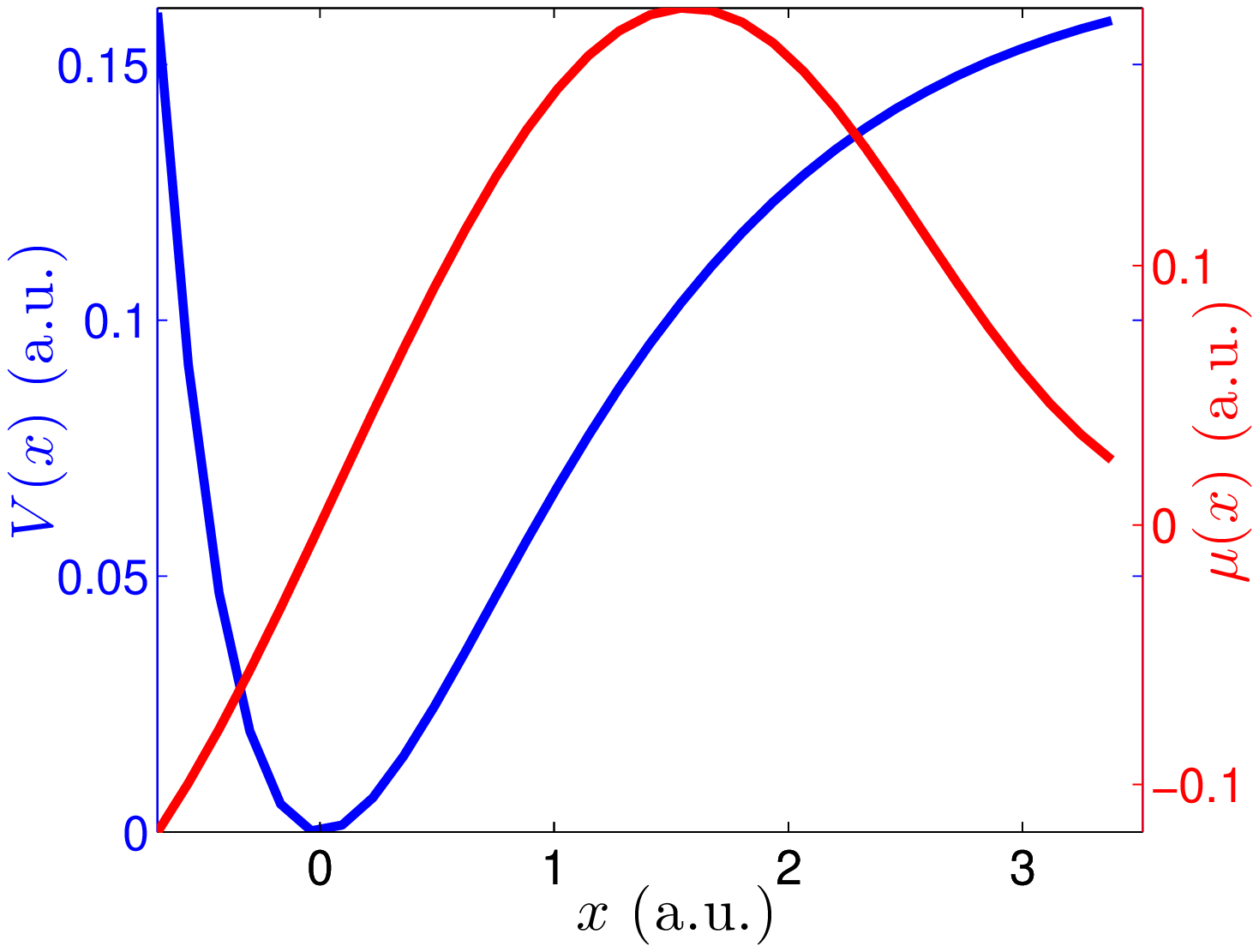}
	\caption{The approximated potential (blue) and dipole function (red) curves, for the HCl molecule; $\mu(x)$ decays to $0$ at the dissociation region of $V(x)$.}\label{fig:Vmux}	
\end{figure}

The resulting potential and dipole function are presented in Fig.~\ref{fig:Vmux}. We can see that $\mu(x)$ indeed decays at the dissociation region of $V(x)$.

In our problem, we want to maximize the emission at the neighbourhood of second harmonic:
\begin{equation}\label{eq:HCl02har}
	\omega_{2,0} = E_2 - E_0 = {2.54\cdot 10^{-2}}_{a.u.}
\end{equation}
$\tfeps$ and $\tfmu$ were chosen to be rectangular functions. We restrict the field frequency not to exceed much $\omega_0$ (see Eq.~\eqref{eq:HClw0}). $\tfmu$ is chosen in a way, that from all the Bohr frequencies of $\opmu{2}$, \ie $\omega_{n+2,n}$, only $\omega_{2,0}$ is contained in the non-zero part of $\tfmu$. The other $\omega_{n+2,n}$ are smaller, because of an anharmonic effect. We want to see if we get an equal 0.5 occupation of the eigenstates: $\eigs 0$, $\eigs 2$.

It was necessary to restrict the allowed eigenstates in order to prevent dissociation and occupation of non-physical states. We use the method described in Subsection~\ref{ssec:modif}.

The details of the problem are summarised in Table~\ref{tab:HCl}.

\begin{figure}
	\centering \includegraphics[width=3in]{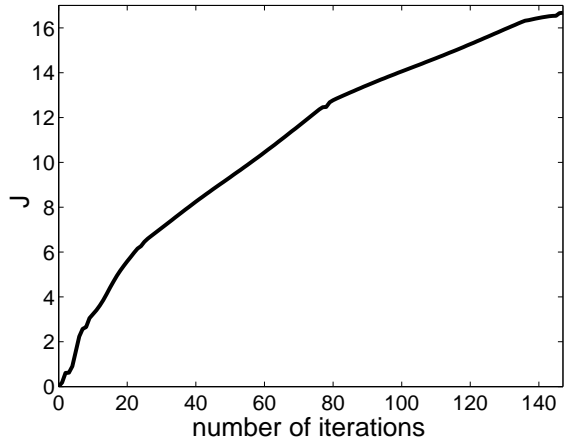}
	\caption{The convergence curve of the HCl problem}\label{fig:HClconv}	
\end{figure}

The resulting convergence curve is presented in Fig.~\ref{fig:HClconv}.

\begin{figure}
	\centering \includegraphics[width=3in]{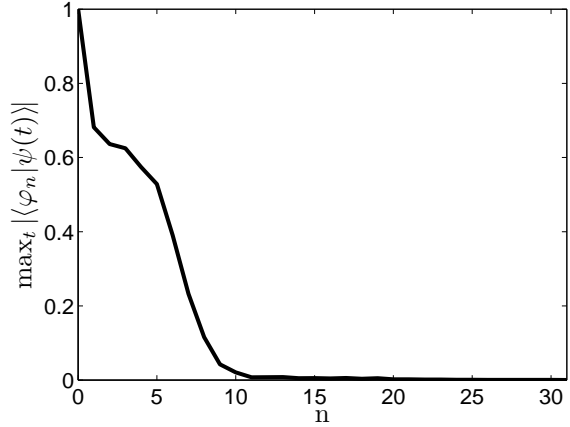}
	\caption{The maximal $|c_n(t)|$ during the propagation for all $n$; the forbidden states are of $n>19$. The restriction of the allowed eigenstates is successful in this case.}\label{fig:HClprojs}	
\end{figure}

In Fig.~\ref{fig:HClprojs}, the maximal $|c_n(t)|$ during the propagation is shown for all eigenstates. The restriction of the allowed eigenstates is shown to be successful in this case.

\begin{figure}
	\centering \includegraphics[width=3in]{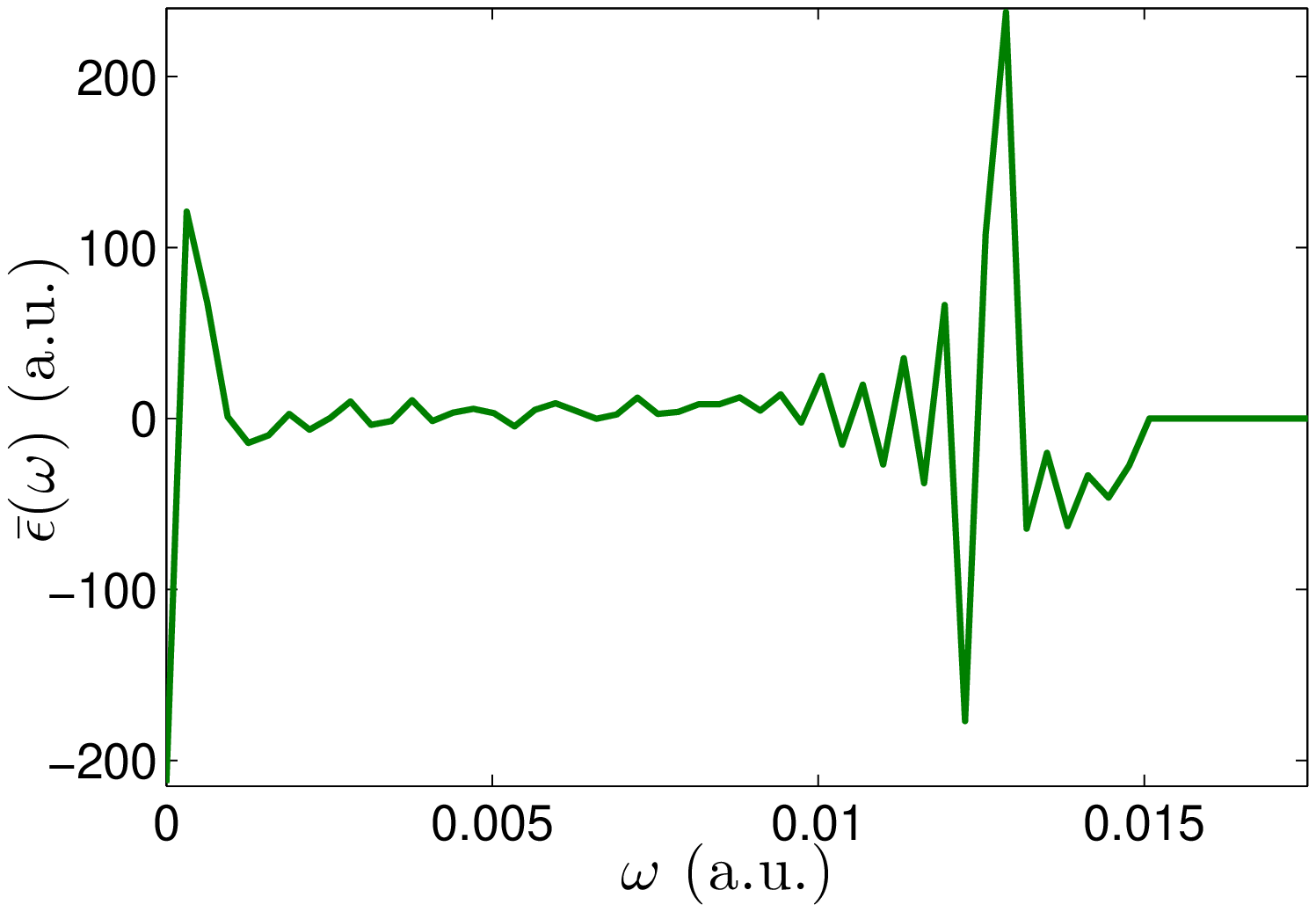} \\
	\centering \includegraphics[width=3.5in]{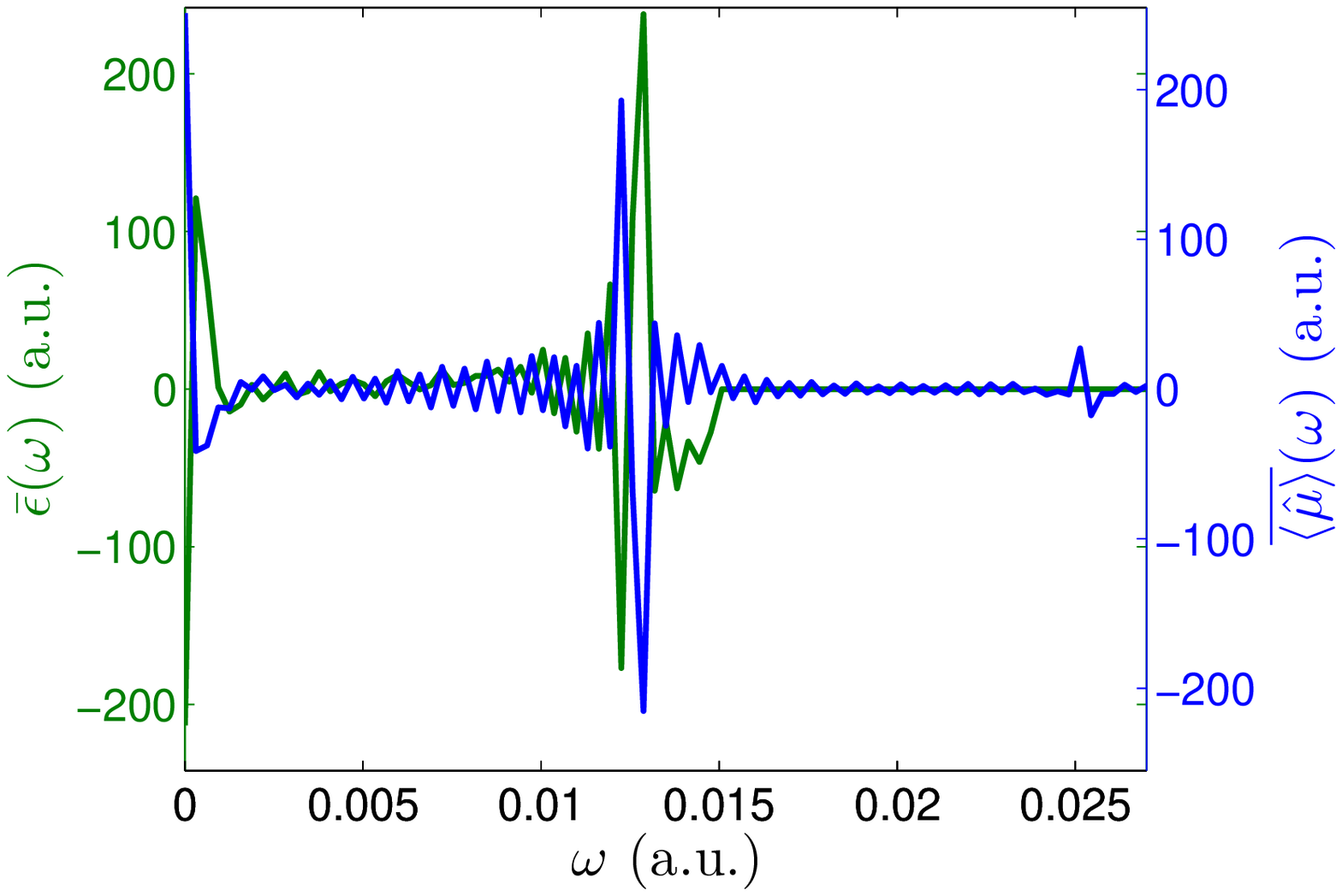}
	\caption{The $\epsw$ (green) and $\muw$ (blue) curves of the HCl problem; the spectrum of the field mainly consists of the region of $\omega_0$, besides a large, negative, constant field ($\omega=0$) component. There is an important component at $\omega=\pi/T$, that represents the tendency of the mean field to attain more negative values during the process (see Fig.~\ref{fig:HClt}). $\muw$ has a large peak at $\omega=0$, due to the positive deviation from $x=0$, typical to chemical bond potentials. There is an important component at $\omega=\pi/T$, that represents the gradual increase in this deviation during the process, due to occupation of higher energy levels. There is a large response in the region of $\omega_0$. Smaller components exist at the neighbourhood of the second harmonic. The most of the important components of the two spectra are out of phase.}\label{fig:HClw}
\end{figure}

\begin{figure}
	\centering \includegraphics[width=3.5in]{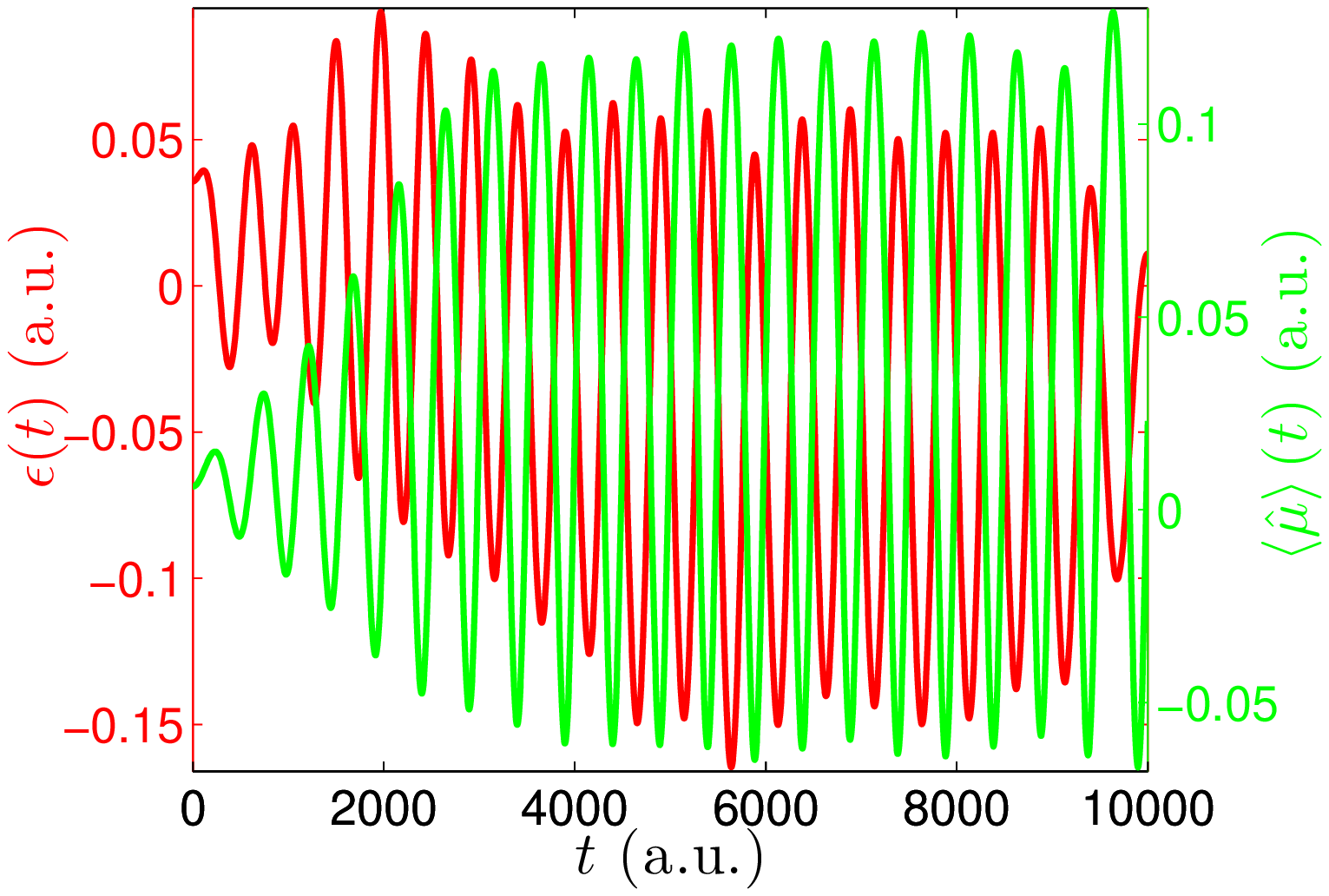} \\
	\centering \includegraphics[width=3in]{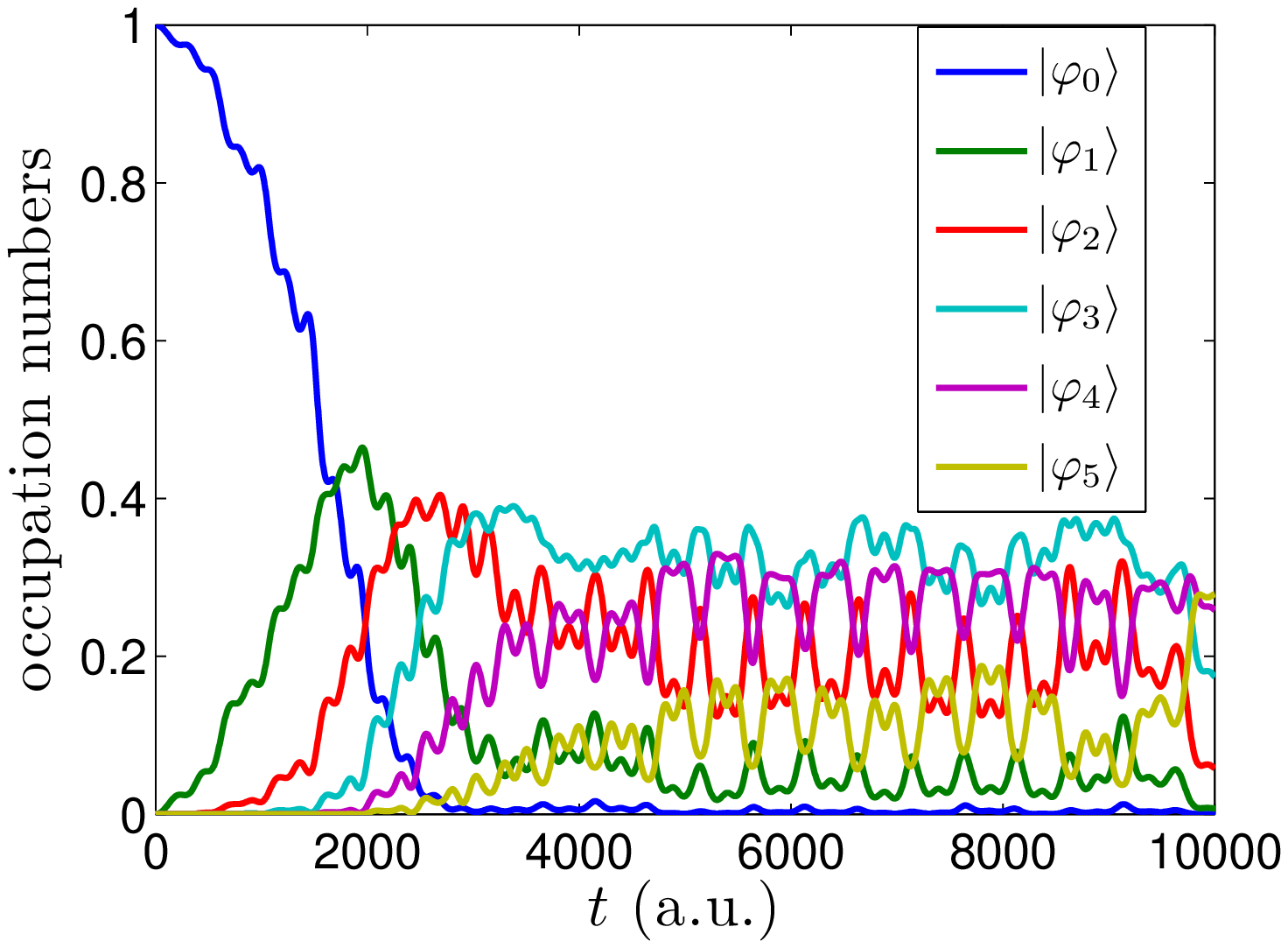}
	\caption{The time picture of the system, for the HCl problem; at the top: The $\epsilon(t)$ (red) and $\evmu$ (green) curves; at the bottom: The occupation vs.~time curves for the 6 first eigenstates; the mean forcing field becomes more negative, as higher energy levels are occupied. The system seems to reach a steady-state around \text{$t=5000_{a.u.}$}. This steady state is interrupted at the end of the propagation, because of boundary effects. At the steady state, $\epsilon(t)$ and $\evmu$ are out-of-phase. The eigenstates are more or less distributed around $\eigs 3$. There is almost no occupation in $\eigs 0$.}\label{fig:HClt}
\end{figure}

In Fig.~\ref{fig:HClw}, the $\epsw$ and $\muw$ curves are presented. The time-picture is presented in Fig.~\ref{fig:HClt}.

The spectrum of the forcing field mainly consists of the region of $\omega_0$, as could be expected. This is along with a large, negative, constant field ($\omega=0$) component. This means, that the mean field has a  negative value. This is also apparent from the time-picture. There is an important positive component at the first non-zero frequency: $\omega=\pi/T$. This term completes half a period at $T$. Since $\epsw$ represents the coefficients of cosine terms, this means that the field tends to more negative values during the propagation. This is also apparent from the time-picture. These observations require an explanation.

$\muw$ also attains large values at the region of $\omega_0$, as could be expected. There is a large, positive peak at \text{$\omega=0$}. This is clearly due to the positive deviations of $\exval{\operator{X}}$ from $x=0$, typical to chemical bond potentials. There is an important negative component at $\omega=\pi/T$. It represents the gradual increase in this deviation during the process, due to occupation of higher energy levels; this can be seen in the time-picture. $\muw$ attains smaller, but significant, values in the neighbourhood of the second harmonic. The smaller values, compared with the first harmonic, are a consequence of the much smaller couplings of $\opmu{2}$, compared with $\opmu{1}$.

The system seems to achieve a more-or-less steady state around \text{$t=5000_{a.u.}$}. This steady state is interrupted at the end of the propagation due to boundary effects; these will be discussed in Sec.~\ref{sec:problems}.

We see that the occupation of the eigenstate at the steady state is more or less distributed around $n=3$. This indicates that a semi-classical view is not very far from reality in this case. This may be verified by following the $x$ shape of $|\psi(x, t)|^2$ during the process, which has a localized character. However, the shape may be very shallow, or with two maxima.

Interestingly, contrary to our expectations, there is almost no occupation in $\eigs 0$ at the steady state. This means, that almost all the contribution to $J_{max}$ comes from couplings between states with lower Bohr frequencies. We see that there are oscillating patterns in the occupations of the states. These are necessary for the production of frequencies other than the Bohr frequencies of the involved eigenstates. This requires that $\epsilon(t)$ will be active at the steady state. We see in the time-picture, that there are large oscillations of $\epsilon(t)$ at the steady state, of frequencies close to $\omega_0$.

We observe that at the steady state, $\epsilon(t)$ is out of phase with $\evmu$. This results in opposite signs of $\epsw$ and $\muw$, at the relevant frequencies. If we follow a semi-classical view, this means that the net work transferred to the system per oscillation cycle is $0$. This is a necessary condition for a steady state, by definition. The $\pi$ phase difference is characteristic to a forced, undamped oscillator, when the forcing frequency is higher than that of the characteristic frequency of the system.

We see that the constant equal occupation of $\eigs 0$ and $\eigs 2$ is not favoured in this case. This is explained by the fact that the coupling $d_{2,0}^\mu$ is rather small in magnitude, compared with $d_{2n}^\mu$ of larger $n$, as was discussed in the beginning of the section. Hence, it is preferable to utilize the Bohr frequencies of pairs of states with larger $n$, while altering the occupation. For comparison: At the end of our optimization process, we have: \text{$J_{max}=16.9$}. If we compute $J_{max}$ using:
\begin{equation}\label{eq:equaloc02}
	\ket{\psi(t)} = \frac{1}{\sqrt{2}}\exp\left(-i\operator{H}_0 t\right)(\eigs 0 + \eigs 2)
\end{equation}
we have: $J_{max}=1.92$.

It is possible to get more insight into the mechanism of the steady state. The large, negative, constant component of the forcing field, has a simple explanation; it has the meaning of an addition of a time-independent term: $k\mu(x)$, to $V(x)$, where $k$ is positive. $\mu(x)$ is not very far from being linear, unless there are large deviations from the bottom of the well. The deviations in this problem are not very large. The addition of a nearly linear term to $V(x)$ narrows the potential well, because of the asymmetry of $V(x)$ around \text{$x=0$}. The effect is of an increase in the separation between the energy levels.

\begin{figure}
	\centering \includegraphics[width=3in]{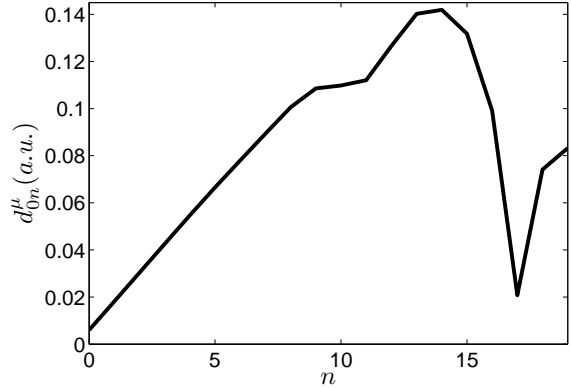}
	\caption{The $d_{0n}^\mu$ vs.~$n$ curve of the HCl dipole operator; for the lower levels ($n\leq 14$), there is a  gradual increase in the diagonal terms of $\operator{\mu}$. Only the lower levels are relevant to this problem --- see Fig.~\ref{fig:HClprojs}.}\label{fig:HCldiag}
\end{figure}

Another view of the same effect is to analyse the system in the terms of the original eigenstates, $\eigs n$. The addition of a constant term $k\operator{\mu}$ to $\operator{H}_0$ has a contribution to the diagonal of $\operator{H}_0$, \ie to eigenenergies $E_n$. According to the discussion at the beginning of the section, the values of the diagonal elements of $\operator{\mu}$ --- the $d_{0n}^{\mu}$, increase with $n$, at least for linear $\mu(x)$. Hence, this contribution to the diagonal increases the separation between the energy levels. This is verified in Fig.~\ref{fig:HCldiag} for $\mu(x)$ of our problem, by plotting the $d_{0n}^{\mu}$ vs.~$n$. We see that for the occupied levels in this problem (see Fig.~\ref{fig:HClprojs}) $d_{0n}^{\mu}$ is an increasing function of $n$. This means that the addition of $k\opmu{0}$ to $\operator{H}_0$ induces a greater separation of the energy levels.

The larger space between the new energy levels causes the new Bohr frequencies to be larger. This is an important part of the mechanism, since the original $\omega_{n+2,n}$ for $n>0$ are not included in $\tfmu$. This may be verified, by solving a similar problem, without the inclusion of the \text{$\omega=0$} region in $\tfeps$. Another option is to use \text{$\operator{\mu}-\opmu{0}$} instead of $\operator{\mu}$. These two options were implemented in the context of a similar problem, with the Toda potential. In both cases, the resulting $J_{max}$ is much smaller (although it still exceeds the $J_{max}$ computed with \eqref{eq:equaloc02}).

The field component of $\omega=\pi/T$ is explained by the fact, that during the process of achieving the steady state, eigenstates with increasing $n$ values are occupied. $\omega_{n+2,n}$ decreases with $n$; this requires more negative fields, to increase the separation between the levels.

As we saw, the components of $\epsw$ with $\omega$ values around $\omega_0$ also play a role in the steady state. We can offer an explanation, using a classical argument: the frequency of the largest component of $\epsw$ in this region is close to $\omega_{1,0}$; this frequency is higher than the other $\omega_{n+1,n}$. $\eigs 0$ is hardly occupied, so the characteristic frequencies of the oscillator are lower than that of the forcing field. The picture is of a forced oscillator, where the forcing field causes the system to oscillate with a slightly higher frequency. The overall system may be viewed as a new effective oscillator, with a higher fundamental frequency. The frequencies of the nonlinear effects of higher harmonics are also increased.

We did not succeed to see higher harmonics than the second harmonic in this system. The couplings of higher harmonics are much smaller. In addition, the possibility of using higher levels, with larger couplings, is restricted by the requirement of the prevention of dissociation. There are problems in the method, which make the production of very small effects a difficult task. These problems will be discussed in Sec.~\ref{sec:problems}.

In the next two examples, we show that it is possible to see higher harmonics, when the system possesses appropriate properties.

\subsection{HCl with fictitious $\mu(x)$}\label{ssec:HClfic}
In this problem, we show that it is possible to see higher harmonics, when using another, fictitious $\mu(x)$, for the HCl problem.

\begin{figure}
	\centering \includegraphics[width=3in]{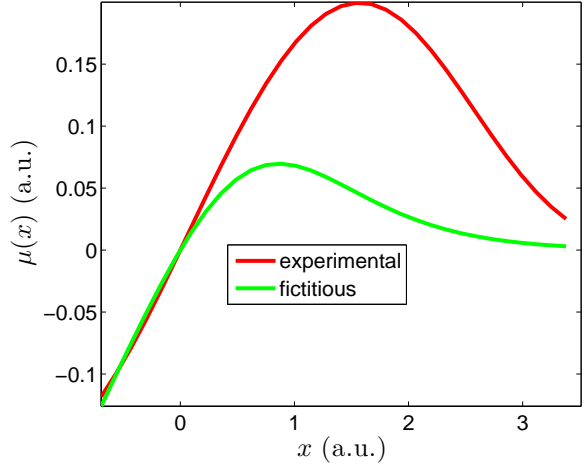}
	\caption{The experimental HCl dipole function, and the fictitious dipole function; the deviation of the fictitious function from linearity starts closer to \text{$x=0$}.}\label{fig:HCl3mu}
\end{figure}

The experimental $\mu(x)$ is not very far from being linear, unless there are large positive deviations from \text{$x=0$}. These deviations are hard to be achieved with the restriction on the allowed states, and only lower energy states are occupied. In order to increase the couplings of higher harmonics, we have to choose $\mu(x)$ that deviates from linearity closer to \text{$x=0$}. We have chosen a function of the form of Eq.~\eqref{eq:mutanh}:
\begin{equation}\label{eq:ficmu}
	\mu(x) = 0.5\left(\deriv{\mu}{x}\right)_{eq}x[1-\tanh(x-0.7)]
\end{equation}
It is designed to be similar to the experimental function at the neighbourhood of \text{$x=0$}, but to decay rapidly closer to \text{$x=0$}. The two functions are plotted in Fig.~\ref{fig:HCl3mu}.

The importance of this example is that molecules with rapidly decaying dipole functions do exist. If we get more successful results for this problem we can conclude that such molecules are preferable for harmonic generation.

\begin{figure}
	\centering \includegraphics[width=3in]{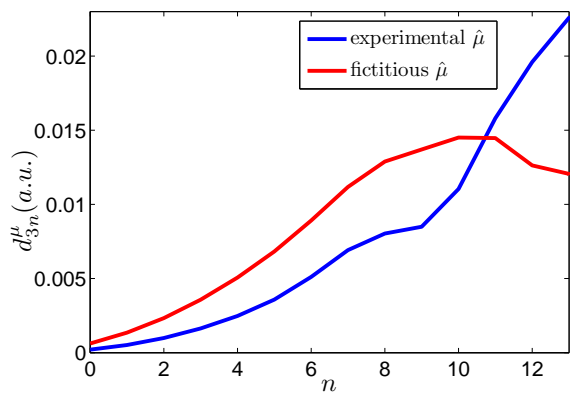}
	\caption{The $d_{3n}^\mu$ vs.~$n$ curves of the realistic HCl dipole operator and the fictitious one; at the lower states, the couplings are larger for the fictitious function.}\label{fig:HCl3d3mu}
\end{figure}

In this problem, we maximize the response in the region of the third harmonic. In Fig.~\ref{fig:HCl3d3mu}, the couplings $d_{3n}^\mu$ are plotted vs.~$n$, for the experimental and fictitious $\operator{\mu}$. We see that at the lower states, the couplings are larger for the fictitious function. As we have already said, we are restricted to the lower states, in order to prevent dissociation.

$\tfmu$ is again chosen in a way that only \text{$\omega_{3,0}=3.73\cdot 10^{-2}_{a.u.}$} is included in the maximized $\omega$ interval, while the other $\omega_{n+3,n}$ are of lower frequencies. $\tfeps$ is identical to that of the previous problem.


The details of the problem are summarised in Table~\ref{tab:HCl3f}.

\begin{table}
	\begin{equation*}
	\renewcommand{\arraystretch}{2}
	\begin{array}{|c||c|}
		\hline
		\operator{H}_0 & \frac{\operator{P}^2}{2\cdot 1785} + 0.171\left[\exp\left(-0.975\,\operator{X}\right)-\operator{I}\right]^2 \\ \hline
		\operator{\mu} &  0.0963\,\operator{X}\,\left[\operator{I}-\tanh\left(\operator{X} - 0.7\,\operator{I}\right)\right]\\ \hline
		\ket{\psi_0} & \ket{\varphi_0} \\ \hline
		T & 10^4 \\ \hline
		\tfeps & 2500\,u(0.015-\omega) \\ \hline
		\tfmu & 100\,u(\omega - 0.036)\,u(0.038-\omega) \\ \hline
		L & 19 \\ \hline
		\gamma_n & (n-19)^2 \\ \hline
		\kappa & 10 \\ \hline
		\bar\epsilon^{0}(\omega) & u(0.015-\omega) \\ \hline
		K_i & 1 \\ \hline
		x \text{ domain} & [-0.69407,\;3.51178) \\ \hline
		N_{grid} & 32 \\ \hline
		\text{tolerance} & 10^{-3} \\ \hline
	\end{array}
\end{equation*}
	\caption{The details of the HCl problem with a fictitious $\mu(x)$}\label{tab:HCl3f}
\end{table}

\begin{figure}
	\centering \includegraphics[width=3in]{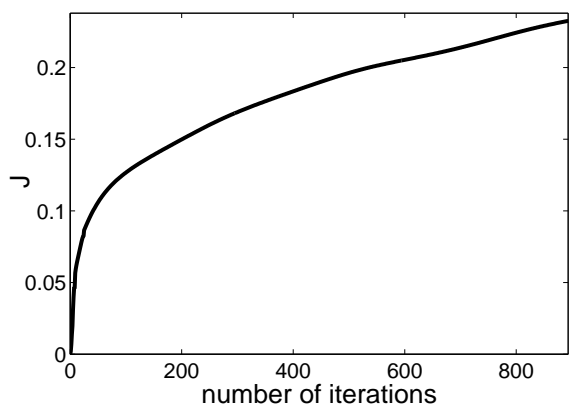}
	\caption{The convergence curve of the HCl problem with fictitious $\mu(x)$}\label{fig:HCl3conv}	
\end{figure}

The convergence curve is shown in Fig.~\ref{fig:HCl3conv}.

\begin{figure}
	\centering \includegraphics[width=3in]{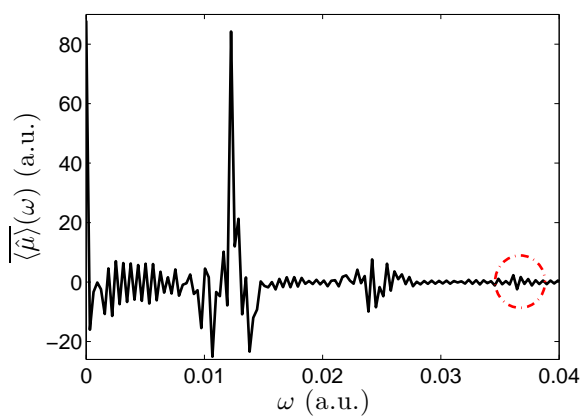}
	\caption{The $\muw$ curve of the HCl problem with fictitious $\mu(x)$; the response in the third harmonic region is marked by a red circle. There are also large components at $\omega=0$, and at the first and second harmonics.}\label{fig:HCl3muw}	
\end{figure}

The main features of this problem are similar to that of the previous one. The resulting $\muw$ curve is shown in Fig.~\ref{fig:HCl3muw}. We can see a small response in the region of the third harmonic. There are large component at the first and second harmonics.

\begin{figure}
	\centering \includegraphics[width=3in]{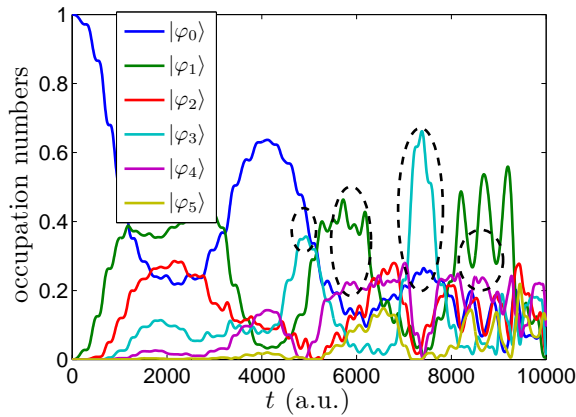}
	\caption{The occupation vs.~$t$ curve of the first 6 eigenstate, for the HCl problem with fictitious $\mu(x)$; the system has not achieved yet a real steady state. The state is far from being semi-classical. We can see regions with higher occupation of pairs of states, with $n=3$ difference. A few of these regions are marked by black circles.}\label{fig:HCl3oc}	
\end{figure}

The occupation vs.~time curves for the first eigenstates, are shown in Fig.~\ref{fig:HCl3oc}. We see that the system did not achieve a real steady state at $t=T$. However, regions of more ordered patterns exist. We can observe that in these regions there are pairs of states with $n=3$ difference, with larger occupation. A few of them are marked by black circles. This time, the occupation is not distributed around a single $n$ value, and the state is far from being semi-classical.

\subsection{The Toda anharmonic oscillator}\label{ssec:Toda}
In this problem we show that when we do not have to be concerned on the possibility of dissociation, it is possible to see higher harmonics.

We use the Toda potential from Eq.~\eqref{eq:Toda} (see Fig.~\ref{fig:Toda}). We still have to put a restriction on the allowed eigenstates, because the higher states are physically meaningless. Nevertheless, there is much more freedom for occupation of higher states in this problem, because the overall number of states is $128$, compared with $32$ in the HCl problems.

When we used a linear dipole function: \text{$\mu(x)=x$}, we succeeded to see the 4'th harmonic. The results will not be shown here.

We present the results for the Toda potential, with a non-linear dipole function:
\begin{equation}\label{eq:Todamu}
	\mu(x) = -4[\exp(-0.25x)-1]
\end{equation}
This function satisfies:
\[
	\left(\deriv{\mu}{x}\right)_{x=0}=1
\]
as for the problem of $\mu(x)=x$. The exponential term is chosen in a way that it will not compete with the exponential term of the potential itself. This makes this choice more realistic.

We maximize the response at the 6'th harmonic. $\tfmu$ is again chosen to include only: \text{$\omega_{6,0}=4.82_{a.u.}$}, and not the other $\omega_{n+6,n}$.

The details of the problem are summarised in Table~\ref{tab:Toda}.

\begin{table}
	\begin{equation*}
	\renewcommand{\arraystretch}{1.7}
	\begin{array}{|c||c|}
		\hline
		\operator{H}_0 & \frac{\operator{P}^2}{2} + \exp\left(-\operator{X}\right) + \operator{X} - \operator{I} \\ \hline
		\operator{\mu} &  -4\,\left[\exp\left(-0.25\,\operator{X}\right)-\operator{I}\right]\\ \hline
		\ket{\psi_0} & \ket{\varphi_0} \\ \hline
		T & 100 \\ \hline
		\tfeps & 100\,u(1.3-\omega) \\ \hline
		\tfmu & 100\,u(\omega - 4.7)\,u(4.9-\omega) \\ \hline
		L & 87 \\ \hline
		\gamma_n & n-87 \\ \hline
		\kappa & 1 \\ \hline
		\bar\epsilon^{0}(\omega) & 5\,u(1.3-\omega) \\ \hline
		K_i & 0.01 \\ \hline
		x \text{ domain} & [-3.8045,\;41.0989) \\ \hline
		N_{grid} & 128 \\ \hline
		\text{iterations} & 1224 \\ \hline
	\end{array}
\end{equation*}
	\caption{The details of the Toda potential problem}\label{tab:Toda}
\end{table}

\begin{figure}
	\centering \includegraphics[width=3in]{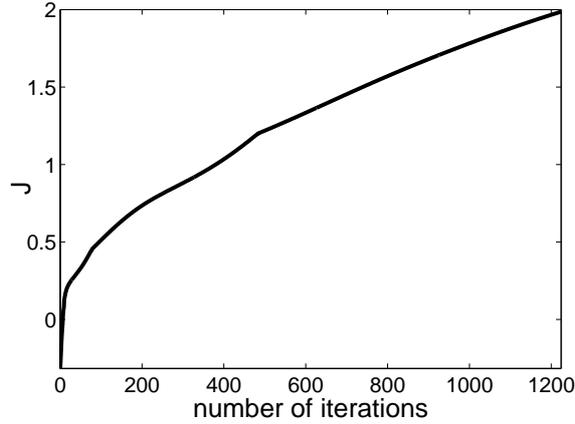}
	\caption{The convergence curve of the Toda potential problem}\label{fig:toda6conv}	
\end{figure}

The convergence was found to be very slow. We observed, that this is typical for large fields, when using the relaxation method. We stopped the optimization process before the picture of the system was stabilized. The convergence curve is shown in Fig.~\ref{fig:toda6conv}.

\begin{figure}
	\centering \includegraphics[width=3in]{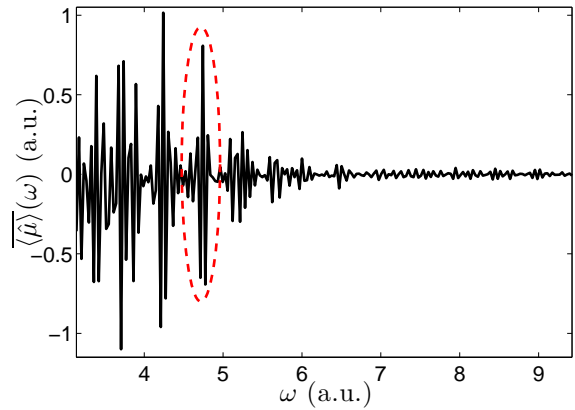}
	\caption{The relevant part of the $\muw$ curve of the Toda potential problem; the response in the maximized region is marked by a red circle.}\label{fig:toda6muw}	
\end{figure}

The relevant part of the $\muw$ curve is shown in Fig.~\ref{fig:toda6muw}. We can see also higher harmonics.

\begin{figure}
	\centering \includegraphics[width=3in]{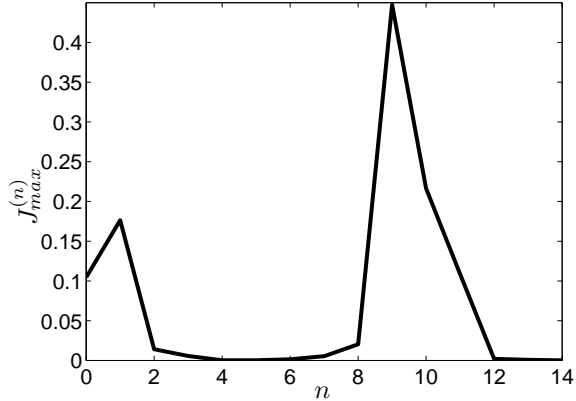}
	\caption{$J_{max}^{(n)}$ vs.~$n$, for the Toda potential problem; the main contribution to $J_{max}$ comes from: $n=9$ (the maximum) and $n=10$.}\label{fig:toda6Jalln}	
\end{figure}

When we examine the harmonic generation process more carefully we find that the results are rather different from our expectations; the origin of the response at the desired frequency region is not from the 6'th harmonic, but from higher harmonics. In Fig.~\ref{fig:toda6Jalln} $J_{max}^{(n)}$ is plotted vs.~$n$. We see that the main contribution comes from \text{$n=9$}. The second important contribution is from \text{$n=10$}.

\begin{figure}
	\centering \includegraphics[width=3in]{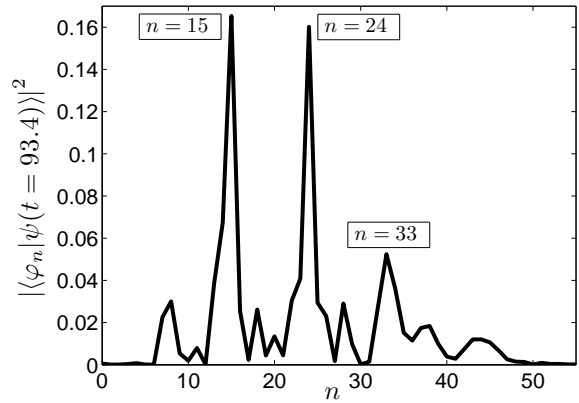}
	\caption{The occupation at $t=93.4$, vs.~$n$, for the Toda potential problem; the $n$'s of the 3 largest components are written beside the maxima. The $n$ difference between these components is $9$. There are other series of smaller components, with $n$ differences of $9$ or $10$.}\label{fig:toda6oc}	
\end{figure}

\begin{figure}
	\centering \includegraphics[width=3in]{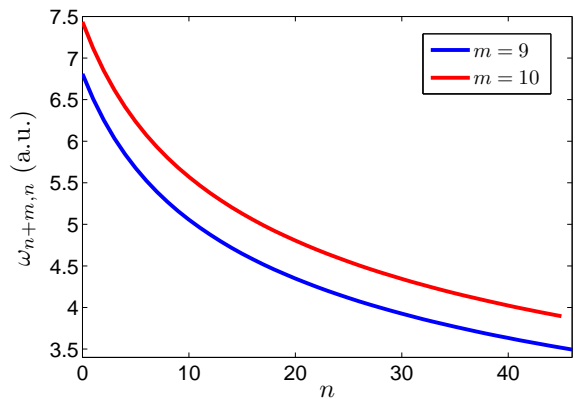}
	\caption{$\omega_{n+9,n}$ and $\omega_{n+10,n}$ vs.~$n$; in the region of the occupied states, these Bohr frequencies are close to the maximized region in the spectrum.}\label{fig:toda6Bohr}	
\end{figure}

The explanation for these observations becomes clear when examining the occupation picture. The occupation vs.~time picture is extremely complicated in this case, and is not shown here; instead, we show the occupation picture at a single time point, close to the end of the propagation (at \text{$T=100_{a.u.}$}, we observe undesirable boundary effects). In Fig.~\ref{fig:toda6oc}, the occupation is plotted vs.~$n$. The occupied states are of large $n$, where the separation between the energy levels is considerably smaller. In this $n$ region, the Bohr frequencies $\omega_{n+9,n}$ and $\omega_{n+10,n}$ are the closest to the maximized region in the spectrum. This may be seen in Fig.~\ref{fig:toda6Bohr}.

It is possible to recognize in Fig.~\ref{fig:toda6oc} several series of states with relatively large occupation, with $n$ differences of $9$ or $10$ between the states. This shows that the main contribution to $J_{max}$ indeed comes from the 9'th and 10'th harmonics, as we concluded from the $J_{max}^{(n)}$ curve.

We can conclude from the occupation picture that the state is very far from being semi-classical. If we follow $|\psi(x, t)|^2$, we can see that the state is split into many separated entities. Indeed, the semi-classical description is definitely inappropriate in this case.

\section{Problems in the new method}\label{sec:problems}

\subsection{The lack of treatment of the emitted field}
One obvious deficiency of our model is the lack of treatment of the emitted field. We assume that when $\evmu$ oscillates, the field with the corresponding spectrum is emitted, but we ignore the effects of this emission.

We can point at least two problems with this approach:
\begin{enumerate}
	\item The emission is a \emph{dissipative process}. The loss of energy affects the system, and has to be taken into account. The lack of treatment of dissipative processes led us to non-physical steady-states, with constant $0$ field, in Sec.~\ref{sec:MLS}. In the realistic steady state of these problems, there must be a permanent, continuous transfer of energy from the forcing field into the system, to keep the optimal occupational state unchanged.
	\item The emitted radiation may interact with the system.
\end{enumerate}

\subsection{Boundary effects}
As we have already mentioned (Subsection~\ref{ssec:modif}), the approximation of the cosine transform by the DCT causes boundary problems. We encounter boundary problems with both $\epsw$ and $\muw$.

We start from $\epsw$. The DCT consists of cosine terms that complete an integer number of half-cycles in $T$. The definition of $\epsw$ using a DCT forces to end the process with:
\[
	\deriv{\epsilon(T)}{t}=0
\]
This problem has a similar origin to that mentioned in Subsection~\ref{ssec:modif}: A non-zero final time derivative  results in a discontinuity in the first derivative of the extended periodic function. This is impossible if we are restricted to low frequency components.

We often observe that this condition causes undesirable effects at the end of the propagation.

This problem is not very disturbing because the very end of the resulting process may be ignored. The effect of this problem on the resulting spectra, $\epsw$ and $\muw$, is rather small.

The problem with $\muw$ is that mentioned in Subsection~\ref{ssec:modif}. The solution that we proposed there to the problem is not ideal. We observe that the field is not likely to just be shifted by an appropriate phase. Instead, undesirable effects appear at the end of the propagation, in order to decrease $d\exval{\operator{\mu}}\!(T)/dt$. This problem is not very disturbing, as the problem mentioned for $\epsw$.

A more serious problem is that the insertion of the boundary term increases the difficulty in the optimization process. This problem is more disturbing when dealing with small harmonic generation effects. This will be discussed in Subsection~\ref{ssec:smallef}.

\subsection{Difficulties when the effect is small}\label{ssec:smallef}
When the harmonic generation effect is small, the problem becomes a difficult task. When the first guess gives a negative $J$, and the maximal possible effect is very small, the optimization process often converges very rapidly to the solution: $\epsilon(t)\equiv 0$, with: $J=0$.

In order to solve the problem, we have to try to make $J$ positive for the first guess. The term $J_{max}$ is always positive. The penalty terms: $J_{penal}$, $J_{forb}$ and $J_{bound}$ are always negative. We can increase $J$ in two ways:
\begin{enumerate}
	\item By increasing $J_{max}$;
	\item By decreasing the magnitude of the penalty terms.
\end{enumerate}

We may suggest to play with the constant coefficients of the penalty terms (altering also the $\lambda$ coefficient of $J_{max}$ is not helpful for this purpose, because we are interested in the \emph{relative} weight of the terms). Here, we have to distinguish between  $J_{penal}$, and $J_{forb}$ or $J_{bound}$. The coefficient $\tilde{\alpha}$ of $J_{penal}$ may be decreased, if necessary. It is impossible to decrease $\gamma_n$ or $\kappa$ (the coefficients of $J_{forb}$ and $J_{bound}$), because it will make these penalty terms ineffective. $J_{bound}$ is always necessary when the effect is small, as was discussed in Subsection~\ref{ssec:modif}. $J_{forb}$ is also frequently necessary for small effect problems; the reason is, that in order to get a significant value of $J_{max}$, we often need large fields. This may lead to the occupation of the forbidden states.

%

In all the examples mentioned in this thesis, we used a rather arbitrary guess. The difficulty may be solved by using a more clever guess. It may be necessary to solve a former control problem, to produce a guess with the desired properties.

We propose the following ideas for a former control problem, that may produce a larger $J_{max}$ value:
\begin{enumerate}
	\item It is possible to solve the problem of achieving an equal occupation of $0.5$ of two states $\eigs m$, $\eigs n$, that their Bohr frequency $\omega_{mn}$ is in the maximized region. We want to achieve this occupational state as fast as possible. We use the regular formulation for time-dependent control problems (Sec.~\ref{sec:OCTt}), with:
	\begin{align}
		& \operator{O}(t) = \operator P_{\phi(t)} = \ket{\phi(t)}\bra{\phi(t)} \nonumber \\
		& \ket{\phi(t)} = \frac{1}{\sqrt{2}}\exp\left(-i\operator{H}_0 t\right)(\eigs m + \eigs n)
	\end{align}
	Of course, $J_{penal}$ is of the form of Eq.~\ref{eq:Jpenalw2}.
	
	We implemented this idea; it is effective only when the resulting effect is significant. The main problem with this approach is that when the new method starts with this guess it tends to get stuck in this mechanism (although the results continue to improve). As we saw, this mechanism is not necessarily the ideal one.
	\item For the anharmonic oscillator problems, we can suggest using a guess that shifts the occupation to another region of $n$ values. When higher levels are occupied there are larger anharmonic effects. This gives a better starting point for the problem. We use the regular formulation of time-dependent problems, using a projection operator into the subspace of $M_1\leq n\leq M_2$:
	\begin{equation}
		\operator{O}(t)=\operator{P}_{subspace} = \sum_{n=M_1}^{M_2}\eigs{n}\bra{\varphi_n}
	\end{equation}
	$J_{penal}$ is of the form of Eq.~\ref{eq:Jpenalw2}.
	
	This method was tried successfully for a problem with the Toda potential.
	\item Sometimes, in the solution of a problem of a low harmonic, we see also higher harmonics (see Fig.~\ref{fig:toda6muw}). Hence, we can use a problem of a lower harmonic as a former problem for higher harmonics. We did not try to implement this suggestion.
\end{enumerate}

The magnitudes of $J_{forb}$ and $J_{bound}$ may be decreased by solving a former simpler control problem, with the insertion of these penalty terms. If we need only the $J_{bound}$ term, its magnitude may be decreased easily by shifting the phase of the field forward. We did not try to implement these suggestions.

%
	\chapter{Conclusion}\label{ch:conc}
%
%
%
In the present work, a new theoretical method of calculation for controlling harmonic generation, was developed in the framework of QOCT\@.

The development of the method involved coping with the more general problem of control requirements that are formulated in the frequency domain. It has been shown that it is possible and, apparently preferable, to formulate these requirements in their natural domain --- the frequency domain. The method that was developed for the harmonic generation problem, was generalized to other control problems with frequency requirements.

The new formulation required the use of the relaxation optimization method, which has not been used yet for QOCT problems. The relaxation method was found to be successful for the new formulation.

The new method was applied to harmonic generation problems in simple systems. It was shown that the method succeeds to deal with these relatively simple problems. However, difficulties were encountered when the harmonic generation effect is very small. These difficulties originate from noise effects. Several ways to deal with the problem were suggested. Nevertheless, it seems that a satisfactory solution for the problem is still missing.

The analysis of the results led to general conclusions on possible mechanisms of harmonic generation. Typical mechanisms in anharmonic oscillator systems were also discussed.

The limits of the ability of the new method are still unknown. The question on the ability of the method may be divided into 3 parts:
\begin{enumerate}
	\item \emph{The complexity of the problem}: The method has not been tested yet for a ``real'' problem, with more than one degree of freedom, and realistic complexities. It is unknown if the method is effective for more complex cases than the simple problems mentioned in this thesis.
	\item \emph{Small harmonic generation effects}: It is unknown how small is the effect that the method will be able to deal with. This depends on the existence and efficiency of a possible solution to the noise problem.
	\item \emph{Production of high harmonics}: The maximal harmonic that was attempted to be achieved using the new method is the 10'th harmonic. It is not certain that the method will be able to produce much higher harmonics, like in the high-harmonic generation mechanism.
\end{enumerate}
  
There are many possible directions for taking the research further. Here are a few:
\begin{itemize}
	\item The method should be tested for real problems.
	\item A satisfactory solution for the noise problem has to be found.
	\item The method should be tested for the high-harmonic generation mechanism.
	\item The effects of emission, missing in the present formulation, have to be taken into account.
	\item The possibilities of improving the ability of the method, using clever guesses, should be investigated.
\end{itemize}
\subsubsection*{Acknowledgements}
I am deeply indebted to my supervisor, Prof.~Ronnie Kosloff, for his invaluable advise, kind support and patient guidance. I would also like to thank my father, Dr.~Paul Schaefer, for his constant assistance and help. I am grateful to Prof.~Hillel Tal-Ezer, for his direction and collaboration in the implementation of the new propagator, and to Dr.~Ilan Degani and Reuven Eitan, for their helpful discussions.
	\appendix
	\chapter{The full derivation of the Euler-Lagrange equations for the new formulation}\label{ap:derivation}
%
%
%
%
%
%
The derivation will be performed for the most general case, with the exception of the case that \text{$\commut{\operator{\mu}}{\operator{O}} \neq \operator{0}$} and \text{$\kappa>0$}. The maximized functional is defined by:
\begin{align}
	& J \equiv J_{max} + J_{bound} + J_{forb} + J_{penal} + J_{con} \label{eq:Jwd}\\ 
	& J_{max} \equiv \frac{1}{2}\int_0^\Omega \tfO\overline{\exval{\operator{O}_a}}^2(\omega)\,d\omega  & \tfO\geq 0 \label{eq:Jmaxd} \\
	& \operator{O}_a \equiv \operator{P}_{a}\operator O\operator{P}_{a} \label{eq:Oad} \\
	& \operator{P}_{a} \equiv \sum_{n=0}^L\ket{\varphi_n}\bra{\varphi_n} \label{eq:Pad} \\
	& \Oaw \equiv \sqrt{\frac{2}{\pi}}\int_0^T \exval{\operator{O}_a}(t)\cos(\omega t)\,dt \label{eq:Oawd} \\
	& J_{bound} \equiv -\frac{1}{2}\kappa\left[\deriv{\exval{\operator{O}}(T)}{t}\right]^2 & \kappa \geq 0 \label{eq:Jboundd}\\
	& J_{forb} \equiv -\int_0^T \bracketsO{\psi(t)}{\operator{P}_f^\gamma}{\psi(t)}\,dt \label{eq:Jforbd}\\
	& \operator{P}_f^\gamma \equiv \sum_{n=L+1}^{N-1}\gamma_n\ket{\varphi_n}\bra{\varphi_n} & \gamma_n>0 \label{eq:Pfd} 
\end{align}
\begin{align}
	& J_{penal} \equiv -\int_0^\Omega\frac{1}{\tfeps}\bar{\epsilon}^2(\omega)\,d\omega & \tfeps>0 \label{eq:Jpenald}\\
	& \epsw \equiv \sqrt{\frac{2}{\pi}}\int_0^T \epsilon(t)\cos(\omega t)\,dt \label{eq:epswd} \\
	& J_{con} \equiv -2\Real{\int_0^T\bracketsO{\chi(t)}{\pderiv{}{t}+i\operator H(t)}{\psi(t)}\,dt} \label{eq:Jcond}\\
	& \operator{H}(t) = \operator{H}_0 - \operator{\mu}\epsilon(t) = \operator{H}_0 - \operator{\mu}\left(\sqrt{\frac{2}{\pi}}\int_0^\Omega \bar{\epsilon}(\omega)\cos(\omega t)\,d\omega\right) \label{eq:Hd}
\end{align}

The constraint equations are:
\begin{align}
	\pderiv{\ket{\psi(t)}}{t} &= -i\operator{H}(t)\ket{\psi(t)} & & \ket{\psi(0)} = \ket{\psi_0}
	\label{eq:Schrd}\\
	\pderiv{\bra{\psi(t)}}{t} &= i\bra{\psi(t)}\operator{H}(t) & & \bra{\psi(0)} = \bra{\psi_0} \label{eq:conjSchrd}
\end{align}
\eqref{eq:conjSchrd} ensures that:
\[
	\bra{\psi(t)} = \left[\ket{\psi(t)} \right]^+
\]
Assuming this, all the computations can be performed using \eqref{eq:Schrd} only.

The extremum conditions are:
\begin{align}
	&\fnlderiv{J}{\bar{\epsilon}(\omega)} = 0 \label{eq:dJdepswd}\\
	&\fnlderiv{J}{\ket{\psi(t)}} = 0 \label{eq:dJdpsitkd} \\ 
	&\fnlderiv{J}{\bra{\psi(t)}} = 0 \label{eq:dJdpsitbd} \\	
	&\fnlderiv{J}{\ket{\psi(T)}} = 0 \label{eq:dJdpsiTkd} \\ 
	&\fnlderiv{J}{\bra{\psi(T)}} = 0 \label{eq:dJdpsiTbd}
\end{align}

$J_{con}$ is more easily handled after integrating by parts the expression:
\[
	\int_0^T\bracketsbiggm{\chi(t)}{\pderiv{\psi(t)}{t}}\,dt
\]
We obtain:
\begin{align}
	J_{con} = -2\Real&\left[\bracketsbiggm{\chi(T)}{\psi(T)}-\bracketsbiggm{\chi(0)}{\psi(0)} - \int_0^T\bracketsbiggm{\left(\pderiv{}{t}+i\operator H(t)\right)\chi(t)}{\psi(t)}\,dt\right]
	  \label{eq:Jconp}
\end{align}

The expression for the LHS of \eqref{eq:dJdepswd}, is obtained using \eqref{eq:Jpenald}, \eqref{eq:Jconp}, \eqref{eq:Hd}:
\begin{align}
	\fnlderiv{J}{\epsw} =& \fnlderiv{J_{penal}}{\epsw} + \fnlderiv{J_{con}}{\epsw} \label{eq:dJdepscomp}\\
	\fnlderiv{J_{penal}}{\epsw} =& -\frac{2}{\tfeps}\epsw \label{eq:dJpenaldeps}\\
	\fnlderiv{J_{con}}{\epsw} =& 2\,\Real\left[-i\int_0^T\bracketsO{\chi(t)}{\fnlderiv{\operator{H}(t)}{\epsw}}{\psi(t)}\,dt\right] \nonumber \\
	=&-2\,\Imag\left[\sqrt{\frac{2}{\pi}}\int_0^T\bracketsO{\chi(t)}{\operator{\mu}}{\psi(t)}\cos(\omega t)\,dt\right] \nonumber \\
	=&-2\,\Imag\left\lbrace \mathcal{C}\left[\bracketsO{\chi(t)}{\operator{\mu}}{\psi(t)}\right]\right\rbrace \label{eq:dJcondeps}
\end{align}
From \eqref{eq:dJdepswd}, \eqref{eq:dJdepscomp}, \eqref{eq:dJpenaldeps}, \eqref{eq:dJcondeps}, we get the following expression for $\epsw$:
\begin{equation}\label{eq:epswresult}
	\epsw = \tfeps\mathcal{C}\left[-\Imag{\bracketsO{\chi(t)}{\operator{\mu}}{\psi(t)}}\right] 
\end{equation}

In order to derive the LHS of \eqref{eq:dJdpsitkd}, we first write the explicit expression of $J_{max}$ as a functional of $\ket{\psi(t)}$:
\begin{align}
	&J_{max} = \frac{1}{\pi}\!\int_0^\Omega\!\int_0^T\!\int_0^T \tfO \bracketsO{\psi(t)}{\operator{O}_a}{\psi(t)} \bracketsO{\psi(t')}{\operator{O}_a}{\psi(t')}\cos(\omega t)\cos(\omega t')\,dt\,dt'\,d\omega 
	\label{eq:Jmaxex}
\end{align}
The expression for the LHS of \eqref{eq:dJdpsitkd}, is obtained using \eqref{eq:Jmaxex}, \eqref{eq:Jforbd}, \eqref{eq:Jconp}:
\begin{align}
	\fnlderiv{J}{\ket{\psi(t)}} =& \fnlderiv{J_{max}}{\ket{\psi(t)}} + \fnlderiv{J_{forb}}{\ket{\psi(t)}} + \fnlderiv{J_{con}}{\ket{\psi(t)}} \label{eq:dJdpsicomp}\\
	\fnlderiv{J_{max}}{\ket{\psi(t)}} =& \frac{2}{\pi}\int_0^\Omega\!\int_0^T \tfO \bra{\psi(t)}\operator{O}_a \bracketsO{\psi(t')}{\operator{O}_a}{\psi(t')}\cos(\omega t)\cos(\omega t')\,dt'\,d\omega\nonumber \\
	= &\sqrt{\frac{2}{\pi}}\int_0^\Omega \tfO\Oaw\cos(\omega t)\,d\omega\bra{\psi(t)}\operator{O}_a \nonumber \\
	= &\mathcal{C}^{-1}\left[\tfO\Oaw\right]\bra{\psi(t)}\operator{O}_a \label{eq:dJmaxdpsi}\\
	\fnlderiv{J_{forb}}{\ket{\psi(t)}} = & -\bra{\psi(t)}\operator{P}_f^\gamma \label{eq:dJforbdpsi}\\
	\fnlderiv{J_{con}}{\ket{\psi(t)}} = & \pderiv{\bra{\chi(t)}}{t} + \bra{i\operator{H}(t)\chi(t)} \label{eq:dJcondpsi}
\end{align}
Using \eqref{eq:dJdpsitkd}, \eqref{eq:dJdpsicomp}, \eqref{eq:dJmaxdpsi}, \eqref{eq:dJforbdpsi}, \eqref{eq:dJcondpsi}, we obtain:
\begin{equation}\label{eq:ihSchrbra}
	\pderiv{\bra{\chi(t)}}{t} =  -\bra{i\operator{H}(t)\chi(t)} - \bra{\psi(t)} \left\lbrace\mathcal{C}^{-1}\left[\tfO\Oaw\right]\operator{O}_a - \operator{P}_f^\gamma\right\rbrace
\end{equation}
Eq.~\eqref{eq:dJdpsitbd} gives the adjoint of \eqref{eq:ihSchrbra}:
\begin{equation}\label{eq:ihSchrket}
	\pderiv{\ket{\chi(t)}}{t} = -i\operator{H}(t)\ket{\chi(t)} - \left\lbrace\mathcal{C}^{-1}\left[\tfO\overline{\exval{\operator{O}_a}}(\omega)\right]\operator{O}_a - \operator{P}_f^\gamma\right\rbrace\ket{\psi(t)}
\end{equation}

In order to derive the expression of the LHS of \eqref{eq:dJdpsiTkd}, we write \eqref{eq:Jboundd} in a more useful form. Taking the expectation value of both sides of the Heisenberg equation, we have:
\begin{equation}\label{eq:Heisd}
	\deriv{\exval{\operator{O}}(T)}{t} = i\exval{\left[\operator{H}(T), \operator{O}\right]}(T)
\end{equation}
In the special case that \text{$\commut{\operator{\mu}}{\operator{O}}=\operator{0}$}, we have:
\begin{equation}\label{eq:Heis0d}
	\deriv{\exval{\operator{O}}(T)}{t} = i\exval{\left[\operator{H}_0, \operator{O}\right]}(T)
\end{equation}
In this case, $J_{bound}$ becomes:
\begin{equation}
	J_{bound} = \frac{\kappa}{2}\bracketsO{\psi(T)}{\left[\operator{H}_0, \operator{O}\right]}{\psi(T)}^2
\end{equation}
The LHS of \eqref{eq:dJdpsiTkd} is:
\begin{align}
	&\fnlderiv{J}{\ket{\psi(T)}}=\fnlderiv{J_{bound}}{\ket{\psi(T)}} + \fnlderiv{J_{con}}{\ket{\psi(T)}} \label{eq:dJdpsiTcomp} \\
	&\fnlderiv{J_{bound}}{\ket{\psi(T)}} = \kappa\bracketsO{\psi(T)}{\left[\operator{H}_0, \operator{O}\right]}{\psi(T)}\bra{\psi(T)}\left[\operator{H}_0, \operator{O}\right] \label{eq:dJbounddpsiT} \\
	&\fnlderiv{J_{con}}{\ket{\psi(T)}} = -\bra{\chi(T)} \label{eq:dJcondpsiT}
\end{align}
Using \eqref{eq:dJdpsiTkd}, \eqref{eq:dJdpsiTcomp}, \eqref{eq:dJbounddpsiT}, \eqref{eq:dJcondpsiT}, we obtain:
\begin{align}
	\bra{\chi(T)} =& \kappa\bracketsO{\psi(T)}{\left[\operator{H}_0, \operator{O}\right]}{\psi(T)}\bra{\psi(T)}\left[\operator{H}_0, \operator{O}\right] \nonumber \\
	=& \kappa\exval{\left[\operator{H}_0, \operator{O}\right]}(T)\bra{\psi(T)}\left[\operator{H}_0, \operator{O}\right] \label{eq:chiTbra}
\end{align}
Eq.~\eqref{eq:dJdpsiTbd} gives the adjoint of \eqref{eq:chiTbra}:
\begin{equation} \label{eq:chiTket}
	\ket{\chi(T)} = \kappa\exval{\left[\operator{H}_0, \operator{O}\right]}(T) \left[\operator{H}_0, \operator{O}\right]\ket{\psi(T)}
\end{equation}
Eqs.~\eqref{eq:ihSchrbra}, \eqref{eq:ihSchrket}, \eqref{eq:chiTbra}, \eqref{eq:chiTket}, ensure that:
\[
	\bra{\chi(t)} = \left[\ket{\chi(t)} \right]^+
\]
Assuming this, all the computations can be performed using \eqref{eq:ihSchrket}, \eqref{eq:chiTket} only.

We collect the resulting equations, \eqref{eq:epswresult}, \eqref{eq:ihSchrket}, \eqref{eq:chiTket}, together with the constraint \eqref{eq:Schrd}:
\begin{align}
	&\pderiv{\ket{\psi(t)}}{t} = -i\operator{H}(t)\ket{\psi(t)}, \nonumber\\
	& \hspace{2cm}\ket{\psi(0)} = \ket{\psi_0} \\ 
	&\pderiv{\ket{\chi(t)}}{t} = -i\operator{H}(t)\ket{\chi(t)} - \left\lbrace\mathcal{C}^{-1}\left[\tfO\overline{\exval{\operator{O}_a}}(\omega)\right]\operator{O}_a - \operator{P}_f^\gamma\right\rbrace\ket{\psi(t)}, \nonumber\\
	&\hspace{2cm} \ket{\chi(T)} = \kappa\exval{\left[\operator{H}_0, \operator{O}\right]}(T) \left[\operator{H}_0, \operator{O}\right]\ket{\psi(T)} \\
	& \operator{H}(t) = \operator{H}_0 - \operator{\mu}\mathcal{C}^{-1}[\epsw] \nonumber \\
	& \epsw = \tilde{f}_{\epsilon}(\omega)\mathcal{C}\left[-\Imag{\bracketsO{\chi(t)}{\operator{\mu}}{\psi(t)}}\right]
\end{align}
These are the Euler-Lagrange equations of the problem.


	\chapter{Numerical details}\label{ap:num}
\section{The propagator for the Schr\"odinger equation}\label{ap:Hillel}
The propagator for the Schr\"odinger equation is based on a new, efficient and highly accurate algorithm, by Hillel Tal-Ezer (see~\cite{Hillel}).

\section{The performance of the Hamiltonian operations}
The Hamiltonian operations in the problems that depend on a spatial variable $x$, were performed using the Fourier grid method (see~\cite{TDmethods}). The operation of the $x$ dependent terms in the Hamiltonian is performed in the $x$ domain, and the operation of the $p$ dependent term, \ie the kinetic energy, is performed in the $p$ domain.

\section{The choice of the $x$ grid}
The $x$ grids of the various problems are equidistant grids. The distance between neighbouring points in the $x$ domain \text{$[x_{min}, x_{max})$} is:
\[
	\Delta x = \frac{x_{max}-x_{min}}{N_{grid}}
\]
where $N_{grid}$ is the number of grid points.

The $x$ domain is chosen to satisfy: 
\begin{align}
	& V(x_{min}) = V(x_{max}) \label{eq:Vmm}\\
	& V_{max} = \frac{p_{max}^2}{2m} = \frac{\pi^2}{2m\,\Delta x^2} \label{eq:VmTm}
\end{align}
where $V_{max}$ is the maximal $V(x)$, $p_{max}$ is the maximal momentum, and $m$ is the mass, or reduced mass.  \eqref{eq:VmTm} makes the maximal $V(x)$ possible, equal to the maximal kinetic energy possible.

\section{The approximation of the cosine transform}
The cosine transform was approximated using a discrete-cosine-transform that include the boundaries of the domain. It is sometimes referred as the DCT of the first kind (DCT-I, see~\cite{DCT}). The DCT-I of $N_t$ equidistant time points:
\[
	t_i, \qquad i=0,1,\ldots,N_t-1
\]
is defined as:
\begin{align}
	& \bar{g}(\omega_j) = \sqrt{\frac{2}{N_t-1}}\sum_{i=0}^{N_t-1}\frac{1}{h_i} g(t_i)\cos\left(\frac{ij\pi}{N_t-1}\right), & &j=0,1,\ldots,N_t-1 \label{eq:DCT}\\
	& h_i =
	\begin{cases}
		2 & \qquad i=0 \text{ or } i=N_t-1 \\
		1 & \qquad 1\leq i\leq N_t-2
	\end{cases} \label{eq:hdef}
\end{align}

Using this convention, DCT-I is its own inverse. In order to be consistent with the continuous formulation, the direct transform is multiplied by the factor: $T/\sqrt{(N_t-1)\pi}$. The inverse transform is divided by this factor. This is also necessary for another reason: when using Eq.~\eqref{eq:DCT} as is, the definition of the spectral function $\bar{g}(\omega)$, represented by the transform, varies with the sampling frequency. This is corrected by the insertion of the factor. 

The computation of the DCT-I is performed by the FFT of the ``folded'' function (see~\cite{DCT}).

\section{The computation of $J$}
The gradient and relaxation methods involve the computation of the functional $J$, during the search for an appropriate parameter $K$. In order that these methods will be successful, it is important to compute the integrals in $J$ as accurately as possible. The accuracy of the numerical integration should be the same as that of the propagator.

The integration of the $\omega$ dependent functions is performed like in the DCT-I. This is consistent with the accuracy of the $\omega$ grid.

The integration of the $t$ dependent functions is performed by utilizing the internal Chebyshev time points of every time step. We describe here the integration method.

Consider a time step with the boundaries: \text{$[t_n, t_{n+1}]$}. Within the time interval, there are $N_c$ internal points --- the boundary including Chebyshev points:
\begin{align}
	& t_k = t_n + \frac{\Delta t}{2}(1-y_k) &  \label{eq:Chebpt}\\
	& y_k = \cos\left(\frac{k\pi}{N_c - 1}\right) & \qquad k=0,1,\ldots ,N_c - 1 \label{eq:Chebpy}
\end{align}
where $\Delta t = t_{n+1}-t_n$.

An arbitrary time-dependent function $g(t)$ defined in the interval may be approximated by a truncated Chebyshev series:
\begin{equation}\label{eq:chebser}
	g(t) \approx \sum_{m=0}^{N_c-1}a_m T_m(y) \qquad\qquad y=\frac{-2t+t_n+t_{n+1}}{\Delta t}
\end{equation}
where the $T_m(y)$ are the Chebyshev polynomials. The $a_m$ are given by:
\begin{equation}\label{eq:chebcoef}
	a_m = -\frac{2}{(N_c-1)h_m}\sum_{k=0}^{N_c-1}\frac{1}{h_k}g(t_k)\cos\left(\frac{km\pi}{N-1}\right)
\end{equation}
where $h_i$ is defined by Eq.~\eqref{eq:hdef}.

The integration of $g(t)$ over the interval may be approximated using \eqref{eq:chebser}:
\begin{align}
	& \int_{t_n}^{t_{n+1}}g(t)\,dt\approx-\frac{\Delta t}{2}\sum_{m=0}^{N_c-1}a_m\int_{-1}^{1}T_m(y)\,dy=-\Delta t \sum_{m=0}^{N_c-1}a_m c_m \label{eq:sumac}\\
	& c_m = 
	\begin{cases}
		-\frac{1}{m^2-1} & \qquad m \text{ even} \\
		0 & \qquad m \text{ odd}
	\end{cases}
\end{align}
The integration of the $T_m(y)$ is performed analytically.

Inserting \eqref{eq:chebcoef} into \eqref{eq:sumac}, we obtain:
\begin{align}
	& \int_{t_n}^{t_{n+1}}g(t)\,dt\approx \Delta t \sum_{k=0}^{N_c-1}w_k g(t_k) \label{eq:chebint}\\
	& w_k \equiv \frac{2}{(N_c-1)h_k}\sum_{m=0}^{N_c-1}\frac{1}{h_m}c_m\cos\left(\frac{km\pi}{N-1}\right) \label{eq:wdef}
\end{align}
The $w_k$ are the integration weights of the $t_k$, for $N_c$ Chebyshev points. They are independent of $\Delta t$. The expression for $w_k$ is just a DCT-I of $c_m$, within a factor, and an additional small difference (compare Eq.~\eqref{eq:DCT}). It may be computed efficiently, using a FFT. If all time steps have the same internal structure, the computation of the $w_k$ has to be performed only once.

\section{The tolerance parameters}
The tolerance parameter of the convergence in the optimization procedures is the maximal allowed relative difference of the field at the end of the process. Let us denote the tolerance parameter of the optimization as $\tau$. For the problems with the regular $J_{penal}$ (Eq.~\eqref{eq:Jpenals2s}), the convergence condition is:
\begin{align}
	& \frac{\Vert \vec{\epsilon}^{\text{ }new}-\vec{\epsilon}^{\text{ }old} \Vert}{\Vert \vec{\epsilon}^{\text{ }new} \Vert}<\tau \label{eq:tolt}\\
	& \vec{\epsilon} \equiv
	\begin{bmatrix}
		\epsilon(t_0)\\
		\epsilon(t_1)\\
		\vdots\\
		\epsilon(t_{N_t-1})
	\end{bmatrix}
\end{align}
For the problems with the new $J_{penal}$ (Eq.~\eqref{eq:Jpenalw}), the condition is:
\begin{align}
	& \frac{\Vert \vec{\bar\epsilon}^{\text{ }new}-\vec{\bar\epsilon}^{\text{ }old} \Vert}{\Vert \vec{\bar\epsilon}^{\text{ }new} \Vert}<\tau \label{eq:tolw}\\
	& \vec{\bar\epsilon} \equiv
	\begin{bmatrix}
		\bar\epsilon(\omega_0)\\
		\bar\epsilon(\omega_1)\\
		\vdots\\
		\bar\epsilon(\omega_{N_t-1})
	\end{bmatrix}
\end{align}

We also used a tolerance parameter for the propagator, unlike in the attached article. We allowed the possibility of more than one iteration in all time-steps, and not only in the first one (for efficiency, the average number of iterations should not exceed $1$ too much; this may be achieved by choosing a sufficiently small $\Delta t$, and a sufficiently large $N_c$). Let us denote the tolerance parameter of the propagator as $\zeta$. The convergence condition is:
\begin{equation}
	\frac{\Vert \mathbf{u}^{new}-\mathbf{u}^{old} \Vert}{\Vert \mathbf{u}^{old} \Vert}<\zeta
\end{equation}
where $\mathbf{u}$ is the solution vector at the edge of the time step interval. We take: \text{$\zeta=10^{-3}\,\tau$}.


\begin{thebibliography}{999}
\bibitem{David}
	D. J. Tannor, \emph{Quantum Mechanics, A Time Dependent Perspective}, University Science Books, 2003 (Chapter 16)
\bibitem{tutorial}
	J. Werschnik and E. K. U. Gross, \emph{``Quantum optimal control theory"}, J. Phys. B. At. Mol. Opt. Phys., 40 (2007)
\bibitem{Lanczos}
	C. Lanczos, \emph{The Variational Principles of Mechanics}, Dover Publications, 4'th ed., 1970 (Chapter 2)
\bibitem{AOC}
	A. E. Bryson and Y. C. Ho, \emph{Applied Optimal Control}, Ginn and Company, 1969
\bibitem{mathworks}
	\verb"http://www.mathworks.com/help/pdf_doc/optim/optim_tb.pdf"
\bibitem{Rabitz}
	A. P. Peirce, M. A. Dahleh, H. Rabitz, \emph{``Optimal control of quantum-mechanical systems: Existence, numerical approximation, and application''}, Phys. Rev. A, 37 (1988)
\bibitem{Ronnie89}
	R. Kosloff, S. A. Rice, P. Gaspard, S. Tersigni, D. J. Tannor, \emph{``Wavepacket dancing: achieving chemical selectivity by shaping light pulses''}, Chem. Phys., 139 (1989)
\bibitem{Jose}
	J. P. Palao, R. Kosloff, and C. P. Koch, \emph{``Protecting coherence in optimal control theory: State-dependent constraint approach"}, Phys. Rev. A, 77 (2008)
\bibitem{Orlov}
	D. J. Tannor, V. Kazakov, V. Orlov, \emph{``Control of photochemical branching: novel procedures for finding optimal pulses and global upper bounds''}, in \emph{Time-Dependent Quantum Molecular Dynamics}, ed. by J. Broeckhove, L. Lathouwers, Plentum press
\bibitem{genKrotov}
	Y. Maday, G. Turinici, \emph{``New formulations of the monotonically convergent quantum control algorithms"}, J. Chem. Phys., 118 (2003)
\bibitem{Ruvi}
	R. Eitan, M. Mundt, D. J. Tannor, \emph{``Optimal control with accelerated convergence: Combining the Krotov and quasi-Newton methods"}, Phys. Rev. A, 83 (2011)
\bibitem{Degani}
	I. Degani, A. Zanna, L. S{\ae}len, and R. Nepstad, \emph{``Quantum control with piecewise constant control functions"}, SIAM J. Sci. Comput., Vol 31, No. 5 (2009)
\bibitem{Skinner}
	T. E. Skinner, N. I. Gershenzon, \emph{``Optimal control design of pulse shapes as analytic functions"}, J. Magnetic Resonance, 204 (2010)
\bibitem{Serban}
	I. Serban, J. Werschnik and E. K. U. Gross, \emph{``Optimal control of time-dependent targets"}, Phys. Rev. A, 71 (2005)
\bibitem{HHGcol}
	C. Winterfeldt, C. Spielmann, G. Gerber, \emph{``Colloquium: Optimal control of high-harmonic generation''}, Reviews of Modern Physics, 80 (2008)
\bibitem{Toda}
	M. Toda, \emph{``Nonlinear lattice and soliton theory''}, IEEE Transactions on circuits and systems, cas-30 No. 8 (1983) 
\bibitem{HCl}
	E. W. Kaiser, \emph{``Dipole moment and hyperfine parameters of }H$\,^{35}$Cl\emph{ and }D$\,^{35}$Cl\emph{''}, J. Chem. Phys., 53 (1970)
\bibitem{Hillel}
	H. Tal-Ezer, R. Kosloff, I. Schaefer, \emph{``New, highly accurate propagator for the linear and non-linear Schr\"odinger equation''}, J. Sci. Comput., DOI 10.1007/s10915-012-9583-x
\bibitem{TDmethods}
	R. Kosloff, \emph{``Time-dependent quantum-mechanical methods for molecular dynamics''}, J. Phys. Chem., 92 (1988)
\bibitem{DCT}
	\verb"http://en.wikipedia.org/wiki/Discrete_cosine_transform"
\end{thebibliography}
\end{document}